\documentclass[11pts,onecolumn]{article}
\usepackage[dvips]{graphics}
\usepackage{epsfig}
\usepackage[centertags]{amsmath}
\usepackage{amsfonts}
\usepackage{amssymb}
\usepackage{amsthm}
\usepackage{newlfont}
\usepackage{hyperref}

\begin{document}
\title{Graphical law beneath each written natural language}
\author{
Anindya Kumar Biswas, Department of Physics;\\
North-Eastern Hill University, Mawkynroh-Umshing, Shillong-793022.\\
email:anindya@nehu.ac.in}
\date{\today}
\maketitle
\begin{abstract}
We study twenty four written natural languages. We draw in the log scale, 
number of words starting with a letter vs rank of the letter, both normalised. 
We find that all the graphs are of the similar type. 
The graphs are tantalisingly closer to the curves of reduced 
magnetisation vs reduced temperature for magnetic materials. 
We make a weak conjecture that a curve of magnetisation
underlies a written natural language.
\end{abstract}

\begin{section}{Introduction}
Our world is alive with languages. Some are spoken by many. Some 
are spoken by few. English is a language known to large part of the 
world. Languages are there yet to be discovered.

\noindent
Like English there are languages which use written scripts. Many 
others exist in spoken form. A subset of languages of the world use 
alphabets. A subsubset have dictionaries.
Official language in most part of the 
world is English. Dictionaries from English to a natural language
is of common availability.

\noindent
A Dictionary is where words beginning with letters of an alphabet are recorded 
in their relative abundances. Few letters are heavily used.
Few are scantily done so. Number of letters in an alphabet varies from a language 
to another language.

\noindent
Linguistics about Zipf's law and its variants has been a subject 
of interesting intense computational physics study for quite 
some time \cite{Stanley}. 
Moreover, Drożdż etal, \cite{Dro}, have observed trace of scaling
in verbs, in English and Polish languages respectively.
In this work we study the word contents along the letters of an alphabet in 
a language. We arrange the letters in an alphabet in ascending order of their ranks.
The letter with the highest number of words starting with is of rank one.
For a natural language, a dictionary from it to English, is a natural choice for this 
type of study.  

\noindent
In the next section we describe our method of study. In the following section,
we describe the standard curves of magnetisation of Ising model.
In the ensuing section, section IV, we describe our graphical results. 
Then we comment about the generality and then pre-conclude about the graphical law,
the tantalising similarity of the curves to that of curves of magnetisation, in the 
in the section V. We go on to add two more languages to 
our study in the section VI, followed by a section, section VII, on successive 
normalisations.
We end up through conclusion, acknowledgement and an appendix 
providing language datas of this module in the sections VIII, IX and X. In 
the next five modules, we extend our conjecture of graphical law to verbs, 
adverbs and adjectives, before proceeding to verify the existence of the 
graphical law in the chinese usages and in all components of 
the Lakher(Mara) language. 
\end{section}

\begin{section}{Method of study}
We take bilingual dictionaries of different languages 
(\cite{French}-\cite{Arabian}), say Spanish to 
English,  English to English, Sanskrit to English etc. Then we count pages 
associated with each letter and multiply with average number of words belonging
to a page. In the case of Webster's dictionary the variation of words from
page to page is more. For other dictionaries, variation is very less, of two to 
three words. We have counted each dictionary only once. Hence, our quantitative 
method is semi-rigorous.

\noindent
For each language, we assort the letters according to their rankings. 
We take natural logarithm of both number of words, denoted by $f$ 
and the respective rank, denoted by $k$. $k$ is a positive 
integer starting from one. Since each language has a 
letter, number of words initiating with it being very close to one, we attach 
a limiting rank, $k_{lim}$ or, $k_{d}$, and a limiting number of word to each language. 
The limiting rank is just maximum rank plus one and 
the limiting number of word is one. 
As a result both $\frac{lnf}{lnf_{max}}$ and $\frac{lnk}{lnk_{lim}}$ varies from
zero to one. Then we plot $\frac{lnf}{lnf_{max}}$ against $\frac{lnk}{lnk_{lim}}$.

\end{section}
\section{Magnetisation}
\subsection{Bragg-Williams approximation}
\noindent
Let us consider an arbitrary lattice, with each up spin assigned a value one and a down spin a value minus 
one, with an unspecified long-range order parameter defined as above by 
$L=\frac{1}{N}\Sigma_{i}\sigma_{i}$, where $\sigma_{i}$ is i-th spin, N being total number of spins. 
L can vary from minus one to one.
$N=N_{+}+N_{-}$, where $N_{+}$ is the number of up spins, $N_{-}$ is the number of down spins. 
$L=\frac{1}{N}(N_{+}-N_{-})$. As a result, $N_{+}=\frac{N}{2}(1+L)$ and $N_{-}=\frac{N}{2}(1-L)$. 
Magnetisation or, net magnetic moment , $M$ is $\mu \Sigma_{i}\sigma_{i}$ or, $\mu (N_{+}-N_{-})$ or, $\mu NL$, 
$M_{max}=\mu N$. $\frac{M}{M_{max}}=L$. $\frac{M}{M_{max}}$ is referred to as reduced magnetisation.
Moreover, the Ising Hamiltonian,\cite{Ising}, for the lattice of spins, setting $\mu$ to one, is
$-\epsilon\Sigma_{n.n}\sigma_{i}\sigma_{j}- H \Sigma_{i}\sigma_{i}$, where n.n refers to nearest neighbour pairs.

\noindent
The difference $\bigtriangleup E$ of energy if we flip an up spin to down spin is, \cite{Pathria}, 
$2\epsilon \gamma\bar{\sigma}+2 H$, where $\gamma$ is the number of nearest neighbours of a spin. 
According to Boltzmann principle,
$\frac{N_{-}}{N_{+}}$ equals $exp(-\frac{\bigtriangleup E}{k_{B}T})$, \cite{Kittel}. 
In the Bragg-Williams approximation,\cite{Bragg}, $\bar{\sigma}=L$, considered in the thermal average sense. 
Consequently, 
\begin{equation}\label{bragg}
ln\frac{1+L}{1-L}=2\frac{\gamma \epsilon L+ H}{k_{B}T} 
= 2\frac{L+ \frac{ H}{\gamma \epsilon}}{\frac{T}{\gamma \epsilon/k_{B}}}
=2\frac{L+c }{\frac{T}{T_{c}}}   
\end{equation}
 where, $c=\frac{ H}{\gamma \epsilon}$ , $T_{c}=\gamma \epsilon/k_{B}$, \cite{Lubo}.
 $\frac{T}{T_{c}}$ is referred to as reduced temperature.

\noindent
Plot of $L$ vs $\frac{T}{T_{c}}$ or, reduced magentisation vs. reduced temperature 
is used as reference curve. 
In the presence of magnetic field, $c\neq0$, the curve bulges outward. 
Bragg-Williams is a Mean Field approximation. 
This approximation holds when number of neighbours interacting with a site 
is very large, reducing the importance of local fluctuation or, local order, 
making the long-range order or, average degree of freedom as the only degree 
of freedom of the lattice. To have a feeling how this approximation leads to 
matching between experimental and Ising model prediction one can refer to 
FIG.12.12 of \cite{Pathria}.  W. L. Bragg was a professor of Hans Bethe. 
Rudlof Peierls was a friend of Hans Bethe. At the suggestion of W. L. Bragg, 
Rudlof Peierls following Hans Bethe improved the approximation scheme, applying quasi-chemical 
method.
\subsection{Bethe-peierls approximation in presence of four nearest neighbours, in absence of external magnetic field}
\begin{figure}
\centering
\includegraphics[width=13cm,height=4cm]{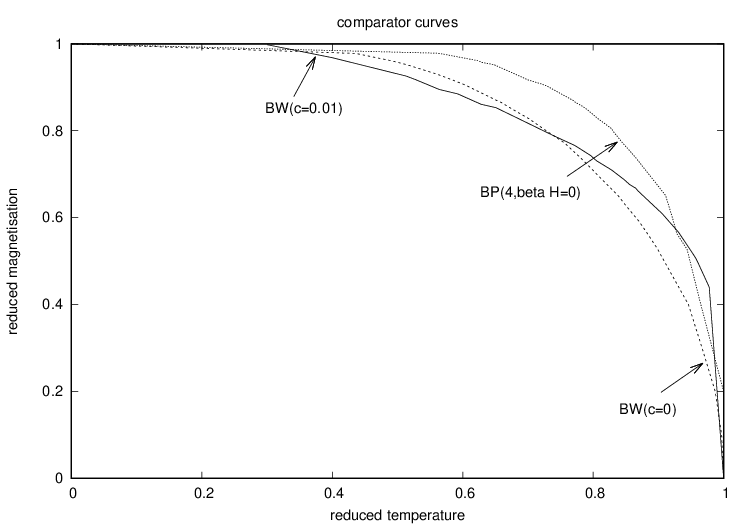}
\caption{Reduced magnetisation vs reduced temperature curves for 
Bragg-Williams approximation, in absence(dark) of and presence(inner in the top) of magnetic field, 
$c=\frac{ H}{\gamma \epsilon}=0.01$, 
and Bethe-Peierls approximation in absence of magnetic field, for four nearest neighbours (outer in the top).}
\label{Figure1}
\end{figure}

\noindent
In the approximation 
scheme which is improvement over the Bragg-Williams,
\cite{Ising},\cite{Pathria},\cite{Kittel},\cite{Bragg},\cite{Lubo}, due to Bethe-Peierls, 
\cite{Kerson}, 
reduced magnetisation 
varies with reduced temperature, for $\gamma$ neighbours, in absence of external magnetic field, as
\begin{eqnarray}\label{bethe}
\frac{ln\frac{\gamma}{\gamma-2}}{ln\frac{factor -1}
{factor^{\frac{\gamma -1}{\gamma}}-factor^{\frac{1}{\gamma}}}}=\frac{T}{T_{c}};
factor = \frac{\frac{M}{M_{max}}+1}{1-\frac{M}{M_{max}}}.
\end{eqnarray}
$ln\frac{\gamma}{\gamma-2}$ for four nearest neighbours i.e. for $\gamma=4$ is 0.693.
For a 
snapshot of different kind of magnetisation curves for magnetic materials
the reader is urged to give a google search ''reduced magnetisation vs 
reduced temperature curve''.
In the following, we describe datas generated from 
the equation(\ref{bragg}) and 
the equation(\ref{bethe}) in the table, \ref{BW}, and curves of magnetisation plotted on the 
basis of those datas.
\noindent
BW stands for reduced temperature in Bragg-Williams approximation, 
calculated from the equation(\ref{bragg}). BP(4) represents reduced temperature in the 
Bethe-Peierls approximation, for four nearest neighbours,
computed from the equation(\ref{bethe}). 
The data set is used to plot fig.\ref{Figure1}. 
Empty spaces in the table, \ref{BW}, mean corresponding point pairs were not used for plotting 
a line.
\begin{table}
\begin{center}
\resizebox{12cm}{6cm}{
\begin{tabular}{|l|l|l|l|}\hline
BW & BW(c=0.01) & BP(4,$\beta H=0$) & reduced magnetisation\\\hline
0     & 0      & 0      & 1\\\hline
0.435 & 0.439  & 0.563  & 0.978\\\hline
0.439 & 0.443  & 0.568  & 0.977\\\hline
0.491 & 0.495  & 0.624  & 0.961\\\hline
0.501 & 0.507  & 0.630  & 0.957\\\hline
0.514 & 0.519  & 0.648  & 0.952\\\hline
0.559 & 0.566  & 0.654  & 0.931\\\hline
0.566 & 0.573  & 0.7    &  0.927\\\hline
0.584 & 0.590  & 0.7    & 0.917\\\hline
0.601 & 0.607  & 0.722  & 0.907\\\hline
0.607 & 0.613  & 0.729  & 0.903\\\hline
0.653 & 0.661  & 0.770  & 0.869\\\hline
0.659 & 0.668  & 0.773  & 0.865\\\hline
0.669 & 0.676  & 0.784  & 0.856\\\hline
0.679 & 0.688  & 0.792  & 0.847\\\hline
0.701 & 0.710  & 0.807  & 0.828\\\hline
0.723 & 0.731  & 0.828  & 0.805\\\hline
0.732 & 0.743  & 0.832  & 0.796\\\hline
0.756 & 0.766  & 0.845  & 0.772\\\hline
0.779 & 0.788  & 0.864  & 0.740\\\hline
0.838 & 0.853  & 0.911  & 0.651\\\hline
0.850 & 0.861  & 0.911  & 0.628\\\hline
0.870 & 0.885  & 0.923  & 0.592\\\hline
0.883 & 0.895  & 0.928  & 0.564\\\hline
0.899 & 0.918  & ×      & 0.527\\\hline
0.904 & 0.926 & 0.941  & 0.513\\\hline
0.946 & 0.968 & 0.965  & 0.400\\\hline
0.967 & 0.998 & 0.965  & 0.300\\\hline
0.987 & ×     & 1     & 0.200\\\hline
0.997 & ×     & 1     & 0.100\\\hline
1     & 1     & 1      & 0 \\\hline
\end{tabular}
}
\end{center}
\caption{Reduced magnetisation vs reduced temperature datas for 
Bragg-Williams approximation, in absence of and in presence of magnetic field, 
$c=\frac{ H}{\gamma \epsilon}=0.01$, 
and Bethe-Peierls approximation in absence of magnetic field, for four nearest neighbours .}
\label{BW}
\end{table}
\subsection{Bethe-peierls approximation in presence of four nearest neighbours, in presence of external magnetic field}
\begin{figure}
\centering
\includegraphics[width=13cm,height=4.5cm]{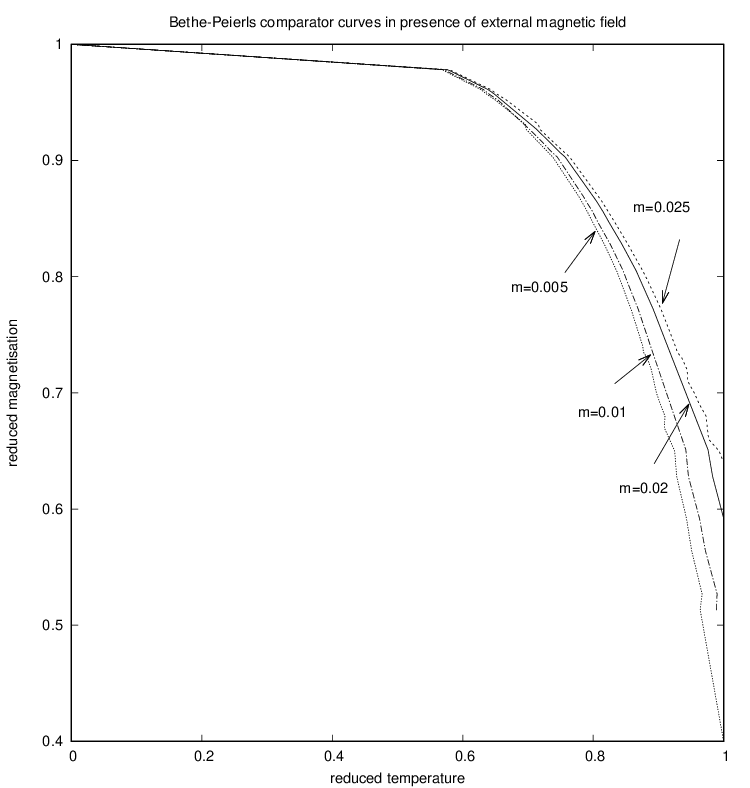}
\caption{Reduced magnetisation vs reduced temperature curves for 
Bethe-Peierls approximation in presence of little external magnetic fields, for four nearest neighbours, 
with $\beta H=2m$.}
\label{Figure2}
\end{figure}

\noindent
In the Bethe-Peierls approximation 
scheme , 
\cite{Kerson}, 
reduced magnetisation 
varies with reduced temperature, for $\gamma$ neighbours, in presence of external magnetic field, as
\begin{eqnarray}
\frac{ln\frac{\gamma}{\gamma-2}}{ln\frac{factor -1}
{e^{\frac{2\beta H}{\gamma}} factor^{\frac{\gamma -1}{\gamma}}-e^{-\frac{2\beta H}{\gamma}}factor^{\frac{1}{\gamma}}}}=\frac{T}{T_{c}};
factor = \frac{\frac{M}{M_{max}}+1}{1-\frac{M}{M_{max}}}.
\end{eqnarray} 
$ln\frac{\gamma}{\gamma-2}$ for four nearest neighbours i.e. for $\gamma=4$ is 0.693. For four neighbours, 
\begin{eqnarray}\label{bethem}
\frac{0.693}{ln\frac{factor -1}
{e^{\frac{2\beta H}{\gamma}} factor^{\frac{\gamma -1}{\gamma}}-e^{-\frac{2\beta H}{\gamma}}factor^{\frac{1}{\gamma}}}}=\frac{T}{T_{c}};
factor = \frac{\frac{M}{M_{max}}+1}{1-\frac{M}{M_{max}}}.
\end{eqnarray} 
\noindent
In the following, we describe datas in the table, \ref{BP}, generated from  
the equation(\ref{bethem}) and curves of magnetisation plotted on the 
basis of those datas.
BP(m=0.03) stands for reduced temperature in Bethe-Peierls approximation, 
for four nearest neighbours, in presence of a variable external magnetic field, H, such that $\beta H =0.06$.
calculated from the equation(\ref{bethem}). 
BP(m=0.025) stands for reduced temperature in Bethe-Peierls approximation, 
for four nearest neighbours, in presence of a variable external magnetic field, H, such that $\beta H =0.05$.
calculated from the equation(\ref{bethem}). 
BP(m=0.02) stands for reduced temperature in Bethe-Peierls approximation, 
for four nearest neighbours, in presence of a variable external magnetic field, H, such that $\beta H =0.04$.
calculated from the equation(\ref{bethem}). 
BP(m=0.01) stands for reduced temperature in Bethe-Peierls approximation, 
for four nearest neighbours, in presence of a variable external magnetic field, H, such that $\beta H =0.02$.
calculated from the equation(\ref{bethem}).
BP(m=0.005) stands for reduced temperature in Bethe-Peierls approximation, 
for four nearest neighbours, in presence of a variable external magnetic field, H, such that $\beta H =0.01$.
calculated from the equation(\ref{bethem}).
The data set is used to plot fig.\ref{Figure2}. 
Empty spaces in the table, \ref{BP}, mean corresponding point pairs were not used for plotting 
a line.

\begin{table}
\begin{center}
\resizebox{15cm}{6cm}{
\begin{tabular}{|l|l|l|l|l|l|}\hline
BP(m=0.03)&BP(m=0.025)&BP(m=0.02) &BP(m=0.01)&BP(m=0.005) & reduced magnetisation\\\hline
0     &0  &0     & 0      & 0      & 1\\\hline
0.583 &0.580   &0.577 & 0.572  & 0.569  & 0.978\\\hline
0.587 &0.584& 0.581 & 0.575  & 0.572  & 0.977\\\hline
0.647 &0.643& 0.639 & 0.632  & 0.628  & 0.961\\\hline
0.657 &0.653& 0.649 & 0.641  & 0.637  & 0.957\\\hline
0.671 &0.667&      & 0.654  & 0.650  & 0.952\\\hline
      &0.716&      &        & 0.696  & 0.931\\\hline
0.723 &0.718&0.713 & 0.702  & 0.697  & 0.927\\\hline
0.743 &0.737&0.731 & 0.720  & 0.714  & 0.917\\\hline
0.762 &0.756&0.749 & 0.737  & 0.731  & 0.907\\\hline
0.770 &0.764&0.757 & 0.745  & 0.738  & 0.903\\\hline
0.816 &0.808&0.800 & 0.785  & 0.778  & 0.869\\\hline
0.821 &0.813&0.805 & 0.789  & 0.782  & 0.865\\\hline
0.832 &0.823&0.815 & 0.799  & 0.791  & 0.856\\\hline
0.841 &0.833&0.824 & 0.807  & 0.799  & 0.847\\\hline
0.863 &0.853&0.844 & 0.826  & 0.817  & 0.828\\\hline
0.887 &0.876&0.866 & 0.846  & 0.836  & 0.805\\\hline
0.895 &0.884&0.873 & 0.852  & 0.842  & 0.796\\\hline
0.916 &0.904&0.892 & 0.869  & 0.858  & 0.772\\\hline
0.940 &0.926&0.914 & 0.888  & 0.876  & 0.740\\\hline
      &0.929&&      & 0.877  & 0.735\\\hline
      &0.936 &&      & 0.883  & 0.730\\\hline
      &0.944&&        & 0.889  & 0.720\\\hline
      &0.945&&         &        & 0.710\\\hline
      &0.955&&         & 0.897   & 0.700\\\hline
      &0.963&&         & 0.903   & 0.690\\\hline
      &0.973&&         & 0.910   & 0.680\\\hline
      &&&         & 0.909   & 0.670\\\hline
      &0.993&&         & 0.925  & 0.650\\\hline
      &&0.976 &0.942 &        & 0.651\\\hline
      &1.00& &         &        & 0.640\\\hline
      &&0.983& 0.946  & 0.928  & 0.628\\\hline
      &&1.00& 0.963  & 0.943  & 0.592\\\hline
      &&& 0.972  & 0.951  & 0.564\\\hline
      &&& 0.990  & 0.967  & 0.527\\\hline
      &&&  & 0.964  & 0.513\\\hline
      &&&1.00  &        &0.500\\\hline
      &&&  & 1.00   & 0.400\\\hline
      &&& &        & 0.300\\\hline
      &&& ×     &     & 0.200\\\hline
      &&& ×     &      & 0.100\\\hline
      &&&      &       & 0 \\\hline
\end{tabular}
}
\end{center}
\caption{Bethe-Peierls approx. in presence of little external magnetic fields}
\label{BP}
\end{table}
\clearpage

\begin{figure}
\centering
\includegraphics[width=13cm,height=4.5cm]{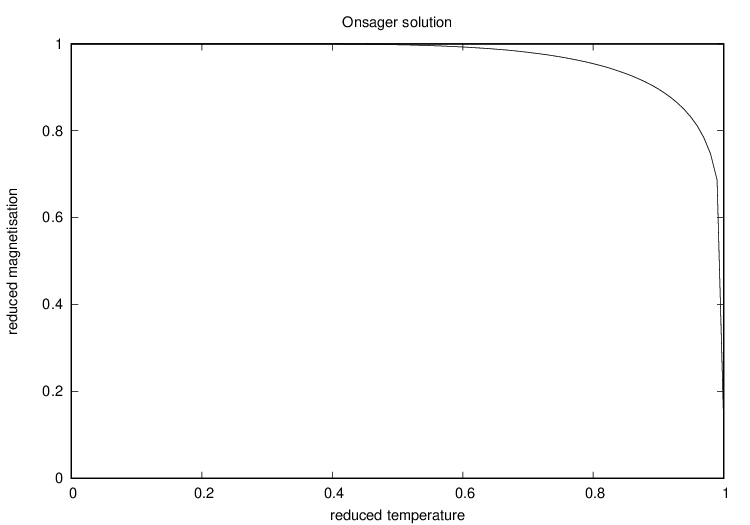}
\caption{Reduced magnetisation vs reduced temperature curves for 
exact solution of two dimensional Ising model, due to Onsager, in absence of external 
magnetic field}
\label{Figure3}
\end{figure}

\noindent
In the case of exact, unapproximated, solution of two dimensional Ising model, by
Onsager, reduced spontaneous magnetisation varies with reduced temperature as,
\cite{Yang},  
\begin{equation}
\frac{M}{M_{max}}=
[1-(sinh\frac{.8813736}{\frac{T}{T_{c}}})^{-4}]^{1/8}. \nonumber
\end{equation}
Graphically, the Onsager solution appears as in fig.\ref{Figure3}.
\clearpage
\subsection{Spin-Glass}
\noindent
In the case coupling between( among) the spins, not necessarily n.n, for the Ising model is( are) random, we get Spin-Glass,
\cite{SpinGlass1, SpinGlass7, SpinGlass2, SpinGlass3, SpinGlass4, SpinGlass5, SpinGlass6}. 
\noindent
When a lattice of spins randomly coupled and in an external magnetic field, goes over to the Spin-Glass 
phase, magnetisation increases steeply like $\frac{1}{T-T_{c}}$ upto the the phase transition temperature, 
followed by very little increase,\cite{SpinGlass1, SpinGlass6}, in magnetisation, as the ambient temperature continues to drop.
This happens at least in the replica approach of the Spin-Glass theory, \cite{SpinGlass3, SpinGlass4}.

\noindent
For a 
snapshot of different kind of magnetisation curves for magnetic materials
 the reader is urged to give a google search ''reduced magnetisation vs 
reduced temperature curve''. Moreover, whenever we write Bragg, it will 
stand for Bragg-Williams approximation; Bethe will represent Bethe-Peierls 
approximation in absence of external magnetic field.

\begin{section}{Results}
English, German, French, Italian, Spanish, Russian are some languages from Europe.
Turkmen is the language of Turkmenistan. In South-Africa, a version of English
termed South-African English is in vogue. Arabic is one major language in the
Middle-East. In South-East Asia, Sanskrit, Urdu, Hindi, Kachin,
Tibetian, Sinhalese, Nepali, Kannada, Assamese are among the main languages. 
In India few tribal languages are Onge in little Andaman; Taraon, Abor-Miri 
in Arunachal Pradesh; Lushai(Mizo) in Mizoram; Khasi, Garo in Meghalaya.
 
\noindent
We study dictionaries(\cite{French}-\cite{Arabian}) of these languages. 
We put our results in the form of plots in separate 
panels,(\ref{Figure4}-\ref{Figure7}). 
On each plot we superimpose the best fit curve of magnetisation 
as a comparator. On the reference curve by reference curve basis, 
we put our results 
in graphical form, here, in different subsections. In the panel, fig.\ref{Figure8}, 
we superimpose Onsager solution on six selected languages for closer 
scrutiny. Datas related to analysis and for the plots of the languages are 
appended in tabular forms in the appendix section.
\clearpage
\begin{subsection}{Bragg-Williams approximation}
We start our results with languages very close to 
Bragg-Williams approximation line, fig.\ref{Figure4} and fig.\ref{Figure5}.
\begin{figure}
\centering
\includegraphics[width=13cm,height=10cm]{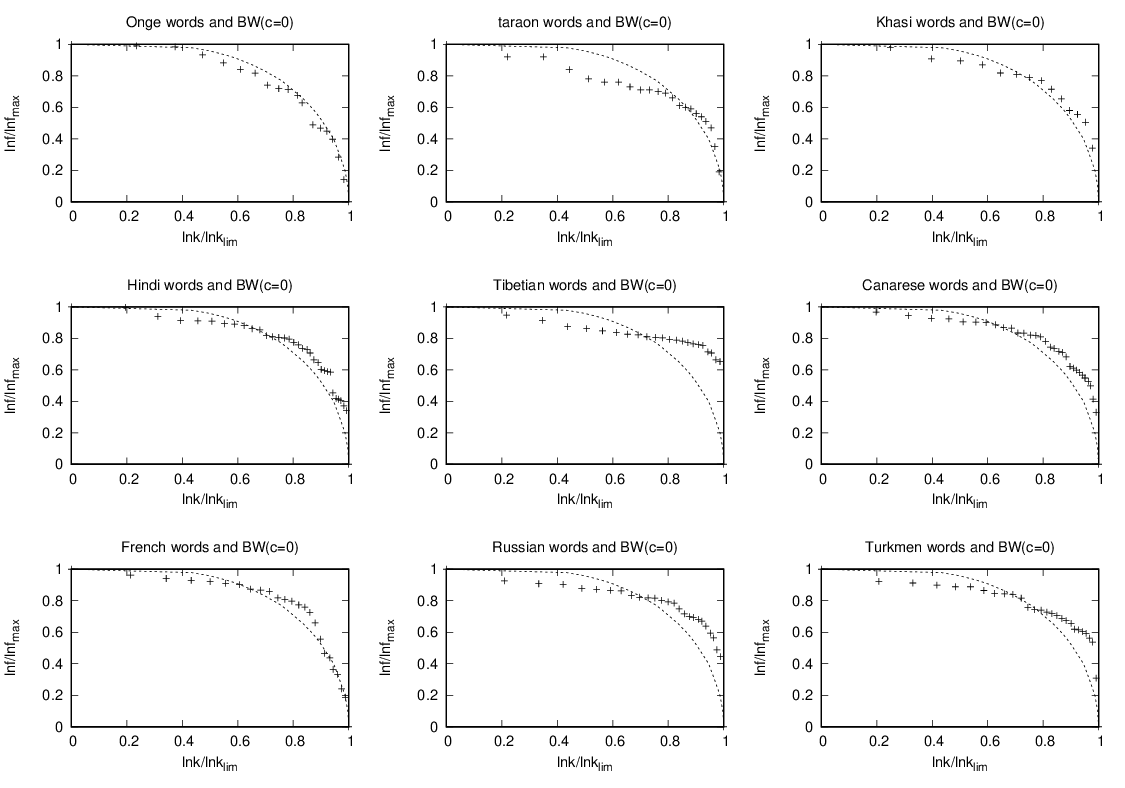}
\caption{Vertical axis is $\frac{lnf}{lnf_{max}}$ and horizontal 
axis is $\frac{lnk}{lnk_{lim}}$. The $+$ points represent the words of the languages 
in the titles. 
The dashed line is the Bragg-Williams line, BW(c=0), in absence of external magnetic field, 
used as the reference curve.}
\label{Figure4}
\end{figure}

\begin{figure}
\centering
\includegraphics[width=13cm,height=8cm]{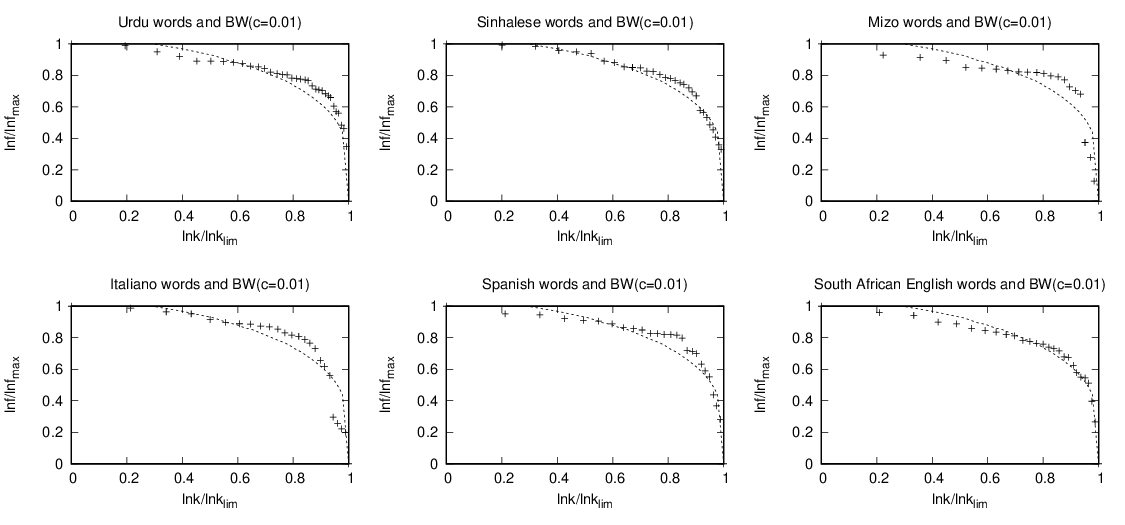}
\caption{Vertical axis is $\frac{lnf}{lnf_{max}}$ and horizontal 
axis is $\frac{lnk}{lnk_{lim}}$. The $+$ points represent words of the languages 
in the titles. The best fit curve, BW(c=0.01), is 
the Bragg-Williams curve, in presence of external magnetic field, 
$c = \frac{H}{\gamma \epsilon}=0.01$.
}
\label{Figure5}
\end{figure}
\end{subsection}
\clearpage
\begin{subsection}{Bethe-Peierls approximation}
In this subsection, we present the languages which fall on the Bethe lines, 
for different nearest neighbours, fig.\ref{Figure6}and fig.\ref{Figure7}.
\begin{figure}
\centering
\includegraphics[width=14cm,height=5cm]{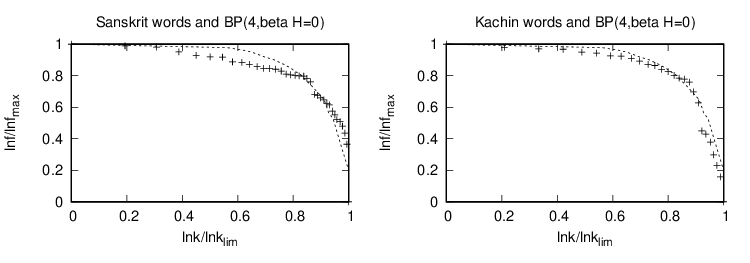}
\caption{Vertical axis is $\frac{lnf}{lnf_{max}}$ and horizontal 
axis is $\frac{lnk}{lnk_{lim}}$. The $+$ points represent the words of the languages, 
in the titles. BP(4,beta H=0) is the Bethe-Peierls line, for four nearest neighbours,  
in absence of external magnetic field.
For Kachin and sanskrit, 
the best fit curve is BP(4, $\beta H=0$).
}
\label{Figure6}
\end{figure}
We note, Bethe curve for four nearest neighbours, is falling on the 
Kachin language completely.
\begin{figure}
\centering
\includegraphics[width=14cm,height=10cm]{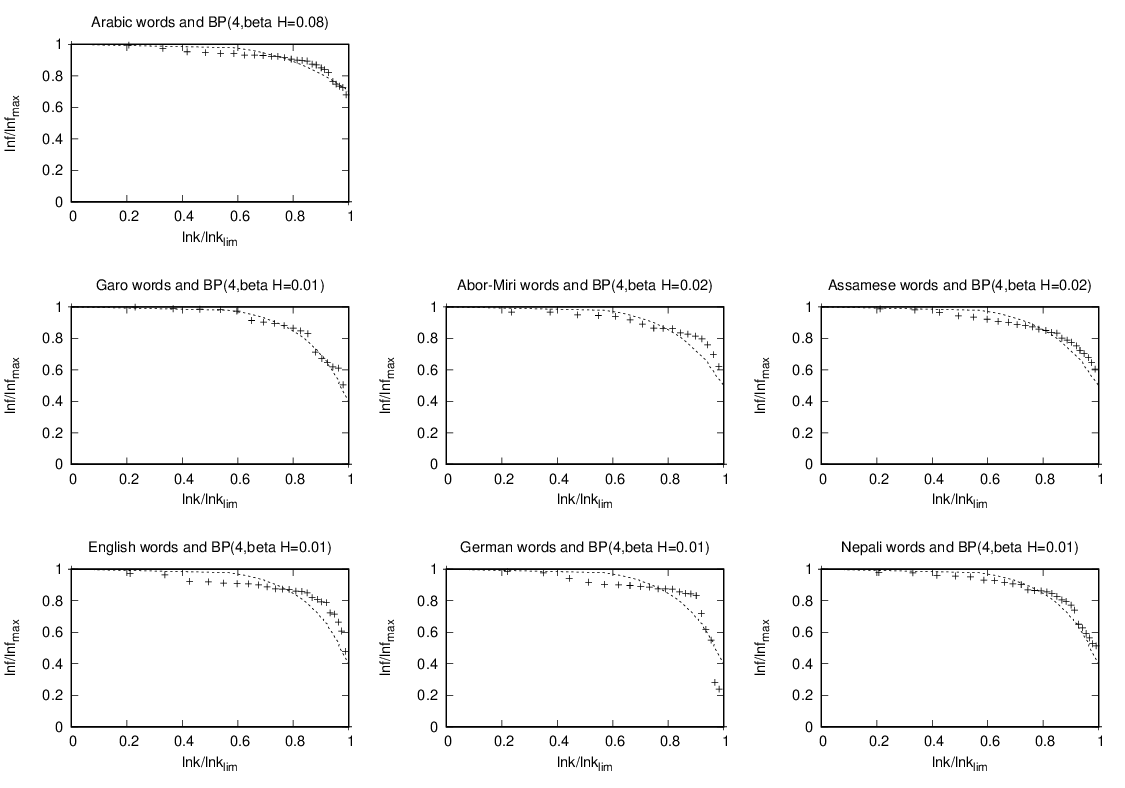}
\caption{Vertical axis is $\frac{lnf}{lnf_{max}}$ and horizontal 
axis is $\frac{lnk}{lnk_{lim}}$. 
The $+$ points represent the words of the languages in the title. 
BP(4, $\beta H=0.01$) is the Bethe-Peierls curve in
presence of four nearest neighbours and little magnetic field, $\beta H = 0.01$. 
For Garo, Nepali, English and German words, the best fit curve is BP(4, $\beta H=0.01$).
For Abor-Miri and Assamese words, the best fit curve is BP(4, $\beta H=0.02$).
For arabian, the best fit curve is BP(4, $\beta H=0.08$). }
\label{Figure7}
\end{figure}
\end{subsection}

\clearpage
\begin{subsection}{Exact, Onsager Solution}
In this subsection, we present the curves of the languages in the fig.\ref{Figure8}
which are almost tending to the exact solution of Ising model, in  
absence of magnetic field.
 \begin{figure}
\centering
\includegraphics[width=13cm,height=11cm]{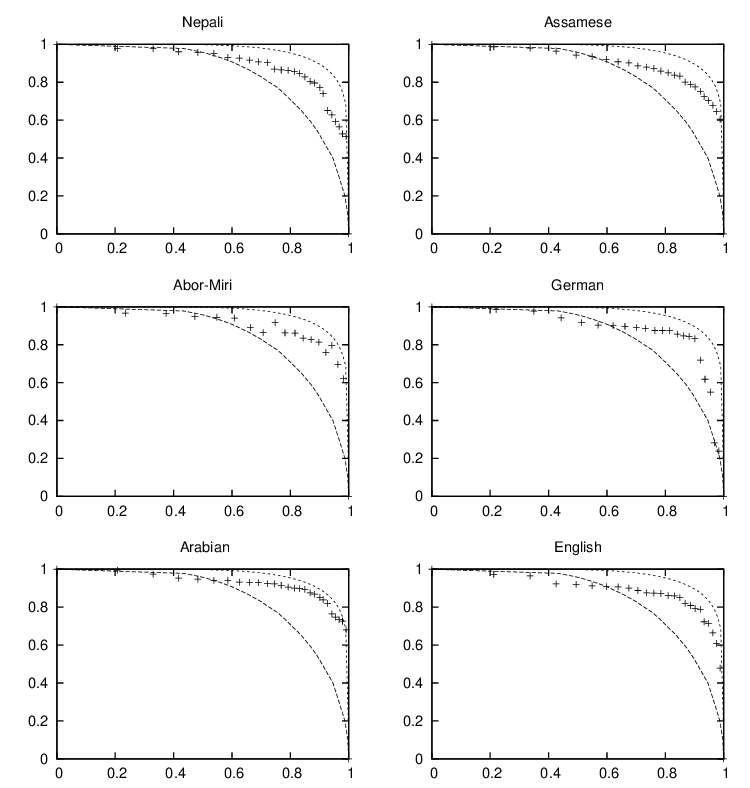}
\caption{The $+$ points represent the languages in the title. 
Vertical axis is $\frac{lnf}{lnf_{max}}$ and horizontal 
axis is $\frac{lnk}{lnk_{lim}}$. The upper line
refers to the Onsager solution. 
The lower line is for the Bragg-Williams approximation.}
\label{Figure8}
\end{figure}
We observe that Arabian comes closest to the Onsager solution.
\end{subsection}
\clearpage
\begin{subsection}{Spin glass}
In this subsection, we collect the languages, the graphical behaviours 
of which come  closer to the disordered Ising model in presence of magnetic 
field, rather than ordered Ising model in absence of magnetic field.
\begin{figure}
\centering
\includegraphics[width=13cm,height=10cm]{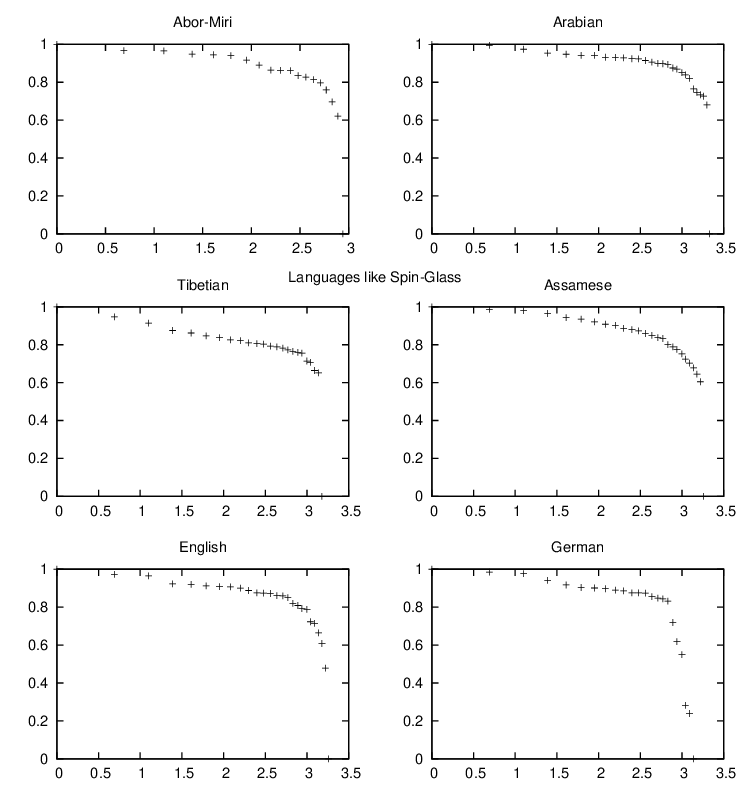}
\caption{The $+$ points represent the languages in the title. 
Vertical axis is $\frac{lnf}{lnf_{max}}$ and horizontal 
axis is $lnk$.}
\label{Figure9}
\end{figure}

\noindent
The plots in the figure fig.\ref{Figure9} show interesting features. Coming
from the far end of $lnk$ it shows steep rise upto a point, followed by very little 
increase as we decrease $lnk$. The steep rise part can be thought of as segment 
of rectangular hyperbolic growth, as is characteristic of paramagnetic 
magnetisation as we decrease temperature, in presence of a 
constant external magnetic field. The near-constant part is the 
analogue of Spin-Glass phase,(\cite{SpinGlass1}-\cite{SpinGlass6}). 
The turning point in each plot is the analogue of 
Spin-Glass phase transition temperature.
Hence, $\frac{lnf}{lnf_{max}}$ vs $lnk$ plot for each language above 
appears to be 
equivalent to a magnetisation vs temperature plot for a Spin-Glass in 
an external magnetic field.
\clearpage
\end{subsection}
\begin{subsection}{Classification}
From our results, we can tentatively make a classification of the 
twenty four languages we have studied,
\begin{itemize}
\item[a.]
The languages which underlie the Bragg-williams approximation magnetisation curve, BW(c=0),  
in absence of external magnetic field are 
French, Khasi, Hindi, Onge, Taraon, Russian, Turkmen, 
Canarese and Tibetian respectively.
\item[b.]
The languages which underlie the Bragg-williams approximation magnetisation curve, BW(c=0.01),  
in presence of external magnetic field are 
Urdu, South-African English, Spanish, sinhalese, Mizo and Italiano respectively.
\item[c.]
The language which fall under Bethe-Peierls approximation magnetisation curve for four nearest neighbours in 
absence of external magnetic field, BP(4, $\beta H=0$) are Sanskrit and Kachin.
\item[d.]
The languages which underlie  Bethe-Peierls approximation magnetisation curve, BP(4, $\beta H=0.01$), 
for four nearest neighbours in 
presence of external magnetic field, $\beta H=0.01$,  are 
English, German, Nepali and Garo.
\item[e.]
Assamese and Abor-Miri underlie  Bethe-Peierls approximation magnetisation curve, BP(4, $\beta H=0.02$), 
for four nearest neighbours in 
presence of external magnetic field, $\beta H=0.02$.
\item[f.]
Arabian is the only language with $\frac{lnf}{lnf_{max}}$ 
vs $\frac{lnk}{lnk_{lim}}$ curve having 
Bethe-Peierls approximation magnetisation curve, BP(4, $\beta H=0.08$), 
for four nearest neighbours in 
presence of external magnetic field, $\beta H=0.08$
as the best fit.
\item[g.]
It seems German, English languages qualify better to be like Spin-Glass, rather
than the Bethe-Peierls approximation magnetisation curve. We
require a serious study for this point.
\end{itemize}
\end{subsection}
\clearpage
\begin{subsection}{Towards error analysis}
An unabridged dictionary is where full stock of words of a language 
at an instant in its evolution is expected to be fully recorded.
There are two kinds of errors about a data, here number of words 
starting with an alphabet. Lexicographical uncertainity and counting 
error. This two adds up. Lexicographical uncertainity is the least 
, ideally zero, in the case of an unabridged dictionary. 
It's one sided, a systematic error. On the 
otherhand, counting errors, random in nature, are of two types. If one counts all 
the words, then it is personal random error. It tends to zero, 
if the counting is done large number of times. If words are 
counted the way we have done, by counting pages and multiplying by 
the average number of words per page, then one has to find most 
probable error from a sample of counted number of words per page.

\noindent
We have taken dictionaries of various formats, hence we do not have 
any control over Lexicographical uncertainity. Moreover, we have 
counted average number of words per page from a small sample set.
Hence, dispersion and therefrom most probable error is not 
reliable. In case of few Webster dictionaries dispersion is high, 
but for others it's very small. But those being of abridged or, concise or,
pocket varities, Lexicographical uncertainities are expected to be higher. 
Hence, we dispense with error analysis and putting error bar and remain within the 
contour of semi-rigorous approach.
But since we take fraction of natural logarithms, these uncertainities 
smoothen out to some extent. 
\noindent
In the case of rigorous treatment, one will go in the following way.
If we denote $\frac{lnf}{lnf_{max}}$ by $y$, then for $y\neq1$, 
\begin{equation}
\delta y= \pm \delta y_{counting} +\delta y_{Lexicographical}, \nonumber\\
\end{equation}
\begin{equation}
\delta y_{counting}=  \frac{p}{average} \frac{1}{lnf_{max}}(1+y), \nonumber\\
\end{equation}
where,
the probable error, $p$, in the average is equal to 
0.8453 (dispersion/square root of 
the number of different pages counted to get to the average),(\cite{prac}).

\end{subsection}

\end{section}

\begin{section}{Pre-conclusion}
Languages seem to follow the  one or, another curve of magnetisations to 
a high degree.
Hence, we tend to conjecture, behind each written language 
there is a curve of magnetisation.
At least for the languages, we have studied, with the proviso,
\begin{eqnarray}
\frac{lnf}{lnf_{max}} \longleftrightarrow \frac{M}{M_{max}},\nonumber\\
lnk\longleftrightarrow T.\nonumber
\end{eqnarray}
\end{section}

\clearpage
\begin{section}{Two more languages}
In Nagaland, one state in North-Eastern India, there are two tribes among many,
distinct from each other linguistically, are Lotha and Sema. We study their 
dictionaries(\cite{Lotha},\cite{Sema}) and show our results in the following 
panel. We subject our results vis a vis our conjecture.

\noindent
In the panel to follow we see that our conjecture, in the 
strong form, is valid for the Lotha
language. But the Sema language deviates too much from the reference curve, in the 
topmost row  of the figure, fig.\ref{Figure10}. We then ignore the letter with 
the highest number of words and redo the plot, normalising the $lnf$'s with 
next-to-maximum $lnf$, and starting from $k=2$. We see, to our surprise, that 
the resulting curve, second row first column, almost falls on the Bragg-Williams line.
We do little more graphical study of the Sema language and the results are
shown in the rest three plots of the panel, in case it helps a curious reader.

We surmise from here two things.
The letter with the rank one, here it is $a$, is simply inadequate to describe 
all the words beginning with it. The language requires two letters there. Or,
the conjecture, in the strong form, is valid provided we go up to rank2 and normalise 
with next-to-maximum, $lnf_{next-to-maximum}$. 

\begin{figure}
\centering
\includegraphics{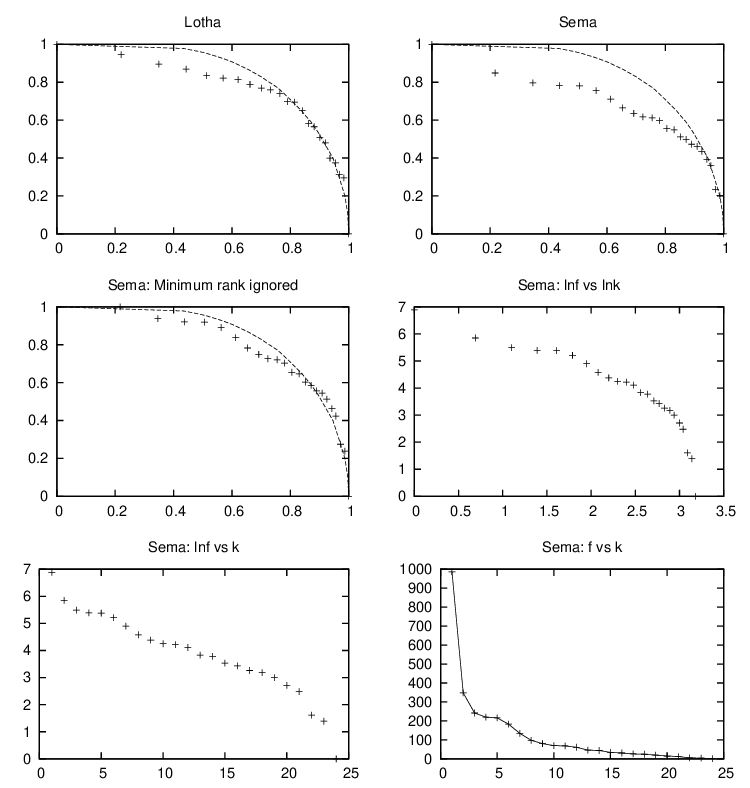}
\caption{The $+$ points represent the languages in the title, 
(\cite{Lotha},\cite{Sema}). 
The dashed line is for the Bragg-Williams approximation. In the first row 
in the two unmentioned
plots, vertical axis is $\frac{lnf}{lnf_{max}}$ and horizontal axis 
is $\frac{lnk}{lnk_{lim}}$. In the second row, first column,  
unmentioned plot, vertical axis is $\frac{lnf}{lnf_{nextmax}}$ 
and horizontal axis is $\frac{lnk}{lnk_{lim}}$.}
\label{Figure10}
\end{figure}
\end{section}

\clearpage
\begin{section}{Successive Normalisation}
\begin{figure}
\centering
\includegraphics[width=12cm,height=10cm]{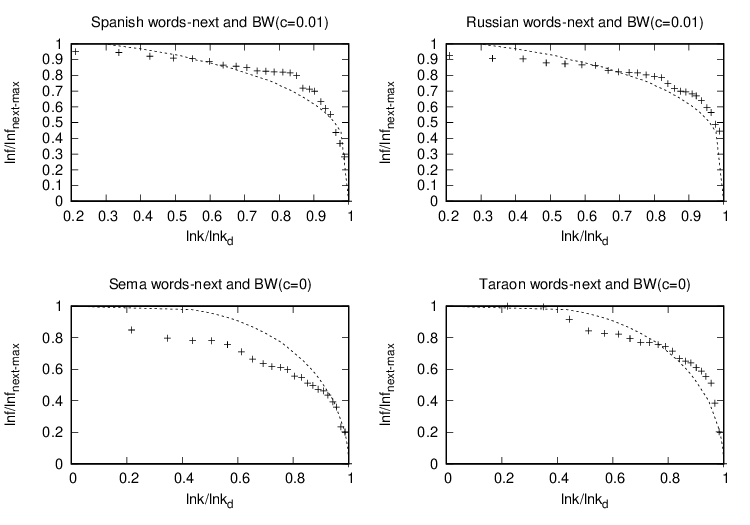}
\caption{The $+$ points represent the words of the languages in the title.
$k_{d}=k_{lim}$. BW(c=0.01) is Bragg-Williams curve in presence of magnetic field, $c = \frac{H}{\gamma \epsilon}=0.01$.
Vertical axis is $\frac{lnf}{lnf_{nextmax}}$ 
and horizontal axis is $\frac{lnk}{lnk_{lim}}$. }
\label{Figure11}
\end{figure}
\noindent
We note that in case of languages other than Sema, $lnf$ and $lnf_{next-to-maximum}$
are almost the same. 
The four languages, among the twenty four plus two languages 
we have studied, which get better fit, see fig.\ref{Figure11}.
For the languages, in which the fit with curves of magnetisation is 
not that well for normalisation with $f_{max}$ or, $f_{next-max}$, we 
do higher order normalisation with $f_{nnmax}$ and $f_{nnnmax}$ and 
try to see whether fit improves and if so whether two consecutive fits 
converge. We notice that this does happen. We describe our results in 
fig.\ref{Figure12}-fig.\ref{Figure14} and the inferences 
in the tables \ref{mag1}-\ref{mag3}.

\begin{figure}
\centering
\includegraphics[width=13cm,height=10cm]{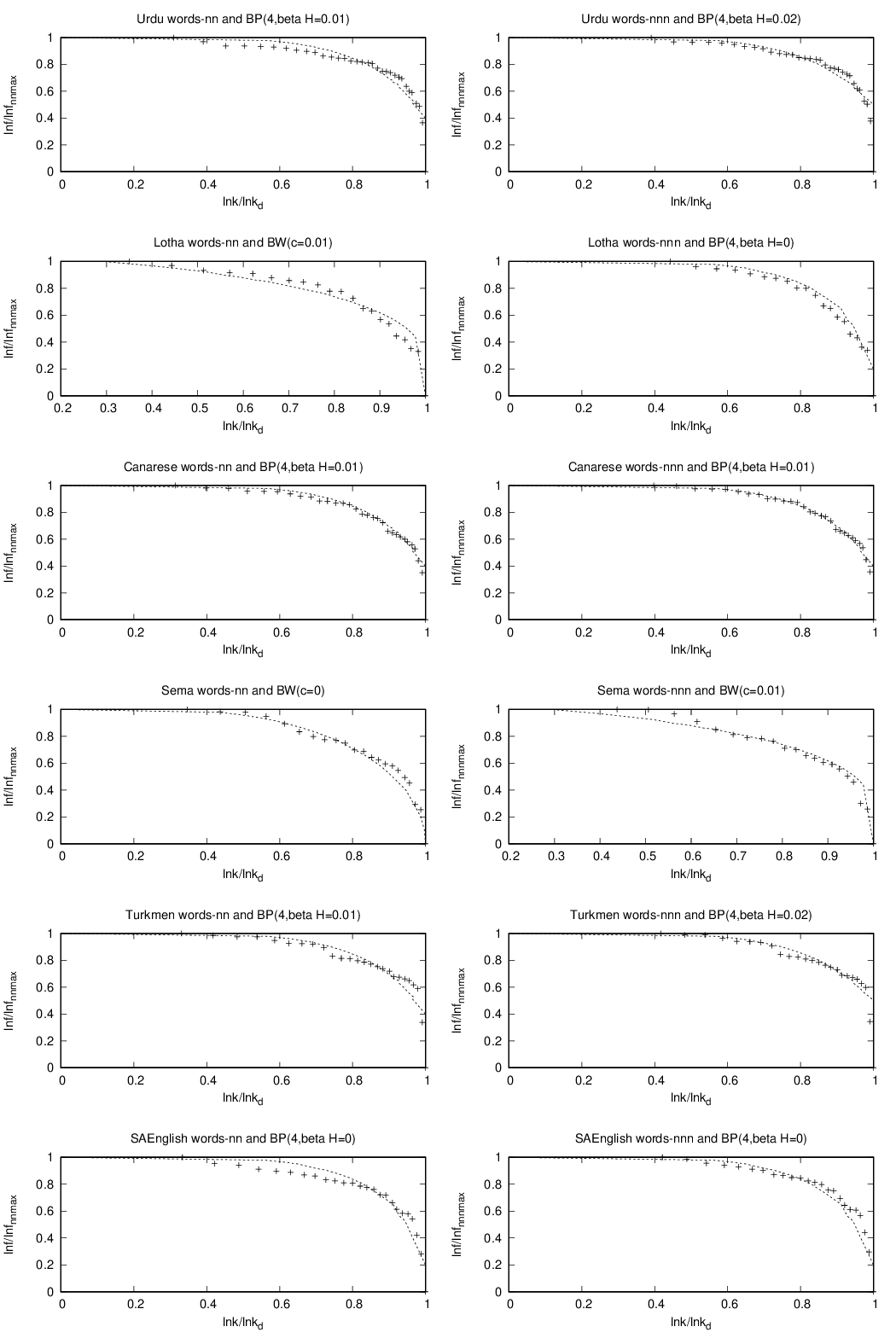}
\caption{The $+$ points represent the words of the languages in the title. 
$k_{d}=k_{lim}$. BP(4, $\beta H=0.01$) is the Bethe-Peierls curve in
presence of four nearest neighbours and little magnetic field, $\beta H = 0.01$. 
BW(c=0.01) is Bragg-Williams curve in presence of magnetic field , $c = \frac{H}{\gamma \epsilon}=0.01$.
For Urdu words, among the two plots, best fit curve is BP(4, $\beta H=0.02$) and for 
$\frac{lnf}{lnf_{nnnmax}}$ vs $\frac{lnk}{lnk_{d}}$. 
For Lotha words, among the two plots, best fit curve is BW(c=0.01) and for 
$\frac{lnf}{lnf_{nnmax}}$ vs $\frac{lnk}{lnk_{d}}$.
For Canarese words, among the two plots, best fit curve is BP(4, $\beta H=0.01$) and for 
$\frac{lnf}{lnf_{nnnmax}}$ vs $\frac{lnk}{lnk_{d}}$. 
For Sema words, among the two plots, best fit curve is BW(c=0.01) and for 
$\frac{lnf}{lnf_{nnnmax}}$ vs $\frac{lnk}{lnk_{d}}$.
For Turkmen words, among the two plots, best fit curve is BP(4, $\beta H=0.02$) and for 
$\frac{lnf}{lnf_{nnnmax}}$ vs $\frac{lnk}{lnk_{d}}$.
For SAEnglish words, among the two plots, best fit curve is BP(4, $\beta H=0$) and for 
$\frac{lnf}{lnf_{nnmax}}$ vs $\frac{lnk}{lnk_{d}}$.}
\label{Figure12}
\end{figure}
\begin{figure}
\centering
\includegraphics[width=13cm,height=10cm]{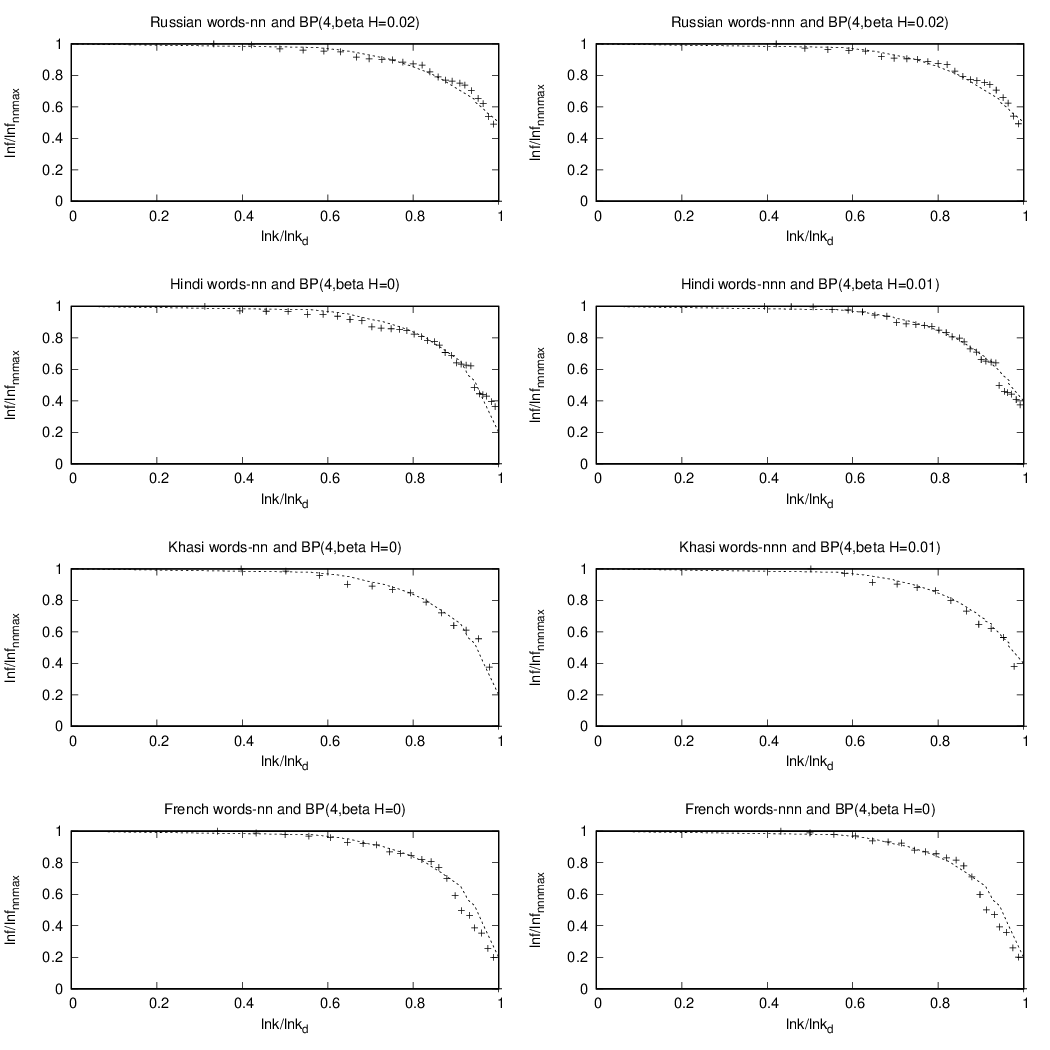}
\caption{The $+$ points represent the words of the languages in the title. 
$k_{d}=k_{lim}$. BP(4, $\beta H=0.01$) is the Bethe-Peierls curve in
presence of four nearest neighbours and little magnetic field, $\beta H = 0.01$. 
For Russian words, among the two plots, best fit curve is BP(4, $\beta H=0.02$) and for 
$\frac{lnf}{lnf_{nnmax}}$ vs $\frac{lnk}{lnk_{d}}$. 
For Hindi words, among the two plots, best fit curve is BP(4, $\beta H=0$) and for 
$\frac{lnf}{lnf_{nnmax}}$ vs $\frac{lnk}{lnk_{d}}$.
For Khasi words, among the two plots, best fit curve is BP(4, $\beta H=0.01$) and for 
$\frac{lnf}{lnf_{nnnmax}}$ vs $\frac{lnk}{lnk_{d}}$. 
For French words, among the two plots, best fit curve is BP(4, $\beta H=0$) and for 
$\frac{lnf}{lnf_{nnnmax}}$ vs $\frac{lnk}{lnk_{d}}$. 
}
\label{Figure13}
\end{figure}

\begin{figure}
\centering
\includegraphics[width=13cm,height=2.7cm]{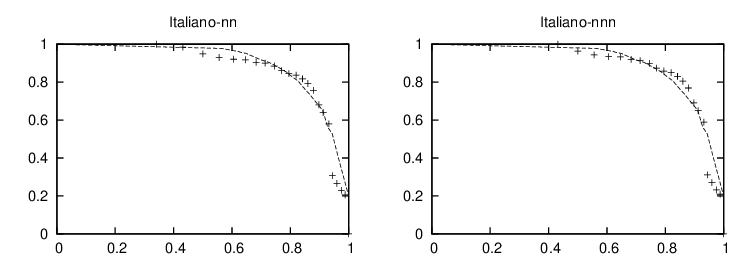}
\caption{The $+$ points represent the Italian language.
Vertical axis is  $\frac{lnf}{lnf_{nnmax}}$ ($\frac{lnf}{lnf_{nnnmax}}$)
and horizontal axis is $\frac{lnk}{lnk_{lim}}$. The fit curve is 
the Bethe line, for $\gamma=4$.}
\label{Figure14}
\end{figure}
\end{section}
\clearpage
\section{looking more for Onsager solution}
In this section, we go more to look for Onsager solution into the Arabian, Abor-Miri, Tibetian, 
English, Lushai(Mizo), German, Nepali languages and present our results graphically in the 
following figures, \ref{Figure15}-\ref{Figure21}.
 \begin{figure}
\centering
\includegraphics[width=13cm,height=11cm]{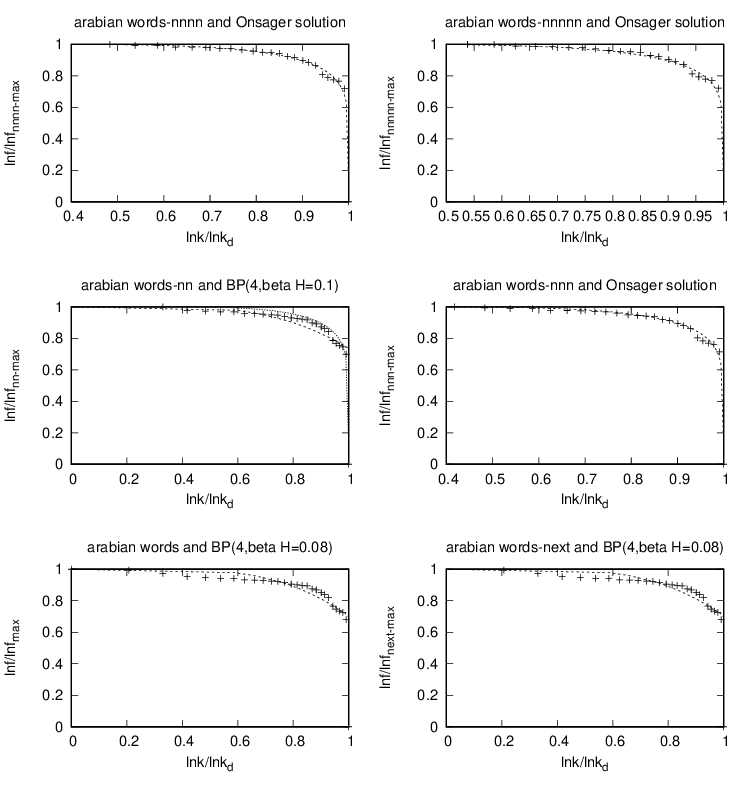}
\caption{The $+$ points represent the words of the arabian language. 
$k_{d}=k_{lim}$. BP(4, $\beta H=0.08$) is the Bethe-Peierls curve in
presence of four nearest neighbours and little magnetic field, $\beta H = 0.08$.
The upper line
refers to the Onsager solution. The pointsline goes over 
almost exactly to the Onsager solution in the topmost left figure.}
\label{Figure15}
\end{figure}
\clearpage
 \begin{figure}
\centering
\includegraphics[width=13cm,height=11cm]{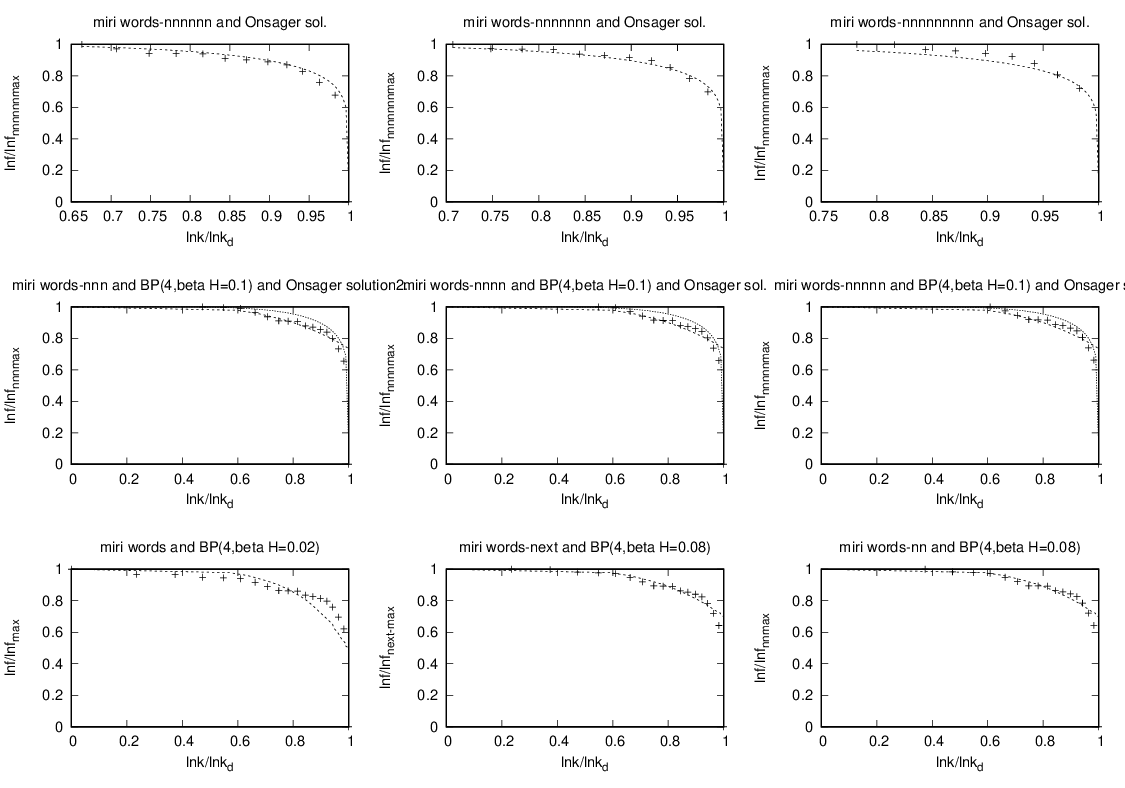}
\caption{The $+$ points represent the words of the Abor-Miri language. 
$k_{d}=k_{lim}$. 
BP(4, $\beta H=0.08$) is the Bethe-Peierls curve in
presence of four nearest neighbours and little magnetic field, $\beta H = 0.08$.
The upper lines
refer to the Onsager solution. The pointsline goes over 
almost exactly to the Onsager solution in the topmost middle figure.}
\label{Figure16}
\end{figure}
\clearpage
 \begin{figure}
\centering
\includegraphics[width=13cm,height=11cm]{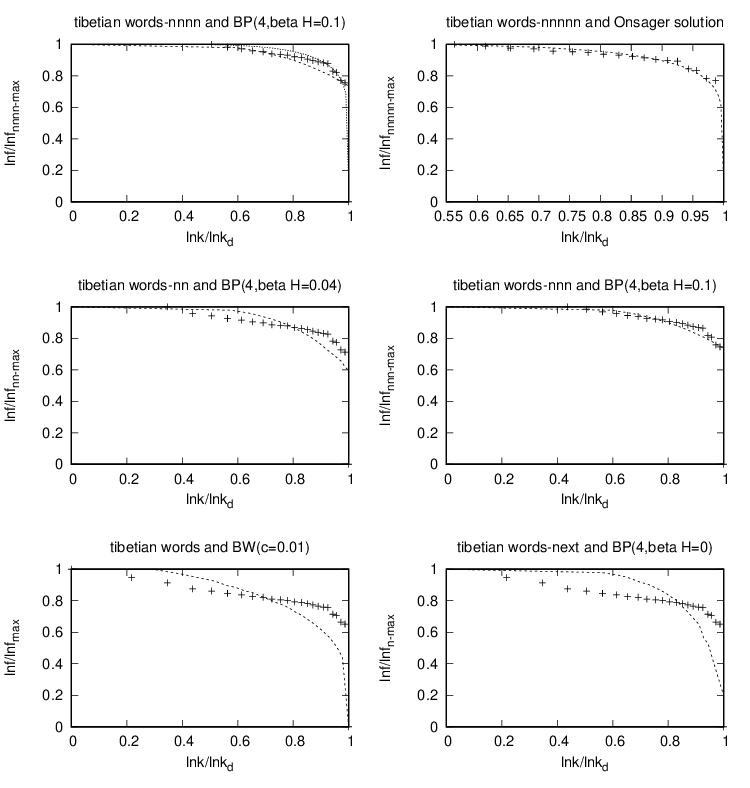}
\caption{The $+$ points represent the words of the tibetian language. 
$k_{d}=k_{lim}$. 
BW(c=0.01) is Bragg-Williams curve in presence of magnetic field, $c = \frac{H}{\gamma \epsilon}=0.01$.
BP(4, $\beta H=0.04$) is the Bethe-Peierls curve in
presence of four nearest neighbours and little magnetic field, $\beta H = 0.04$.
The upper line
refers to the Onsager solution. The pointsline goes over 
almost to the Onsager solution in the topmost right figure.}
\label{Figure17}
\end{figure}
\clearpage
\begin{figure}
\centering
\includegraphics[width=13cm,height=11cm]{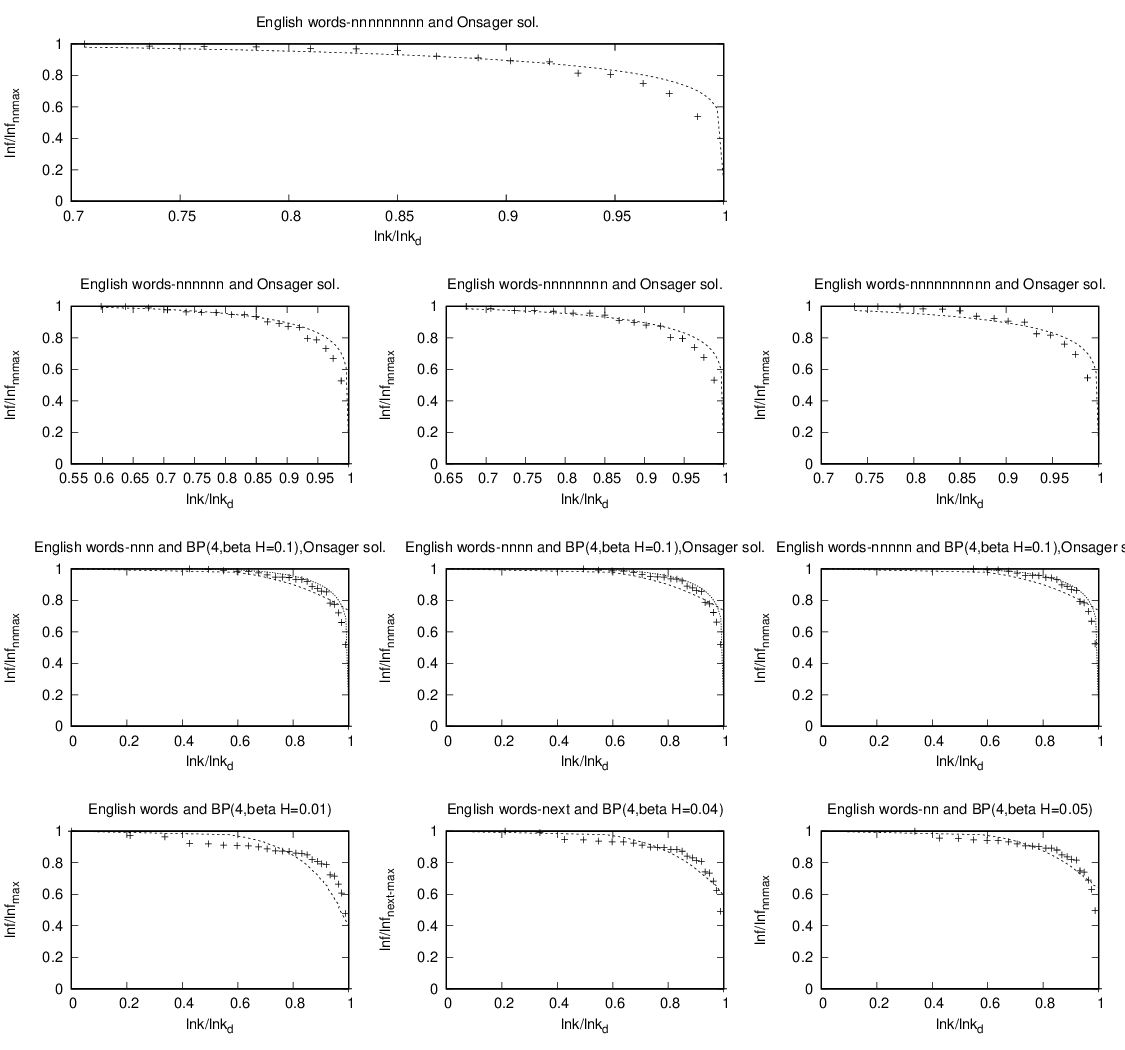}
\caption{The $+$ points represent the words of the English language. 
$k_{d}=k_{lim}$. 
BP(4, $\beta H=0.01$) is the Bethe-Peierls curve in
presence of four nearest neighbours and little magnetic field, $\beta H = 0.01$.
The upper line
refers to the Onsager solution. The pointsline never goes over 
to the Onsager solution.  The Onsager solution becomes the best fit curve for the points 
in the topmost right figure.}
\label{Figure18}
\end{figure}
\clearpage
\begin{figure}
\centering
\includegraphics[width=13cm,height=11cm]{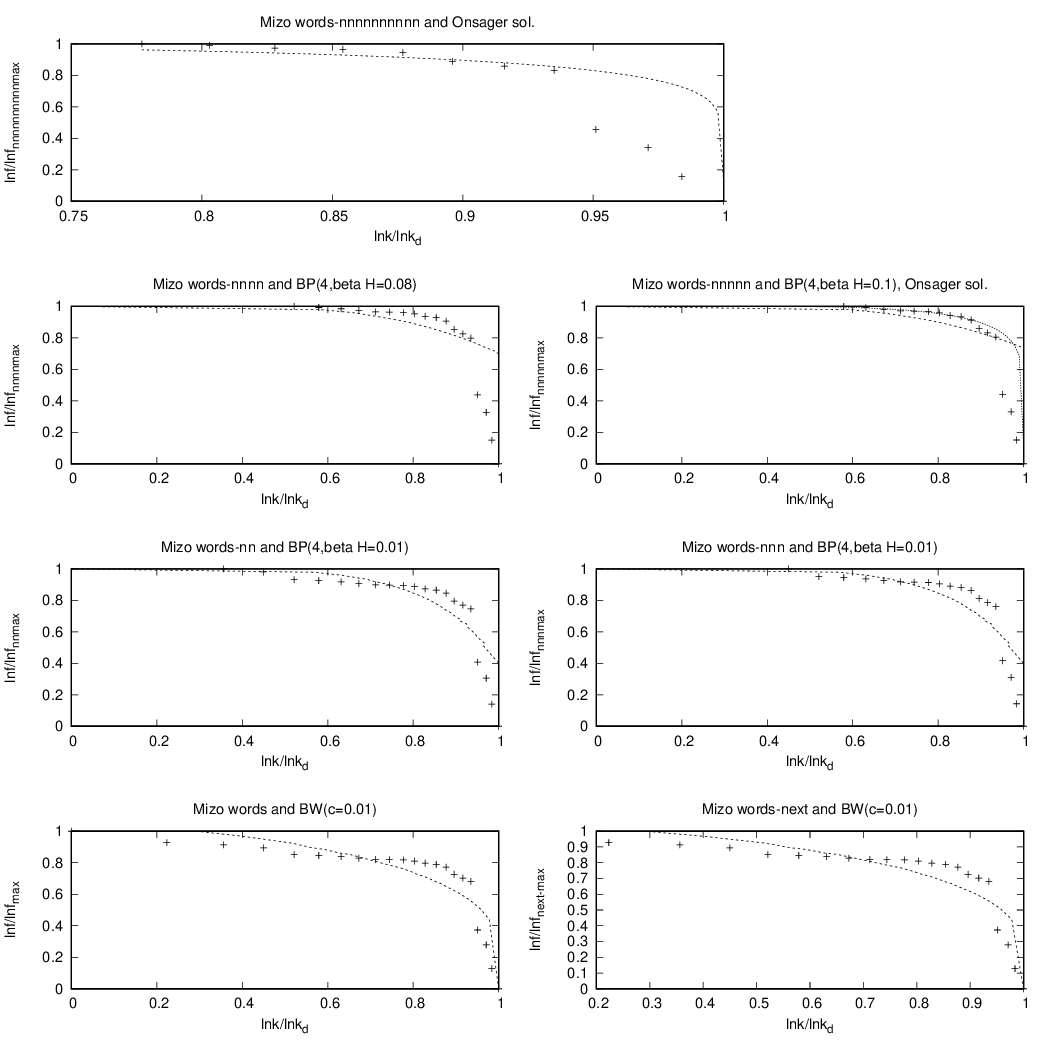}
\caption{The $+$ points represent the words of the Lushai or, Mizo language. 
$k_{d}=k_{lim}$. 
BW(c=0.01) is Bragg-Williams curve in presence of magnetic field, $c = \frac{H}{\gamma \epsilon}=0.01$.
BP(4, $\beta H=0.01$) is the Bethe-Peierls curve in
presence of four nearest neighbours and little magnetic field, $\beta H = 0.01$.
The upper line
refers to the Onsager solution. The pointsline never goes over 
to the Onsager solution. The Onsager solution does not become the best fit curve for the points.
BP(4, $\beta H=0.01$) comes closest to the points, among all the plots, in the plot of $\frac{lnf}{lnf_{nnmax}}$ vs 
$\frac{lnk}{lnk_{d}}$.}
\label{Figure19}
\end{figure}
\clearpage
\begin{figure}
\centering
\includegraphics[width=13cm,height=11cm]{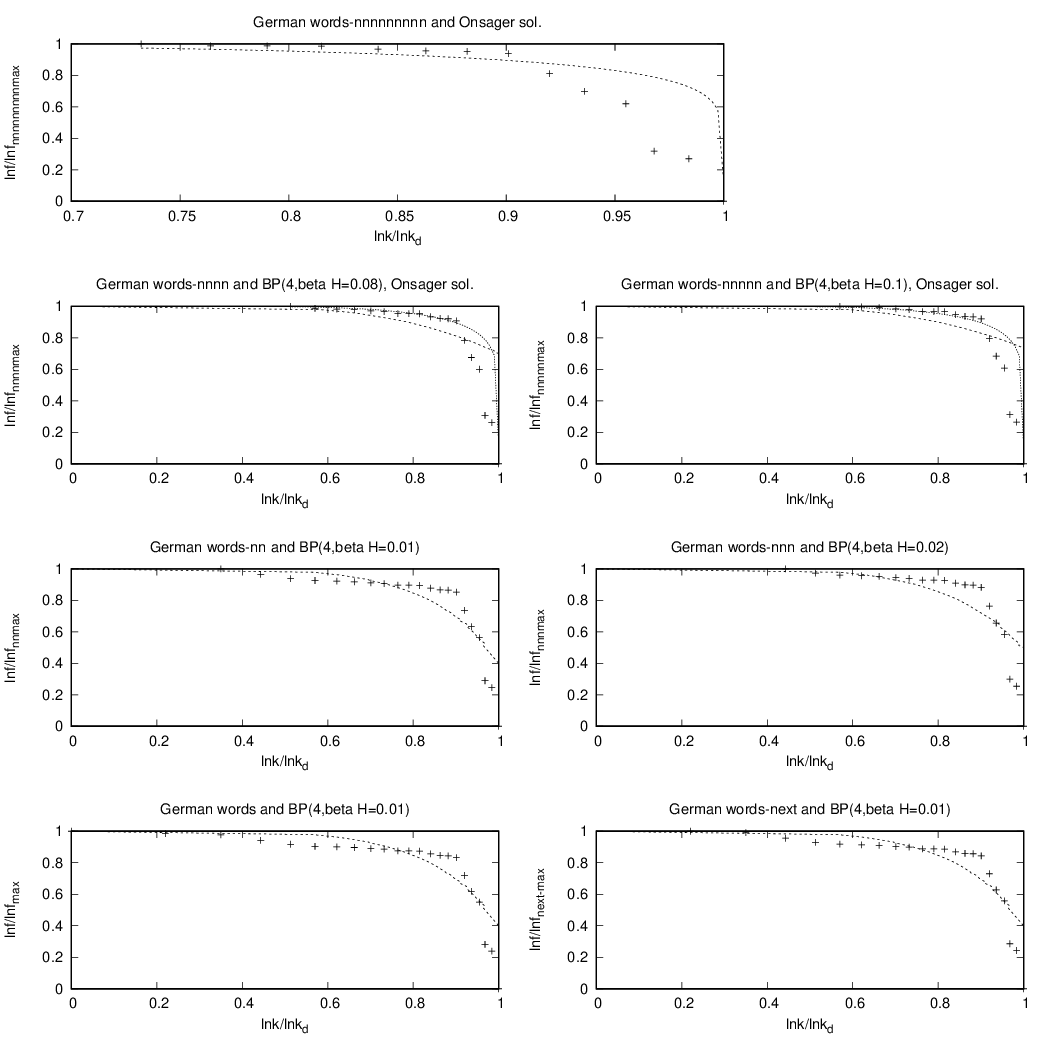}
\caption{The $+$ points represent the words of the German language. 
$k_{d}=k_{lim}$. 
BP(4, $\beta H=0.01$) is the Bethe-Peierls curve in
presence of four nearest neighbours and little magnetic field, $\beta H = 0.01$.
The upper line
refers to the Onsager solution. The pointsline never goes over 
to the Onsager solution. The Onsager solution does not become the best fit curve for the points.
BP(4, $\beta H=0.02$) comes closest to the points, among all the plots,
in the plot of $\frac{lnf}{lnf_{nnnmax}}$ vs 
$\frac{lnk}{lnk_{d}}$.}
\label{Figure20}
\end{figure}
\clearpage
\begin{figure}
\centering
\includegraphics[width=13cm,height=11cm]{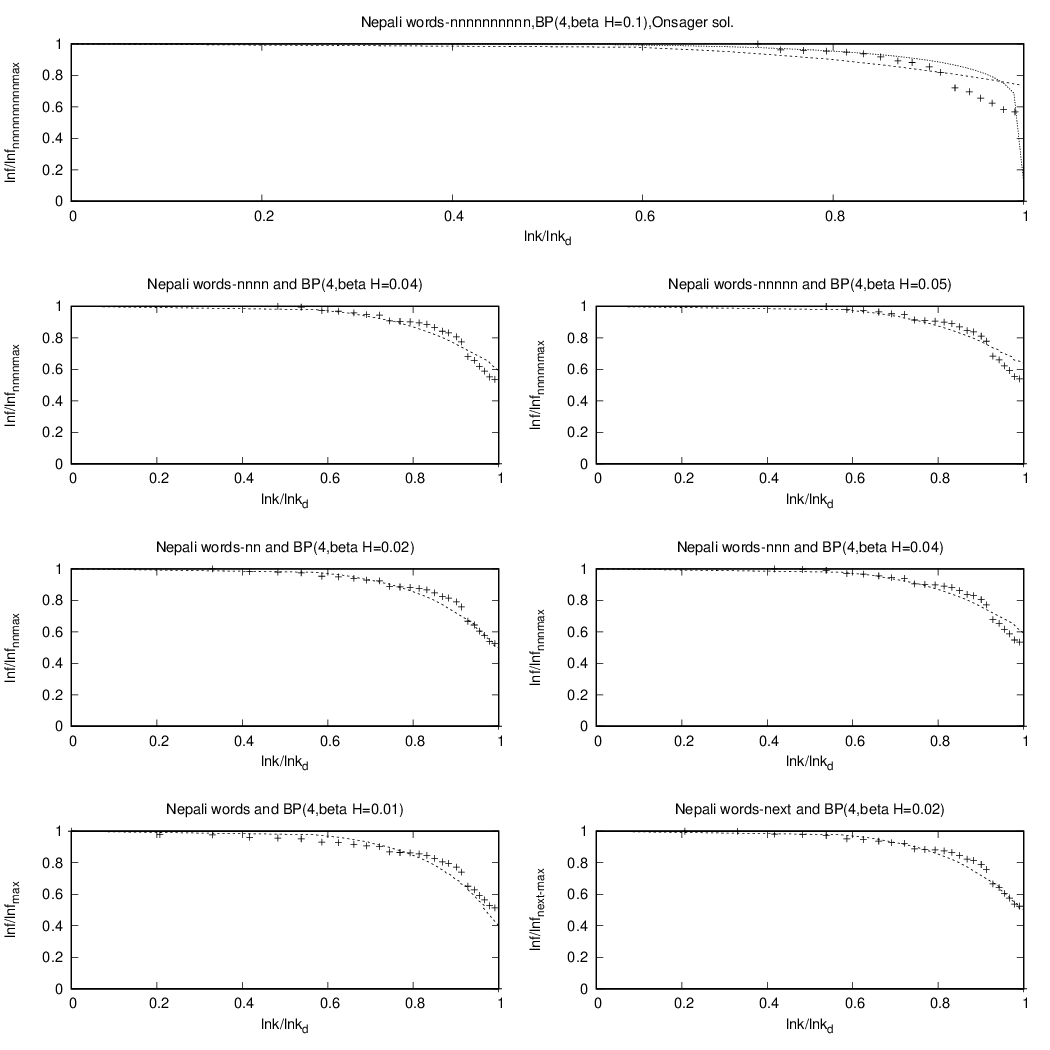}
\caption{The $+$ points represent the words of the Nepali language. 
$k_{d}=k_{lim}$. 
BP(4, $\beta H=0.01$) is the Bethe-Peierls curve in
presence of four nearest neighbours and little magnetic field, $\beta H = 0.01$.
The upper line
refers to the Onsager solution. The pointsline never goes over 
to the Onsager solution. The Onsager solution does not become the best fit curve for the points.
BP(4, $\beta H=0.04$) comes closest to the points, among all the plots,
in the plot of $\frac{lnf}{lnf_{nnnmax}}$ vs 
$\frac{lnk}{lnk_{d}}$.}
\label{Figure21}
\end{figure}

\clearpage
\begin{section}{Conclusion}
Hence we conjecture, 
behind each natural written language, there is a curve of magnetisation.
A correspondance of the following form also exists,
\begin{eqnarray}
\frac{lnf}{Normaliser} \longleftrightarrow \frac{M}{M_{max}},\nonumber\\
lnk\longleftrightarrow T.\nonumber
\end{eqnarray}
\noindent
where, the $Normaliser$ may be $lnf_{max}$ or, $lnf_{next-to-maximum}$ or, \\
$lnf_{next-to-next-maximum}$ or, higher one. 
\noindent
Moreover, the languages, on the basis of this preliminary study, can be characterised tentatively by the 
magnetisation curves of Ising model as tabulated in the tables, \ref{mag1}-\ref{mag4}. 
On the top of it, realisation of Onsager's solution of the two dimensional Ising model, \cite{Somen},\cite{keda}, 
in nature is rare to come acroos. Here, at least, 
in this preliminary study, we have found that the Onsager's solution of two dimensional Ising model to 
underlie the words of the four languages, namely, Arabian, Abor-Miri, Tibetian and English respectively.
\clearpage
\begin{table}
\begin{center}
\resizebox{12cm}{1cm}{
\begin{tabular}{|l|l|l|l|l|l|}\hline
Sanskrit & Kachin & Assamese & Garo & Sinhalese & Onge\\\hline
 BP(4,$\beta H=0$) & BP(4,$\beta H=0$) &BP(4,$\beta H=0.02$) &BP(4,$\beta H=0.01$) &BW(c=0.01) &BW(c=0) \\\hline
lnf/$lnf_{max}$  & lnf/$lnf_{max}$ &lnf/$lnf_{max}$ &lnf/$lnf_{max}$&lnf/$lnf_{max}$ &lnf/$lnf_{max}$\\\hline
\end{tabular}
}
\end{center}
\caption{The second row represents the underlying magnetisation curves and the third row corresponding columns stand for the 
y-axis labels of the corresponding graphs for the words of the languages in the corresponding columns of the first row.}
\label{mag1}
\end{table}
\begin{table}
\begin{center}
\resizebox{14cm}{1cm}{
\begin{tabular}{|l|l|l|l|l|l|l|l|}\hline
French & Hindi & Khasi & Spanish & Italiano & Turkmen & Russian & Cannarese  \\\hline
 BP(4,$\beta H=0$) & BP(4,$\beta H=0$) &BP(4,$\beta H=0.01$)  & BW(c=0.01) & BP(4,$\beta H=0$) & BP(4,$\beta H=0.02$) & BP(4,$\beta H=0.02$) & BP(4,$\beta H=0.01$) \\\hline
lnf/$lnf_{nnnmax}$  & lnf/$lnf_{nnmax}$ &lnf/$lnf_{nnnmax}$  &lnf/$lnf_{nmax}$ & lnf/$lnf_{nnnmax}$ & lnf/$lnf_{nnnmax}$ & lnf/$lnf_{nnmax}$ &lnf/$lnf_{nnnmax}$   \\\hline
\end{tabular}
}
\end{center}
\caption{The second row represents the underlying magnetisation curves and the third row corresponding columns stand for the 
y-axis labels of the corresponding graphs for the words of the languages in the corresponding columns of the first row.}
\label{mag2}
\end{table}
\begin{table}
\begin{center}
\resizebox{14cm}{1cm}{
\begin{tabular}{|l|l|l|l|l|l|l|l|}\hline
Mizo & Nepali & German & SA English & Urdu& Taraon & Lotha & Sema \\\hline
 BP(4,$\beta H=0.01$) & BP(4,$\beta H=0.04$) &BP(4,$\beta H=0.02$)& BP(4,$\beta H=0$) & BP(4,$\beta H=0.02$) & BW(c=0) &BW(c=0.01) & BW(c=0.01)   \\\hline
lnf/$lnf_{nnmax}$  & lnf/$lnf_{nnnmax}$ &lnf/$lnf_{nnnmax}$ & lnf/$lnf_{nnmax}$ &lnf/$lnf_{nnnmax}$ & lnf/$lnf_{nmax}$ &lnf/$lnf_{nnmax}$&lnf/$lnf_{nnnmax}$ \\\hline
\end{tabular}
}
\end{center}
\caption{The second row represents the underlying magnetisation curves and the third row corresponding columns stand for the 
y-axis labels of the corresponding graphs for the words of the languages in the corresponding columns of the first row.}
\label{mag3}
\end{table}

\begin{table}
\begin{center}
\resizebox{10cm}{1cm}{
\begin{tabular}{|l|l|l|l|}\hline
Arabian & Abor-Miri & Tibetian &English \\\hline
 Onsager & Onsager & Onsager& Onsager \\\hline
lnf/$lnf_{nnnnmax}$  & lnf/$lnf_{nnnnnnnmax}$ &lnf/$lnf_{nnnnnmax}$  &lnf/$lnf_{nnnnnnnnnnmax}$  \\\hline
\end{tabular}
}
\end{center}
\caption{The second row represents the underlying magnetisation curve and the third row corresponding columns stand for the 
y-axis labels of the corresponding graphs for the words of the languages in the corresponding columns of the first row.}
\label{mag4}
\end{table}
\end{section}

\clearpage
\begin{section}{Acknowledgement}
We would like to thank various persons for lending us 
bilingual dictionaries say 
Langjaw Kyang Ying for Kachin to English; Anthropological Survey of India, Shillong
branch, for allowing us to use Taraon to English, Onge to English; 
State Central Library, Shillong, for Spanish to English, Italian to English,
Sinhalese to English; NEHU library for many other dictionaries. We would like 
to thank Physics Department members of NEHU and many others for discussion. The 
author's special thanks goes to  Jayant Kumar Nayak, Center for Anthropological
Study, Central University of Orissa, for discussions. 
We have used gnuplot for drawing the figures.
\end{section}

\begin{section}{appendix}
\clearpage
\begin{table}
\resizebox{14cm}{0.5cm}{
\begin{tabular}{|l|l|l|l|l|l|l|l|l|l|l|l|l|l|l|l|l|l|l|l|l|l|l|l|l|l|}\hline
 A&B&C&D&E&F&G&H&I&J&K&L&M&N&O&P&Q&R&S&T&U&V&W&X&Y&Z \\ \hline 
 1097&630&1701&882&1008&630&441&284&599&136&15&378&819&221&315&1292&63&945&882&662&32&410&12&4&6&26 \\ \hline
\end{tabular}
}
\caption{French words: the first row represents letters of the French alphabet in the serial order}
\end{table}
\begin{table}
\begin{center}
\resizebox{14cm}{4.5cm}{
\begin{tabular}{|l|l|l|l|l|l|l|l|l|l|}\hline
k & lnk & lnk/$lnk_{lim}$ & f & lnf & lnf/$lnf_{max}$ &lnf/$lnf_{next-max}$&lnf/$lnf_{nnmax}$&lnf/$lnf_{nnnmax}$\\\hline
1 & 0    & 0     & 1701 & 7.44 & 1 & Blank&Blank&Blank\\\hline
2 & 0.69 & 0.214 & 1292 & 7.16 & 0.962& 1&Blank&Blank \\\hline
3 & 1.10 & 0.342 & 1097 & 7.00 & 0.941& 0.978&1&Blank \\\hline
4 & 1.39 & 0.432 &1008  & 6.92 & 0.930& 0.966&.989 &1 \\\hline
5 & 1.61 & 0.500 & 945 & 6.85 & 0.921& 0.957 &.979&.990\\\hline
6 & 1.79 & 0.556 & 882 & 6.78& 0.911& 0.947 &.969&.980\\\hline
7 & 1.95 & 0.606 & 819 & 6.71 & 0.902& 0.937&.959&.970 \\\hline
8 & 2.08 & 0.646 & 662 & 6.50 & 0.874& 0.908&.929&.939 \\\hline
9 & 2.20 & 0.683 & 630 & 6.45 & 0.867& 0.901&.921&.932 \\\hline
10 & 2.30 & 0.714 & 599 & 6.40 & 0.860& 0.894&.914&.925 \\\hline
11 & 2.40 & 0.745 & 441 & 6.09 & 0.819& 0.851&.870&.880 \\\hline
12 & 2.48 & 0.770 & 410 & 6.02 & 0.809& 0.841&.860&.870 \\\hline
13 & 2.56 & 0.795 & 378 & 5.93 & 0.797& 0.828&.847&.857 \\\hline
14 & 2.64 & 0.820 & 315 & 5.75 & 0.773& 0.803&.821&.831 \\\hline
15 & 2.71 & 0.842 & 284 & 5.65 & 0.759& 0.789&.807&.816 \\\hline
16 & 2.77 & 0.860 & 221 & 5.40 & 0.726& 0.754&.771&.780 \\\hline
17 & 2.83 & 0.879 & 136 & 4.91 & 0.660& 0.686&.701&.710 \\\hline
18 & 2.89 & 0.898 & 63 & 4.14 & 0.556& 0.578&.591&.598 \\\hline
19 & 2.94 & 0.913 & 32 & 3.47 & 0.466& 0.485&.496&.501 \\\hline
20 & 3.00 & 0.932 & 26 & 3.26 & 0.438& 0.455&.466&.471 \\\hline
21 & 3.04 & 0.944 & 15 & 2.71 & 0.364& 0.378&.387&.392 \\\hline
22 & 3.09 & 0.960 & 12 & 2.48 & 0.333& 0.346&.354&.358 \\\hline
23 & 3.14 & 0.975 & 6  & 1.79 & 0.241& 0.250&.256&.259 \\\hline
24 & 3.18 & 0.988 & 4  & 1.39   & 0.187& 0.194&.199&.201    \\\hline
25 & 3.22 & 1     & 1  & 0     & 0 & 0&0&0\\\hline
\end{tabular}
}
\end{center}
\caption{French words: ranking, natural logarithm, normalisations}
\end{table}
\clearpage
\begin{table}
\resizebox{14cm}{.5cm}{
\begin{tabular}{|l|l|l|l|l|l|l|l|l|l|l|l|l|l|l|l|l|l|l|l|l|l|l|l|l|l|l|l|l|l|l|l|l|l|l|l|l|l|l|l|l|}\hline
 1&2&3&4&5&6&7&8&9&10&11&12&13&14&15&16&17&18&19&20&21&22&23&24&25&26&27&28&29&30&31&32&33&34&35&36&37&38&39&40&41 \\ \hline 
 750&280&72&15&280&19&15&80&20&27&21&750&324&660&250&740&210&500&170&200&210&125&75&525&70&600&200&650&1400&250&925&340&760&110&360&380&600&350&12&1430&350 \\ \hline
\end{tabular}
}
\caption{Hindi words: the first row represents letters of the Hindi alphabet in the serial order}
\end{table}
\begin{table}
\begin{center}
\resizebox{14cm}{4.5cm}{
\begin{tabular}{|l|l|l|l|l|l|l|l|l|l|}\hline
k & lnk & lnk/$lnk_{lim}$ & f & lnf & lnf/$lnf_{max}$ & lnf/$lnf_{nextmax}$&lnf/$lnf_{nnmax}$&lnf/$lnf_{nnnmax}$\\\hline
1 & 0    & 0             & 1430 & 7.27 & 1 & Blank&Blank&Blank\\\hline
2 & 0.69 & 0.195         & 1400& 7.24 & 0.997 & 1&Blank&Blank\\\hline
3 & 1.10 & 0.312         & 925 & 6.83 & 0.939 & 0.943&1&Blank\\\hline
4 & 1.39 & 0.394         & 760 & 6.63 & 0.912 & 0.916&.971&1\\\hline
5 & 1.61 & 0.456         & 750 & 6.62 & 0.911 & 0.914&.969&.998\\\hline
6 & 1.79 & 0.507         & 740 & 6.61 & 0.909 & 0.913&.968&.997\\\hline
7 & 1.95 & 0.552         & 660& 6.49 & 0.893 & 0.896&.950&.979\\
\hline
\end{tabular}
}
\end{center}
\caption{Hindi words: ranking, natural logarithm, normalisations}
\end{table}
\clearpage
\begin{table}
\resizebox{14cm}{.5cm}{
\begin{tabular}{|l|l|l|l|l|l|l|l|l|l|l|l|l|l|l|l|l|l|l|l|l|}\hline
 1&2&3&4&5&6&7&8&9&10&11&12&13&&P&R&S&T&U&W&Y \\ \hline 
 11&125&129&2&76&7&76&28&76&98&38&9&4&136&34&9&55&33&2&22&10 \\ \hline
\end{tabular}
}
\caption{Khasi words: the first row represents letters of the khasi alphabet in the serial order}
\end{table}
\begin{table}
\begin{center}
\resizebox{14cm}{4.5cm}{
\begin{tabular}{|l|l|l|l|l|l|l|l|l|l|}\hline
k & lnk  & lnk/$lnk_{lim}$ & f & lnf & lnf/$lnf_{max}$ & lnf/$lnf_{next-max}$&lnf/$lnf_{nnmax}$&lnf/$lnf_{nnnmax}$\\\hline
1 & 0    & 0     & 1150 & 7.05 & 1 & Blank&Blank&Blank\\\hline
2 & 0.69 & 0.249 & 1000 & 6.91 & 0.980 & 1&Blank&Blank\\\hline
3 & 1.10 & 0.397 & 600 & 6.40 & 0.908 & 0.926&1&Blank\\\hline
4 & 1.39 & 0.502 & 550 & 6.31 & 0.895 & 0.913&.986&1\\\hline
5 & 1.61 & 0.581 & 460 & 6.13 & 0.870 & 0.887&.958&.971\\\hline
6 & 1.79 & 0.646 & 320 & 5.77 & 0.818 & 0.835&.902&.914\\\hline
7 & 1.95 & 0.704 & 300 & 5.70 & 0.809 & 0.825&.891&.903\\\hline
8 & 2.08 & 0.751 & 260& 5.56 & 0.789 & 0.805&.869&.881\\\hline
9 & 2.20 & 0.794 & 230 & 5.43 & 0.770 & 0.786&.848&.861\\\hline
10 & 2.30 & 0.830 & 154 & 5.04 & 0.715 & 0.729&.788&.799\\\hline
11 & 2.40 & 0.866 & 100 & 4.61 & 0.654 & 0.667&.720&.731\\\hline
12 & 2.48 & 0.895 & 60 & 4.09 & 0.580 & 0.592&.639&.648\\\hline
13 & 2.56 & 0.924 & 50 & 3.91 & 0.555 & 0.566&.611&.620\\\hline
14 & 2.64 & 0.953 & 35 & 3.56 & 0.505 & 0.515&.556&.564\\\hline
15 & 2.71 & 0.978 & 11 & 2.40 & 0.340 & 0.347&.375&.380\\\hline
16 & 2.77 & 1     & 1  & 0    & 0 & 0&0&0\\\hline
\end{tabular}
}
\end{center}
\caption{Khasi words: ranking, natural logarithm, normalisations}
\end{table}
\clearpage
\begin{table}
\resizebox{14cm}{.5cm}{
\begin{tabular}{|l|l|l|l|l|l|l|l|l|l|l|l|l|l|l|l|l|l|l|l|l|}\hline
 1&2&3&4&5&6&7&8&9&10&11&12&13&14&15&16&17&18&19&20&21 \\ \hline 
 11&125&129&62&76&7&76&28&76&98&38&9&4&136&34&9&55&33&2&22&10 \\ \hline
\end{tabular}
}
\caption{Onge words: the first row represents letters of the Onge alphabet in the serial order}
\end{table}
\begin{table}
\begin{center}
\resizebox{14cm}{4.5cm}{
\begin{tabular}{|l|l|l|l|l|l|l|}\hline
k & lnk  & lnk/$lnk_{lim}$ & f & lnf & lnf/$lnf_{max}$ & lnf/$lnf_{next-max}$\\\hline
1 & 0    & 0     & 136 & 4.91 & 1 & Blank\\\hline
2 & 0.69 & 0.235 & 129 & 4.86 & 0.990 & 1\\\hline
3 & 1.10 & 0.374 & 125 & 4.83 & 0.984 & 0.994\\\hline
4 & 1.39 & 0.473 & 98 & 4.58 & 0.933 & 0.942\\\hline
5 & 1.61 & 0.548 & 76 & 4.33 & 0.882 & 0.891\\\hline
6 & 1.79 & 0.609 & 62 & 4.13 & 0.841 & 0.850\\\hline
7 & 1.95 & 0.663 & 55 & 4.01 & 0.817 & 0.825\\\hline
8 & 2.08 & 0.707 & 38 & 3.64 & 0.741 & 0.749\\\hline
9 & 2.20 & 0.748 & 34 & 3.53 & 0.719 & 0.726\\\hline
10 & 2.30 & 0.782 & 33 & 3.50 & 0.713 & 0.720\\\hline
11 & 2.40 & 0.816 & 28 & 3.32 & 0.676 & 0.683\\\hline
12 & 2.48 & 0.832 & 22 & 3.09 & 0.629 & 0.636\\\hline
13 & 2.56 & 0.871 & 11 & 2.40 & 0.489 & 0.494\\\hline
14 & 2.64 & 0.898 & 10 & 2.30 & 0.468 & 0.473\\\hline
15 & 2.71 & 0.922 &9 & 2.20& 0.448 & 0.453\\\hline
16& 2.77  & 0.942 &7 &1.95 &0.397 &0.401 \\\hline
17& 2.83  & 0.963 &4 &1.39 &0.283 &0.286 \\\hline
18& 2.89  & 0.983 &2 &0.69 &0.141 &0.142 \\\hline
19& 2.94  & 1 &1  &0 &0 &0 \\\hline
\end{tabular}
}
\end{center}
\caption{Onge words: ranking, natural logarithm, normalisations}
\end{table}
\clearpage
\begin{table}
\resizebox{14cm}{.5cm}{
\begin{tabular}{|l|l|l|l|l|l|l|l|l|l|l|l|l|l|l|l|l|l|l|l|l|l|l|l|l|l|l|}\hline
 1&2&3&4&5&6&7&8&9&10&11&12&13&14&15&16&17&18&19&20&21&22&23&24&25&26&27 \\ \hline 
 34&143&27&16&3&220&65&60&3&32&24&37&357&48&86&56&96&16&88&224&8&64&74&64&56&20&224  \\ \hline
\end{tabular}
}
\caption{Taraon words: the first row represents letters of the Taraon alphabet in the serial order}
\end{table}
\begin{table}
\begin{center}
\resizebox{14cm}{4.5cm}{
\begin{tabular}{|l|l|l|l|l|l|l|}\hline
k & lnk  & lnk/$lnk_{lim}$ & f & lnf & lnf/$lnf_{max}$ & lnf/$lnf_{next-max}$\\\hline
1 & 0    & 0     & 357 & 5.88 & 1 & Blank\\\hline
2 & 0.69 & 0.220 & 224 & 5.41 & 0.920 & 1\\\hline
3 & 1.10 & 0.350 & 220 & 5.39 & 0.920 & 0.996\\\hline
4 & 1.39 & 0.443 & 143 & 4.96 & 0.840 & 0.917\\\hline
5 & 1.61 & 0.513 & 96 & 4.56 & 0.78 & 0.843\\\hline
6 & 1.79 & 0.570 & 88 & 4.48 & 0.76 & 0.828\\\hline
7 & 1.95 & 0.621 & 86 & 4.45 & 0.76 & 0.823\\\hline
8 & 2.08 & 0.662 & 74 & 4.30 & 0.73 & 0.795\\\hline
9 & 2.20 & 0.701 & 65 & 4.17 & 0.71 & 0.771\\\hline
10 & 2.30 & 0.732 & 64 & 4.16 & 0.71 & 0.769\\\hline
11 & 2.40 & 0.764 & 60 & 4.09 & 0.70 & 0.756\\\hline
12 & 2.48 & 0.790 & 56 & 4.03 & 0.69 & 0.745\\\hline
13 & 2.56 & 0.815 & 48 & 3.87 & 0.66 & 0.715\\\hline
14 & 2.64 & 0.841 & 37 & 3.61 & 0.61 & 0.667\\\hline
15 & 2.71 & 0.863 &34 & 3.53& 0.60 & 0.652\\\hline
16& 2.77  & 0.882 &32 &3.47 &0.59 &0.641 \\\hline
17& 2.83  & 0.901 &27 &3.30 &0.56 &0.610 \\\hline
18& 2.89  & 0.920 &24 &3.18 &0.54 &0.588 \\\hline
19& 2.94  & 0.936 &20  &3.00 &0.51 &0.555 \\\hline
20& 3.00  & 0.955 &16  &2.77 &0.47 &0.512 \\\hline
21& 3.04  & 0.968 &8   &2.08 &0.35 &0.384 \\\hline
22& 3.09  & 0.984 &3   &1.10 &0.19 & 0.203 \\\hline
23& 3.14  & 1     &1   &0    &0    & 0\\\hline
\end{tabular}
}
\end{center}
\caption{Taraon words: ranking, natural logarithm, normalisations}
\end{table}
\clearpage
\begin{table}
\resizebox{14cm}{.5cm}{
\begin{tabular}{|l|l|l|l|l|l|l|l|l|l|l|l|l|l|l|l|l|l|l|l|l|l|l|l|l|l|l|l|l|l|l|l|l|l|l|l|l|}\hline
 1&2&3&4&5&6&7&8&9&10&11&12&13&14&15&16&17&18&19&20&21&22&23&24&25&26&27&28&29&30 \\ \hline 
 1650&1475&3050&650&625&1050&600&900&5&5&4775&1100&825&1200&2300&1100&625&925&425&5&700&700&250&800&950&750&700&1300&400&275 \\ \hline
\end{tabular}
}
\caption{Tibetian words: the first row represents letters of the Tibetian alphabet in the serial order}
\end{table}
\begin{table}
\begin{center}
\resizebox{14cm}{5cm}{
\begin{tabular}{|l|l|l|l|l|l|l|l|l|l|l|}\hline
k&lnk&lnk/$lnk_{lim}$ &f&lnf&lnf/$lnf_{max}$&lnf/$lnf_{nextmax}$&lnf/$lnf_{nnmax}$&lnf/$lnf_{nnnmax}$&lnf/$lnf_{nnnnmax}$&lnf/$lnf_{nnnnnmax}$\\\hline
1 & 0    & 0     & 4775 & 8.47 & 1 & Blank& Blank& Blank& Blank& Blank\\\hline
2 & 0.69 & 0.217 & 3050 & 8.02 & 0.947& 1 & Blank& Blank& Blank& Blank\\\hline
3 & 1.10 & 0.346 & 2300 & 7.74 & 0.914& 0.965& 1& Blank& Blank& Blank \\\hline
4 & 1.39 & 0.437 & 1650 & 7.41 & 0.875& 0.924& 0.957& 1& Blank& Blank \\\hline
5 & 1.61 & 0.506 & 1475 & 7.30 & 0.862& 0.910 &0.943&0.985 &1 & Blank\\\hline
6 & 1.79 & 0.563 & 1300 & 7.17 & 0.847& 0.894 &0.926&0.968 &0.982 &1 \\\hline
7 & 1.95 & 0.613 & 1200 & 7.09 & 0.837& 0.884 &0.916&0.957&0.971&0.989\\
\hline
\end{tabular}
}
\end{center}
\caption{Tibetian words: ranking, natural logarithm, normalisations}
\end{table}
\clearpage
\begin{table}
\resizebox{14cm}{0.5cm}{
\begin{tabular}{|l|l|l|l|l|l|l|l|l|l|l|l|l|l|l|l|l|l|l|l|l|l|l|l|l|l|}\hline
 A&B&C&D&E&F&G&H&I&L&M&N&O&P&Q&R&S&T&U&V&W&X&Y&Z \\ \hline 
 2513&1025&3025&1325&700&1150&844&11&1688&600&1350&375&500&2225&150&1444&3338&1200&206&750&8&6&5&94 \\ \hline
\end{tabular}
}
\caption{Italiano words: the first row represents letters of the Italian alphabet in the serial order}
\end{table}

\begin{table}
\begin{center}
\resizebox{14cm}{5cm}{
\begin{tabular}{|l|l|l|l|l|l|l|l|l|l|}\hline
k & lnk & lnk/$lnk_{lim}$ & f & lnf & lnf/$lnf_{max}$ &lnf/$lnf_{next-max}$&lnf/$lnf_{nnmax}$&lnf/$lnf_{nnnmax}$\\\hline
1 & 0    & 0     & 3338 & 8.11 & 1 & Blank&Blank&Blank\\\hline
2 & 0.69 & 0.214 & 3025 & 8.01 & 0.988& 1 &Blank &Blank\\\hline
3 & 1.10 & 0.342 & 2513 & 7.83 & 0.965& 0.978&1 &Blank \\\hline
4 & 1.39 & 0.432 & 2225  & 7.71 & 0.951& 0.963&.985&1 \\\hline
5 & 1.61 & 0.500 & 1688 & 7.43 & 0.916& 0.928 &.949&.964\\\hline
6 & 1.79 & 0.556 & 1444 & 7.28 & 0.898& 0.909 &.930&.944\\\hline
7 & 1.95 & 0.606 & 1350 & 7.21 & 0.889& 0.900 &.921&.935\\\hline
8 & 2.08 & 0.646 & 1325 & 7.19 & 0.887& 0.898 &.918&.933\\\hline
9 & 2.20 & 0.683 & 1200 & 7.09 & 0.874& 0.885 &.905&.920\\\hline
10 & 2.30 & 0.714 & 1150 & 7.05 & 0.869& 0.880&.900&.914 \\\hline
11 & 2.40 & 0.745 & 1025 & 6.93 & 0.855& 0.865&.885&.899 \\\hline
12 & 2.48 & 0.770 & 844 & 6.74 & 0.831& 0.841 &.861&.874\\\hline
13 & 2.56 & 0.795 & 750 & 6.62 & 0.816& 0.826 &.845&.859\\\hline
14 & 2.64 & 0.820 & 700 & 6.55 & 0.808& 0.818 &.837&.850\\\hline
15 & 2.71 & 0.842 & 600 & 6.40 & 0.789& 0.799 &.817&.830\\\hline
16 & 2.77 & 0.860 & 500 & 6.21 & 0.766& 0.775 &.793&.805\\\hline
17 & 2.83 & 0.879 & 375 & 5.93 & 0.731& 0.740 &.757&.769\\\hline
18 & 2.89 & 0.898 & 206 & 5.33 & 0.657& 0.665 &.681&.691\\\hline
19 & 2.94 & 0.913 & 150 & 5.01 & 0.618& 0.625 &.640&.650\\\hline
20 & 3.00 & 0.932 & 94 & 4.54 & 0.560& 0.567 &.580&.589\\\hline
21 & 3.04 & 0.944 & 11 & 2.40 & 0.296& 0.300 &.307&.311\\\hline
22 & 3.09 & 0.960 & 8 & 2.08 & 0.256& 0.260 &.266&.270\\\hline
23 & 3.14 & 0.975 & 6  & 1.79 & 0.221& 0.223&.229&.232 \\\hline
24 & 3.18 & 0.988 & 5  & 1.61   & 0.199& 0.201&.206 &.209   \\\hline
25 & 3.22 & 1     & 1  & 0     & 0 & 0&0&0\\\hline
\end{tabular}
}
\end{center}
\caption{Italiano words: ranking, natural logarithm, normalisations}
\end{table}
\clearpage
\begin{table}
\resizebox{14cm}{.5cm}{
\begin{tabular}{|l|l|l|l|l|l|l|l|l|l|l|l|l|l|l|l|l|l|l|l|l|l|l|l|l|l|l|l|l|l|l|}\hline
 A&B&C&D&\"{A}&F&G&H&I&J&\pounds&K&L&M&N&O&\"{0}&P&R&S&?&T&U&\"{U}&W&\"{Y}&Y&Z \\ \hline 
 397&473&404&710&46&55&1241&443&222&165&9&561&74&410&177&194&151&332&120&665&200&597&133&68&82&560&107&80 \\ \hline
\end{tabular}
}
\caption{Turkmen words: the first row represents letters of the Turkmen alphabet in the serial order}
\end{table}
\begin{table}
\begin{center}
\resizebox{14cm}{4.5cm}{
\begin{tabular}{|l|l|l|l|l|l|l|l|l|}\hline
k & lnk & lnk/$lnk_{lim}$ & f & lnf & lnf/$lnf_{max}$ & lnf/$lnf_{nextmax}$&lnf/$lnf_{nnmax}$&lnf/$lnf_{nnnmax}$\\\hline
1 & 0    & 0             & 1241 & 7.12 & 1 & Blank &Blank& Blank\\\hline
2 & 0.69 & 0.207         & 710& 6.57 & 0.923 & 1 & Blank& Blank\\\hline
3 & 1.10 & 0.330         & 665& 6.50 & 0.913 & 0.989& 1 &Blank\\\hline
4 & 1.39 & 0.417         & 597 & 6.40 & 0.899 & 0.974&.985&1\\\hline
5 & 1.61 & 0.483         & 561 & 6.33 & 0.889 & 0.963&.974&.989\\\hline
6 & 1.79 & 0.538         & 560 & 6.33 & 0.889 & 0.963&.974&.989\\\hline
7 & 1.95 & 0.586         & 473& 6.16 & 0.865 & 0.938&.948&.963\\
\hline
\end{tabular}
}
\end{center}
\caption{Turkmen words: ranking, natural logarithm, normalisations}
\end{table}
\clearpage
\begin{table}
\resizebox{14cm}{.5cm}{
\begin{tabular}{|l|l|l|l|l|l|l|l|l|l|l|l|l|l|l|l|l|l|l|l|l|l|l|l|l|l|l|l|l|l|l|l|l|l|l|l|l|}\hline
 1&2&3&4&5&6&7&8&9&10&11&12&13&14&15&16&17&18&19&20&21&22&23&24&25&26&27&28&29 \\ \hline 
 775&2232&4557&1457&2046&248&496&2976&1550&9&3069&1023&1705&3472&4340&10602&3224&5301&1984&1922&558&620&248&651&620&93&372&62&186 \\ \hline
\end{tabular}
}
\caption{Russian words: the first row represents letters of the Russian alphabet in the serial order}
\end{table}

\begin{table}
\begin{center}
\resizebox{14cm}{4.5cm}{
\begin{tabular}{|l|l|l|l|l|l|l|l|l|}\hline
k & lnk & lnk/$lnk_{lim}$ & f & lnf & lnf/$lnf_{max}$ &lnf/$lnf_{next-max}$&lnf/$lnf_{nnmax}$&lnf/$lnf_{nnnmax}$\\\hline
1 & 0    & 0     & 10602 & 9.27 & 1 & Blank&Blank&Blank\\\hline
2 & 0.69 & 0.209 & 5301 & 8.58 & 0.926& 1 &Blank&Blank\\\hline
3 & 1.10 & 0.333 & 4557 & 8.42 & 0.908& 0.981&1&Blank \\\hline
4 & 1.39 & 0.421 & 4340 & 8.38 & 0.904& 0.977&.995&1 \\\hline
5 & 1.61 & 0.488 & 3472 & 8.15 & 0.879& 0.950&.968&.973 \\\hline
6 & 1.79 & 0.542 & 3224 & 8.08 & 0.872& 0.942&.960&.964 \\\hline
7 & 1.95 & 0.591 & 3069 & 8.03 & 0.866& 0.936&.954&.958 \\\hline
8 & 2.08 & 0.630 & 2976 & 8.00 & 0.863& 0.932&.950&.955 \\\hline
9 & 2.20 & 0.667 & 2232 & 7.71 & 0.832& 0.899&.916&.920 \\\hline
10 & 2.30 & 0.697 &2046 & 7.62 & 0.822& 0.888&.905&.909 \\\hline
11 & 2.40 & 0.727 & 1984 & 7.59 & 0.819& 0.885&.901&.906 \\\hline
12 & 2.48 & 0.752 & 1922 & 7.56 & 0.816& 0.881&.898&.902 \\\hline
13 & 2.56 & 0.776 & 1705 & 7.44 & 0.803& 0.867&.884&.888 \\\hline
14 & 2.64 & 0.800 & 1550 & 7.35 & 0.793& 0.857&.873&.877 \\\hline
15 & 2.71 & 0.821 & 1457 & 7.28 & 0.785& 0.848&.865&.869 \\\hline
16 & 2.77 & 0.839 & 1023 & 6.93 & 0.748& 0.808&.823&.827 \\\hline
17 & 2.83 & 0.858 & 775 & 6.65 & 0.717& 0.775&.790&.794 \\\hline
18 & 2.89 & 0.876 & 651 & 6.48 & 0.699& 0.755&.770&.773 \\\hline
19 & 2.94 & 0.891 & 620 & 6.43 & 0.694& 0.749&.764&.767 \\\hline
20 & 3.00 & 0.909 & 558 & 6.32 & 0.682& 0.737&.751&.754 \\\hline
21 & 3.04 & 0.921 & 496 & 6.21 & 0.670& 0.724&.738&.741 \\\hline
22 & 3.09 & 0.936 & 372 & 5.92 & 0.639& 0.690&.703&.706 \\\hline
23 & 3.14 & 0.952 & 248  & 5.51 & 0.594& 0.642&.654&.658 \\\hline
24 & 3.18 & 0.964 & 186  & 5.23 & 0.564& 0.610&.621&.624    \\\hline
25 & 3.22 & 0.976 & 93  & 4.53 & 0.489 & 0.528&.538&.541\\\hline
26 & 3.26 & 0.988    & 62   &4.13 &0.446 &0.481&.490&.493\\\hline
27 & 3.30 & 1      & 1    & 0   & 0    & 0&0&0 \\\hline
\end{tabular}
}
\end{center}
\caption{Russian words: ranking, natural logarithm, normalisations}
\end{table}
\clearpage
\begin{table}
\resizebox{15cm}{.5cm}{
\begin{tabular}{|l|l|l|l|l|l|l|l|l|l|l|l|l|l|l|l|l|l|l|l|l|l|l|l|l|l|l|l|l|l|l|l|l|l|l|l|l|l|l|l|l|l|l|l|l|l|l|l|l|l|l|}\hline
 1&2&3&4&5&6&7&8&9&10&11&12&13&14&15&16&17&18&19&20&21&22&23&24&25&26&27&28&29&30&31&32&33&34&35&36&37&38&39&40&41&42&43&44&45&46&47&48&49&50&51 \\ \hline 
 2820&840&420&120&765&60&15&0&360&90&60&420&135&60&0&0&3690&150&1635&135&0&930&105&825&75&0&105&30&105&15&2&1650&15&945&165&1245&2325&60&1440&270&1950&270&450&345&1200&600&30&2010&1665&1&15 \\ \hline
\end{tabular}
}
\caption{Canarese words: the first row represents letters of the Canarese alphabet in the serial order}
\end{table}
\begin{table}
\begin{center}
\resizebox{14cm}{4.5cm}{
\begin{tabular}{|l|l|l|l|l|l|l|l|l|}\hline
k & lnk & lnk/$lnk_{lim}$ & f & lnf & lnf/$lnf_{max}$ & lnf/$lnf_{nextmax}$&lnf/$lnf_{nnmax}$&lnf/$lnf_{nnnmax}$\\\hline
1 & 0    & 0             & 3690 & 8.21 & 1 & Blank&Blank&Blank\\\hline
2 & 0.69 & 0.197         & 2820 & 7.94 & 0.967 & 1&Blank&Blank\\\hline
3 & 1.10 & 0.314         & 2325 & 7.75 & 0.944 & 0.976&1&Blank\\\hline
4 & 1.39 & 0.397         & 2010 & 7.61 & 0.927 & 0.958&.982&1\\\hline
5 & 1.61 & 0.460         & 1950 & 7.58 & 0.923 & 0.955&.978&.996\\\hline
6 & 1.79 & 0.511         & 1665 & 7.42 & 0.904 & 0.935&.957&.975\\\hline
7 & 1.95 & 0.557         & 1650  & 7.41 & 0.903 & 0.933&.956&.974\\
\hline
\end{tabular}
}
\end{center}
\caption{Canarese words: ranking, natural logarithm, normalisations}
\end{table}
\clearpage
\begin{table}
\resizebox{15cm}{.5cm}{
\begin{tabular}{|l|l|l|l|l|l|l|l|l|l|l|l|l|l|l|l|l|l|l|l|l|l|l|l|l|l|l|l|l|l|l|l|l|l|l|l|l|l|l|l|l|l|l|l|l|l|l|l|l|l|l|l|l|}\hline
 1&2&3&4&5&6&7&8&9&10&11&12&13&14&15&16&17&18&19&20&21&22&23&24&25&26&27&28&29&30&31&32&33&34&35&36&37&38&39&40&41&42&43&44&45 \\ \hline 
 12500&3800&650&150&4933&250&400&4&3&2&566&200&133&400&9366&700&3366&500&3&2400&366&2100&100&2&66&10&100&33&3&3400&15&4333&1500&5000&14766&2300&2900&6700&2133&3200&1866&13366&6750&7550&2200 \\ \hline
\end{tabular}
}
\caption{Sanskrit words: the first row represents letters of the Sanskrit alphabet in the serial order}
\end{table}

\begin{table}
\begin{center}
\resizebox{15cm}{4.5cm}{
\begin{tabular}{|l|l|l|l|l|l|l|}\hline
k & lnk & lnk/$lnk_{lim}$ & f & lnf & lnf/$lnf_{max}$ & lnf/$lnf_{nextmax}$\\\hline
1 & 0    & 0             & 14766 & 9.60 & 1 & Blank\\\hline
2 & 0.69 & 0.193         & 13366  & 9.50 & 0.990 & 1\\\hline
3 & 1.10 & 0.307         & 12500  & 9.43 & 0.982 & 0.993\\\hline
4 & 1.39 & 0.388         & 9366 & 9.14 & 0.952 & 0.962\\\hline
5 & 1.61 & 0.450         & 7550 & 8.93 & 0.930 & 0.940\\\hline
6 & 1.79 & 0.500         & 6750 & 8.82 & 0.919 & 0.928\\\hline
7 & 1.95 & 0.545         & 6700&  8.81 & 0.918 & 0.927\\
\hline
\end{tabular}
}
\end{center}
\caption{Sanskrit words: ranking, natural logarithm, normalisations}
\end{table}
\clearpage
\begin{table}
\resizebox{15cm}{.5cm}{
\begin{tabular}{|l|l|l|l|l|l|l|l|l|l|l|l|l|l|l|l|l|l|l|l|l|l|l|l|l|l|l|l|l|l|l|l|l|l|l|l|l|l|l|l|l|l|l|l|l|l|l|l|l|}\hline
 1&2&3&4&5&6&7&8&9&10&11&12&13&14&15&16&17&18&19&20&21&22&23&24&25&26&27&28&29&30&31&32&33&34&35&36&37&38&39&40&41&42&43&44&45&46&47&48&49 \\ \hline 
 5933&928&438&35&543&53&1138&105&35&0&350&158&35&158&70&18&3710&140&1750&158&0&665&718&18&23&35&35&3&35&3&9&1698&6&2433&350&2240&6300&105&1656&4043&665&1348&823&4316&963&70&5506&1400&70 \\ \hline
\end{tabular}
}
\caption{Sinhalese words: the first row represents letters of the Sinhalese alphabet in the serial order}
\end{table}
\begin{table}
\begin{center}
\resizebox{15cm}{4.5cm}{
\begin{tabular}{|l|l|l|l|l|l|l|}\hline
k & lnk & lnk/$lnk_{lim}$ & f & lnf & lnf/$lnf_{max}$ & lnf/$lnf_{nextmax}$\\\hline
1 & 0    & 0             & 6300 & 8.75 & 1 & Blank\\\hline
2 & 0.69 & 0.201         & 5933& 8.69 & 0.993 & 1\\\hline
3 & 1.10 & 0.321        & 5506& 8.61 & 0.984 & 0.991\\\hline
4 & 1.39 & 0.405         & 4316 & 8.37 & 0.957 & 0.963\\\hline
5 & 1.61 & 0.469         & 4043 & 8.30 & 0.949 & 0.955\\\hline
6 & 1.79 & 0.522         & 3710 & 8.22 & 0.939 & 0.946\\\hline
7 & 1.95 & 0.569         & 2433& 7.80 & 0.891 & 0.898\\\hline
8 & 2.08 & 0.606         & 2240 & 7.71 & 0.881 & 0.887\\\hline
9 & 2.20 & 0.641         & 1750 & 7.47 & 0.854 & 0.860\\\hline
10 & 2.30 & 0.671        & 1698 & 7.44 & 0.850 & 0.856\\\hline
11 & 2.40 & 0.700        & 1656& 7.41 & 0.847 & 0.853\\\hline
12 & 2.48 & 0.723        & 1400 & 7.24 & 0.827 & 0.833\\\hline
13 & 2.56 & 0.746        & 1348 & 7.21 & 0.824 & 0.830\\\hline
14&2.64   & 0.770        &1138  & 7.04& 0.805&0.810\\\hline
15&2.71   &0.790         &963  &6.87 &0.785 &0.791\\\hline
16&2.77   &0.808         &928  &6.83 &0.781 &0.786\\\hline
17&2.83   &0.825         &823  &6.71 &0.767 &0.772\\\hline
18&2.89   &0.843         &718  &6.58 &0.752 &0.757\\\hline
19&2.94   &0.857         &665  &6.50 &0.743 &0.748\\\hline
20&3.00   &0.875         &543  &6.30 &0.720 &0.725\\\hline
21&3.04   &0.886         &438  &6.08 &0.695 &0.700\\\hline
22&3.09   &0.901         &350  &5.86  &0.670 &0.674\\\hline
23&3.14   &0.915         &158  &5.06 &0.578 &0.582\\\hline
24&3.18   &0.927         &140  &4.94 &0.565 &0.568\\\hline
25&3.22   &0.939         &105  &4.65 &0.531 &0.535\\\hline
26&3.26   &0.950         &70  &4.25 &0.486 &0.489\\\hline
27&3.30   &0.962         &53  &3.97 &0.454 &0.457\\\hline
28&3.33   &0.971         &35  &3.56 &0.407 &0.410\\\hline
29&3.37   &0.983         &23   &3.14 &0.359 &0.361\\\hline
30&3.40   &0.991         &18   &2.89 &0.330 &0.333\\\hline
31&3.43   & 1            &1  &0 &0 &0\\\hline
\end{tabular}
}
\end{center}
\caption{Sinhalese words: ranking, natural logarithm, normalisations}
\end{table}
\clearpage
\begin{table}
\resizebox{14cm}{.5cm}{
\begin{tabular}{|l|l|l|l|l|l|l|l|l|l|l|l|l|l|l|l|l|l|l|l|l|l|l|l|l|l|l|l|l|l|l|}\hline
 A&B&C&D&E&F&G&H&I&J&K&L&M&N&O&P&Q&R&S&T&U&V&W&X&Y&Z \\ \hline 
 110&330&115&143&28&60&101&135&61&44&297&83&231&88&99&182&11&157&426&215&33&168&77&5&22&27 \\ \hline
\end{tabular}
}
\caption{South African English words: the first row represents letters of the South African English  alphabet in the serial order}
\end{table}

\begin{table}
\begin{center}
\resizebox{14cm}{4.5cm}{
\begin{tabular}{|l|l|l|l|l|l|l|l|l|}\hline
k & lnk & lnk/$lnk_{lim}$ & f & lnf & lnf/$lnf_{max}$ & lnf/$lnf_{nextmax}$&lnf/$lnf_{nnmax}$&lnf/$lnf_{nnnmax}$\\\hline
1 & 0    & 0             & 426 & 6.05 & 1 & Blank&Blank&Blank\\\hline
2 & 0.69 & 0.209         & 330& 5.80 & 0.959 & 1&Blank&Blank\\\hline
3 & 1.10 & 0.333         & 297& 5.70 & 0.942 & 0.983&1&Blank\\\hline
4 & 1.39 & 0.421         & 231 &5.44 & 0.899 & 0.938&.954&1 \\\hline
5 & 1.61 & 0.488         & 215 & 5.37 & 0.888 & 0.926&.942&.987\\\hline
6 & 1.79 & 0.542         & 182 & 5.20 & 0.860 & 0.897&.912&.956\\\hline
7 & 1.95 & 0.591         & 168& 5.12 & 0.846 & 0.883&.898&.941\\
\hline
\end{tabular}
}
\end{center}
\caption{South African English words: ranking, natural logarithm, normalisations}
\end{table}
\clearpage
\begin{table}
\resizebox{14cm}{0.5cm}{
\begin{tabular}{|l|l|l|l|l|l|l|l|l|l|l|l|l|l|l|l|l|l|l|l|l|l|l|l|l|l|l|l|l|}\hline
 A&B&C&D&E&F&G&H&I&J&K&L&Ll&M&N&\^{N}&O&P&Q&R&S&T&U&V&X&Y&Z \\ \hline 
 2310&858&3465&1607&2205&840&770&665&805&298&1400&850&90&1663&333&20&350&1830&123&1085&1015&1143&123&793&10&35&175 \\ \hline
\end{tabular}
}
\caption{Spanish words: the first row represents letters of the Spanish alphabet in the serial order}
\end{table}
\begin{table}
\begin{center}
\resizebox{14cm}{4.5cm}{
\begin{tabular}{|l|l|l|l|l|l|l|}\hline
k & lnk & lnk/$lnk_{lim}$ & f & lnf & lnf/$lnf_{max}$ &lnf/$lnf_{next-max}$\\\hline
1 & 0    & 0     & 3465 & 8.15 & 1 & Blank\\\hline
2 & 0.69 & 0.212 & 2310 & 7.75 & 0.951& 1 \\\hline
3 & 1.10 & 0.337 & 2205 & 7.70 & 0.945& 0.994 \\\hline
4 & 1.39 & 0.426 & 1830 & 7.51 & 0.921& 0.969 \\\hline
5 & 1.61 & 0.494 & 1663 & 7.42 & 0.910& 0.957 \\\hline
6 & 1.79 & 0.549 & 1607 & 7.38 & 0.906& 0.952 \\\hline
7 & 1.95 & 0.598 & 1400 & 7.24 & 0.888& 0.934 \\\hline
8 & 2.08 & 0.638 & 1143 & 7.04 & 0.864& 0.908 \\\hline
9 & 2.20 & 0.675 & 1085 & 6.99 & 0.858& 0.902 \\\hline
10 & 2.30 & 0.706 & 1015& 6.92 & 0.849& 0.893 \\\hline
11 & 2.40 & 0.736 & 858 & 6.75 & 0.828& 0.871 \\\hline
12 & 2.48 & 0.761 & 840 & 6.73 & 0.826& 0.868 \\\hline
13 & 2.56 & 0.785 & 805 & 6.69 & 0.821& 0.863 \\\hline
14 & 2.64 & 0.810 & 793 & 6.68 & 0.820& 0.862 \\\hline
15 & 2.71 & 0.831 & 770 & 6.65 & 0.816& 0.858 \\\hline
16 & 2.77 & 0.850 & 665 & 6.50 & 0.798& 0.839 \\\hline
17 & 2.83 & 0.868 & 350 & 5.86 & 0.719& 0.756 \\\hline
18 & 2.89 & 0.887 & 333 & 5.81 & 0.713& 0.750 \\\hline
19 & 2.94 & 0.902 & 298 & 5.70 & 0.699& 0.735 \\\hline
20 & 3.00 & 0.920 & 175 & 5.16 & 0.633& 0.666 \\\hline
21 & 3.04 & 0.933 & 123 & 4.81 & 0.590& 0.621 \\\hline
22 & 3.09 & 0.948 & 90 & 4.50 & 0.552& 0.581 \\\hline
23 & 3.14 & 0.963 & 35  & 3.56 & 0.437& 0.459 \\\hline
24 & 3.18 & 0.975 & 20  & 3.00 & 0.368& 0.387    \\\hline
25 & 3.22 & 0.988 & 10  & 2.30 & 0.282 & 0.297\\\hline
26 & 3.26 & 1     & 1   & 0    &0 &0\\\hline
\end{tabular}
}
\end{center}
\caption{Spanish words: ranking, natural logarithm, normalisations}
\end{table}
\clearpage
\begin{table}
\resizebox{14cm}{.5cm}{
\begin{tabular}{|l|l|l|l|l|l|l|l|l|l|l|l|l|l|l|l|l|l|l|l|l|l|l|l|l|l|l|l|l|l|}\hline
 A&\v{A}...o&U&Ai&Au&Aw&Oi&B&Chy&D&G&H&J&K&HK&L&M&N&P&HP&R&S&Sh&T&Ts&Ht&V&W&Y&Z \\ \hline 
 570&0&200&8&5&20&23&350&270&640&930&3&440&860&730&760&1080&870&240&420&240&510&630&200&130&320&14&240&230&80 \\ \hline
\end{tabular}
}
\caption{Kachin words: the first row represents letters of the Kachin alphabet in the serial order}
\end{table}
\begin{table}
\begin{center}
\resizebox{14cm}{4.5cm}{
\begin{tabular}{|l|l|l|l|l|l|l|}\hline
k & lnk & lnk/$lnk_{lim}$ & f & lnf & lnf/$lnf_{max}$ & lnf/$lnf_{next-next-max}$\\\hline
1 & 0 & 0        & 1080 & 6.98 & 1  & Blank\\\hline
2 & 0.69 & 0.209 & 930 & 6.84 & 0.980 & 1\\\hline
3 & 1.10 & 0.333 & 870 & 6.77 & 0.970 & 0.990\\\hline
4 & 1.39 & 0.421 & 860 & 6.76 & 0.968 & 0.988\\\hline
5 & 1.61 & 0.488 & 760 & 6.63 & 0.950 &0.969\\\hline
6 & 1.79 & 0.542 & 730 & 6.59 & 0.944 &0.963\\\hline
7 & 1.95 & 0.591 & 640 & 6.46 & 0.926 &0.944\\\hline
8 & 2.08 & 0.630 & 630 & 6.45 & 0.924 &0.943\\\hline
9 & 2.20 & 0.667 & 570 & 6.35 & 0.910 &0.928\\\hline
10 & 2.30 & 0.697 & 510 & 6.23 & 0.893&0.911\\\hline
11 & 2.40 & 0.727 & 440 & 6.09 & 0.872&0.890\\\hline
12 & 2.48 & 0.752 & 420 & 6.04 & 0.865&0.883\\\hline
13 & 2.56 & 0.776 & 350 & 5.86 & 0.840&0.857\\\hline
14 & 2.64 & 0.800 & 320 & 5.77 & 0.827&0.844\\\hline
15 & 2.71 & 0.821 & 270 & 5.60 & 0.802&0.819\\\hline
16 & 2.77 & 0.839 & 240 & 5.48 & 0.785&0.801\\\hline
17 & 2.83 & 0.858 & 230 & 5.44 & 0.779&0.795\\\hline
18 & 2.89 & 0.876 & 200 & 5.30 & 0.759&0.775 \\\hline
19&2.94 &0.891    &130   &4.87&0.698   &0.712\\\hline
20&3.00 & 0.909   &80   &4.38&0.628   &0.640 \\\hline
21&3.04 & 0.921  &23    &3.14 &0.450  &0.459 \\\hline
22&3.09 & 0.936  &20    &3.00 &0.430  &0.439 \\\hline
23&3.14 & 0.952  &14    &2.64 &0.378  &0.386 \\\hline
24&3.18 & 0.964  &8    &2.08 &0.298 &0.304 \\\hline
25&3.22 & 0.976  &5     &1.61 &0.231  &0.235 \\\hline
26&3.26 & 0.988  &3     &1.10    & 0.158 &0.161      \\\hline
27&3.30 & 1      &1     & 0     & 0  & 0  \\\hline
\end{tabular}
}
\end{center}
\caption{Kachin words: ranking, natural logarithm, normalisations}
\end{table}
\clearpage
\begin{table}
\resizebox{14cm}{.5cm}{
\begin{tabular}{|l|l|l|l|l|l|l|l|l|l|l|l|l|l|l|l|l|l|l|l|l|l|l|l|l|l|l|l|l|l|l|l|l|l|l|l|l|l|}\hline
Alef&be&pe&te&be&se&ji&che&he&khe&dal&d$\bar{a}$l&zal&re&re&ze&zhe&Sin&Shin&sad&zad&toe&zoe&ain&ghain&fe&qaf&kaf&gaf&lam&mlm&nun&vao&he&y$\bar{a}$ \\\hline 
3159&2327&1440&1400&307&52&1050&980&517&700&1252&217&97&805&0&324&17&1427&542&332&105&236&44&595&263&560&577&1820&1111&709&3491&1339&394&753&140\\\hline
\end{tabular}
}
\caption{Urdu words: the first row represents letters of the Urdu alphabet in the serial order}
\end{table}
\begin{table}
\begin{center}
\resizebox{14cm}{4.5cm}{
\begin{tabular}{|l|l|l|l|l|l|l|l|l|l|}\hline
k & lnk & lnk/$lnk_{lim}$ & f & lnf & lnf/$lnf_{max}$ & lnf/$lnf_{nextmax}$&lnf/$lnf_{nnmax}$&lnf/$lnf_{nnnmax}$\\\hline
1 & 0    & 0             & 3491 & 8.16 & 1 & Blank&Blank&Blank\\\hline
2 & 0.69 & 0.194         & 3159& 8.06 & 0.988 & 1&Blank&Blank\\\hline
3 & 1.10 & 0.309         & 2327& 7.75 & 0.950 & 0.962&1&Blank\\\hline
4 & 1.39 & 0.390         & 1820 & 7.51 & 0.920 & 0.932&.969&1\\\hline
5 & 1.61 & 0.452         & 1440 & 7.27 & 0.891 & 0.902&.938&.968\\\hline
6 & 1.79 & 0.503         & 1427 & 7.26 & 0.890 & 0.901&.937&.967\\\hline
7 & 1.95 & 0.548         & 1400& 7.24 & 0.887 & 0.898&.934&.964\\
\hline
\end{tabular}
}
\end{center}
\caption{Urdu words: ranking, natural logarithm, normalisations}
\end{table}
\clearpage
\begin{table}
\resizebox{14cm}{.5cm}{
\begin{tabular}{|l|l|l|l|l|l|l|l|l|l|l|l|l|l|l|l|l|l|l|l|l|l|l|l|l|l|l|l|l|l|l|}\hline
 A&B&C&D&E&F&G&H&I&J&K&L&M&N&O&P&Q&R&S&T&U&V&W&X&Y&Z \\ \hline 
 5540&4970&9170&4900&3640&4095&3150&3605&3521&875&816&3196&5180&1960&2170&8400&508&4585&11760&5705&1610&1680&2870&88&298&245 \\ \hline
\end{tabular}
}
\caption{English words: the first row represents letters of the English alphabet in the serial order}
\end{table}
\begin{table}
\begin{center}
\resizebox{14cm}{4.5cm}{
\begin{tabular}{|l|l|l|l|l|l|l|l|l|l|l|l|l|l|l|}\hline
k & lnk & lnk/$lnk_{lim}$ & f & lnf & lnf/$lnf_{max}$ & lnf/$lnf_{nextmax}$&lnf/$lnf_{nnmax}$&lnf/$lnf_{nnnmax}$&lnf/$lnf_{nnnnmax}$&lnf/$lnf_{nnnnnmax}$&lnf/$lnf_{nnnnnnmax}$&lnf/$lnf_{nnnnnnnnmax}$&lnf/$lnf_{nnnnnnnnnmax}$&lnf/$lnf_{nnnnnnnnnnmax}$\\\hline
1 & 0    & 0             & 11760 & 9.37 & 1 & Blank & Blank & Blank& Blank& Blank& Blank& Blank& Blank& Blank\\\hline
2 & 0.69 & 0.212         & 9170& 9.12 & 0.973 & 1  & Blank  & Blank& Blank& Blank& Blank& Blank& Blank& Blank\\\hline
3 & 1.10 & 0.337         & 8400& 9.04 & 0.965 & 0.991& 1    & Blank& Blank& Blank& Blank& Blank& Blank& Blank\\\hline
4 & 1.39 & 0.426         & 5705 & 8.65 & 0.923 & 0.948&0.957 & 1   & Blank& Blank& Blank& Blank& Blank& Blank\\\hline
5 & 1.61 & 0.494         & 5540 & 8.62 & 0.920 & 0.945&0.954 & 0.997& 1   & Blank& Blank& Blank& Blank& Blank\\\hline
6 & 1.79 & 0.549         & 5180 & 8.55 & 0.912 & 0.938&0.946 & 0.988& 0.992& 1   & Blank& Blank& Blank& Blank\\\hline
7 & 1.95 & 0.598         & 4970& 8.51 & 0.908 & 0.933&0.941 & 0.984& 0.987& 0.995& 1     & Blank& Blank& Blank\\
\hline
\end{tabular}
}
\end{center}
\caption{English words: ranking, natural logarithm, normalisations}
\end{table}
\clearpage
\begin{table}
\resizebox{14cm}{.5cm}{
\begin{tabular}{|l|l|l|l|l|l|l|l|l|l|l|l|l|l|l|l|l|l|l|l|l|l|l|l|l|l|l|l|l|l|l|l|l|l|l|l|l|l|l|l|l|l|l|l|l|}\hline
 1&2&3&4&5&6&7&8&9&10&11&12&13&14&15&16&17&18&19&20&21&22&23&24&25&26&27&28&29&30&31&32&33&34&35 \\ \hline 
 1218&434&84&8&525&2&13&105&56&140&63&1904&882&1246&11&1778&1470&0&2590&1358&896&1848&560&2163&924&1512&140&784&840&336&336&168&4&2184&672 \\ \hline
\end{tabular}
}
\caption{Nepali words: the first row represents letters of the Nepali alphabet in the serial order}
\end{table}
\begin{table}
\begin{center}
\resizebox{14cm}{4.5cm}{
\begin{tabular}{|l|l|l|l|l|l|l|l|l|l|l|l|}\hline
k & lnk & lnk/$lnk_{lim}$ & f & lnf & lnf/$lnf_{max}$ & lnf/$lnf_{nextmax}$&lnf/$lnf_{nnmax}$&lnf/$lnf_{nnnmax}$&lnf/$lnf_{nnnnmax}$&lnf/$lnf_{nnnnnmax}$&lnf/$lnf_{nnnnnnnnnmax}$\\\hline
1 & 0    & 0             & 2590 & 7.86 & 1     & Blank & Blank& Blank& Blank& Blank& Blank\\\hline
2 & 0.69 & 0.207         & 2184 & 7.69 & 0.978 & 1     & Blank& Blank& Blank& Blank& Blank\\\hline
3 & 1.10 & 0.330         & 2163 & 7.68 & 0.977 & 0.999 & 1    & Blank& Blank& Blank& Blank\\\hline
4 & 1.39 & 0.417         & 1904 & 7.55 & 0.961 & 0.982 & 0.983& 1    & Blank& Blank& Blank\\\hline
5 & 1.61 & 0.483         & 1848 & 7.52 & 0.957 & 0.978 & 0.979&0.996 & 1    & Blank& Blank \\\hline
6 & 1.79 & 0.538         & 1778 & 7.48 & 0.952 & 0.973 & 0.974&0.991 & 0.995& 1    & Blank\\\hline
7 & 1.95 & 0.586         & 1512 & 7.32 & 0.931 & 0.952 & 0.953&0.970 & 0.973& 0.979& Blank \\
\hline
\end{tabular}
}
\end{center}
\caption{Nepali words: ranking, natural logarithm, normalisations}
\end{table}
\clearpage
\begin{table}
\resizebox{14cm}{.5cm}{
\begin{tabular}{|l|l|l|l|l|l|l|l|l|l|l|l|l|l|l|l|l|l|l|l|l|}\hline
 A&B&C&D&E&G&H&I&J&K&L&M&N&O&P&R&S&T&U&W \\ \hline 
 1278&1558&822&718&115&1365&94&88&508&766&41&1418&718&251&648&1540&1400&570&140&443 \\ \hline
\end{tabular}
}
\caption{Garo words: the first row represents letters of the Garo alphabet in the serial order}
\end{table}
\begin{table}
\begin{center}
\begin{tabular}{|l|l|l|l|l|l|l|}\hline
k & lnk & lnk/$lnk_{lim}$ & f & lnf & lnf/$lnf_{max}$ &lnf/$lnf_{next-max}$\\\hline
1 & 0 & 0                 & 1558 & 7.35 & 1 & Blank\\\hline
2 & 0.69 & 0.230          & 1540 & 7.34 & 0.999& 1 \\\hline
3 & 1.10 & 0.367          & 1418 & 7.26 & 0.988& 0.989 \\\hline
4 & 1.39 & 0.463          & 1400 & 7.24 & 0.985& 0.986 \\\hline
5 & 1.61 & 0.537          & 1365 & 7.22 & 0.982& 0.984 \\\hline
6 & 1.79 & 0.597          & 1278 & 7.15 & 0.973& 0.974 \\\hline
7 & 1.95 & 0.650          & 822 & 6.71 & 0.913& 0.914 \\\hline
8 & 2.08 & 0.693          & 766 & 6.64 & 0.903& 0.905\\\hline
9 & 2.20 & 0.733          & 718 & 6.58 & 0.895&0.896\\\hline
10 & 2.30 & 0.767         & 648 & 6.47 & 0.880&0.881 \\\hline
11 & 2.40 & 0.800         & 570 & 6.35 & 0.864&0.865 \\\hline
12 & 2.48 & 0.827         & 508 & 6.23 & 0.848&0.849 \\\hline
13 & 2.56 & 0.853         & 443 & 6.09 & 0.829&0.830 \\\hline
14 & 2.64 & 0.880         & 251 & 5.23 & 0.712&0.713 \\\hline
15 & 2.71 & 0.903         & 140 & 4.94  & 0.672&0.673 \\\hline
16 & 2.77 & 0.923         & 115 & 4.74 & 0.645& 0.646\\\hline
17 & 2.83 & 0.943         & 94 & 4.54 & 0.618& 0.619\\\hline
18 & 2.89 & 0.963         & 88 & 4.48 & 0.610& 0.610 \\\hline
19 & 2.94 & 0.980         & 41 & 3.71 & 0.505&0.505 \\\hline
20 & 3.00 & 1             & 1 & 0    & 0  &   0 \\\hline
\end{tabular}
\end{center}
\caption{Garo words: ranking, natural logarithm, normalisations}
\end{table}
\clearpage
\begin{table}
\resizebox{14cm}{.5cm}{
\begin{tabular}{|l|l|l|l|l|l|l|l|l|l|l|l|l|l|l|l|l|l|l|l|l|l|}\hline
 A&AW&B&CH&D&E&F&G&H&I&K&L&M&N&O&P&R&S&T&U&V&Z \\ \hline 
 900&350&1450&2225&1075&525&425&11&2610&1520&2950&1260&950&1180&3&1260&1160&1360&5475&25&775&1140 \\ \hline
\end{tabular}
}
\caption{Lushai or, Mizo words: the first row represents letters of the Lushai alphabet in the serial order}
\end{table}
\begin{table}
\begin{center}
\resizebox{14cm}{4.5cm}{
\begin{tabular}{|l|l|l|l|l|l|l|l|l|l|l|l|l|}\hline
k & lnk  & lnk/$lnk_{lim}$ & f & lnf & lnf/$lnf_{max}$ & lnf/$lnf_{next-max}$&lnf/$lnf_{nnmax}$&lnf/$lnf_{nnnmax}$&lnf/$lnf_{nnnnmax}$&lnf/$lnf_{nnnnnmax}$&lnf/$lnf_{nnnnnnnnnnmax}$\\\hline
1 & 0    & 0     & 5475 & 8.61 & 1 & Blank&Blank&Blank  & Blank&Blank&Blank\\\hline
2 & 0.69 & 0.223 & 2950 & 7.99 & 0.928 & 1&Blank&Blank  & Blank&Blank&Blank\\\hline
3 & 1.10 & 0.356 & 2610 & 7.87 & 0.914 & 0.985&1&Blank  & Blank&Blank&Blank\\\hline
4 & 1.39 & 0.450 & 2225 & 7.71 & 0.895 & 0.965&.980&1   & Blank&Blank&Blank\\\hline
5 & 1.61 & 0.521 & 1520 & 7.33 & 0.851 & 0.917&.931&.951& 1    &Blank&Blank\\\hline
6 & 1.79 & 0.579 & 1450 & 7.28 & 0.846 & 0.911&.925&.944& 0.993& 1   &Blank\\\hline
7 & 1.95 & 0.631 & 1360 & 7.22 & 0.839 & 0.904&.917&.936& 0.985&0.992 &Blank\\
\hline
\end{tabular}
}
\end{center}
\caption{Lushai or, Mizo words: ranking,natural logarithm, normalisations}
\end{table}
\clearpage
\begin{table}
\resizebox{14cm}{.5cm}{
\begin{tabular}{|l|l|l|l|l|l|l|l|l|l|l|l|l|l|l|l|l|l|l|l|l|l|l|l|l|l|l|}\hline
 A&B&C&D&E&F&G&H&I&J&K&L&M&N&O&P&Q&R&S&T&U&V&W&X&Y&Z \\ \hline 
 3080&2860&88&997&1496&1247&1584&1408&352&154&1760&1071&1364&880&352&1254&88&1232&3457&968&1540&2156&1364&10&7&1232 \\ \hline
\end{tabular}
}
\caption{German words: the first row represents letters of the German alphabet in the serial order}
\end{table}
\begin{table}
\begin{center}
\resizebox{14cm}{4.5cm}{
\begin{tabular}{|l|l|l|l|l|l|l|l|l|l|l|l|}\hline
k&lnk&lnk/$lnk_{lim}$&f&lnf&lnf/$lnf_{max}$&lnf/$lnf_{next-max}$&lnf/$lnf_{nnmax}$&lnf/$lnf_{nnnmax}$&lnf/$lnf_{nnnnmax}$&lnf/$lnf_{nnnnnmax}$&lnf/$lnf_{nnnnnnnnnmax}$\\\hline
1 & 0 & 0        & 3457 & 8.15 & 1     & Blank& Blank& Blank& Blank& Blank& Blank\\\hline
2 & 0.69 & 0.220 & 3080 & 8.03 & 0.985 & 1    & Blank& Blank& Blank& Blank& Blank\\\hline
3 & 1.10 & 0.350 & 2860 & 7.96 & 0.977 & 0.991& 1    & Blank& Blank& Blank& Blank\\\hline
4 & 1.39 & 0.443 & 2156 & 7.68 & 0.942 & 0.956&0.965& 1     & Blank& Blank& Blank\\\hline
5 & 1.61 & 0.513 & 1760 & 7.47 & 0.917 &0.930 &0.938&0.973  &  1   & Blank& Blank \\\hline
6 & 1.79 & 0.570 & 1584 & 7.37 & 0.904 &0.918 &0.926&0.960  & 0.987& 1    & Blank \\\hline
7 & 1.95 & 0.621 & 1540 & 7.34 & 0.901 &0.914 &0.922&0.956  & 0.983&0.996 & Blank\\
\hline
\end{tabular}
}
\end{center}
\caption{German words: ranking,natural logarithm, normalisations}
\end{table}
\clearpage
\begin{table}
\resizebox{14cm}{.5cm}{
\begin{tabular}{|l|l|l|l|l|l|l|l|l|l|l|l|l|l|l|l|l|l|l|l|l|l|l|l|l|l|l|l|l|l|l|l|l|l|l|l|l|l|l|l|l|}\hline
 1&2&3&4&5&6&7&8&9&10&11&12&13&14&15&16&17&18&19&20&21&22&23&24&25&26&27&28&29&30&31&32&33&34&35&36&37 \\ \hline 
 612&672&72&372&204&8&96&7&1020&504&528&244&1&744&588&2&240&96&144&120&1&360&144&432&168&480&912&264&1080&408&792&5&288&480&1176&504&5 \\ \hline
\end{tabular}
}
\caption{Assamese words: the first row represents letters of the Assamese alphabet in the serial order}
\end{table}
\begin{table}
\begin{center}
\resizebox{14cm}{4.5cm}{
\begin{tabular}{|l|l|l|l|l|l|l|}\hline
k & lnk & lnk/$lnk_{lim}$ & f & lnf & lnf/$lnf_{max}$ &lnf/$lnf_{next-max}$\\\hline
1 & 0    & 0     & 1176 & 7.07 & 1 & Blank\\\hline
2 & 0.69 & 0.212 & 1080 & 6.98 & 0.987& 1 \\\hline
3 & 1.10 & 0.337 & 1020 & 6.93 & 0.980& 0.993 \\\hline
4 & 1.39 & 0.426 & 912 & 6.82 & 0.965& 0.977 \\\hline
5 & 1.61 & 0.494 & 792 & 6.67 & 0.943& 0.956 \\\hline
6 & 1.79 & 0.549 & 744 & 6.61 & 0.935& 0.947 \\\hline
7 & 1.95 & 0.598 & 672 & 6.51 & 0.921& 0.933 \\\hline
8 & 2.08 & 0.638 & 612 & 6.42 & 0.908& 0.920 \\\hline
9 & 2.20 & 0.675 & 588 & 6.38 & 0.902& 0.914 \\\hline
10 & 2.30 & 0.706 & 528& 6.27 & 0.887& 0.898 \\\hline
11 & 2.40 & 0.736 & 504 & 6.22 & 0.880& 0.891 \\\hline
12 & 2.48 & 0.761 & 480 & 6.17 & 0.873& 0.884 \\\hline
13 & 2.56 & 0.785 & 432 & 6.07 & 0.859& 0.870 \\\hline
14 & 2.64 & 0.810 & 408 & 6.01 & 0.850& 0.861 \\\hline
15 & 2.71 & 0.831 & 372 & 5.92 & 0.837& 0.848 \\\hline
16 & 2.77 & 0.850 & 360 & 5.89 & 0.833& 0.844 \\\hline
17 & 2.83 & 0.868 & 288 & 5.66 & 0.801& 0.811 \\\hline
18 & 2.89 & 0.887 & 264 & 5.58 & 0.789& 0.799 \\\hline
19 & 2.94 & 0.902 & 240 & 5.48 & 0.775& 0.785 \\\hline
20 & 3.00 & 0.920 & 204 & 5.32 & 0.752& 0.762 \\\hline
21 & 3.04 & 0.933 & 168 & 5.12 & 0.724& 0.734 \\\hline
22 & 3.09 & 0.948 & 144 & 4.97 & 0.703& 0.712 \\\hline
23 & 3.14 & 0.963 & 120  & 4.79 & 0.678& 0.686 \\\hline
24 & 3.18 & 0.975 & 96  & 4.56 & 0.645& 0.653    \\\hline
25 & 3.22 & 0.988 & 72  & 4.28 & 0.605 & 0.613\\\hline
26 & 3.26 & 1     & 1   & 0    &0 &0\\\hline
\end{tabular}
}
\end{center}
\caption{Assamese words: ranking,natural logarithm, normalisations}
\end{table}
\clearpage
\begin{table}
\resizebox{14cm}{.5cm}{
\begin{tabular}{|l|l|l|l|l|l|l|l|l|l|l|l|l|l|l|l|l|l|l|l|l|}\hline
 A&B&D&E&G&I&J&K&L&M&N&O&P&R&S&T&U&Y \\ \hline 
 824&331&328&235&256&210&107&650&471&552&390&162&585&270&568&656&65&325 \\ \hline
\end{tabular}
}
\caption{Abor-Miri words: the first row represents letters of the Abor-Miri alphabet in the serial order}
\end{table}
\begin{table}
\begin{center}
\resizebox{14cm}{4.5cm}{
\begin{tabular}{|l|l|l|l|l|l|l|l|l|l|l|l|l|l|}\hline
k & lnk & lnk/$lnk_{lim}$ & f & lnf & lnf/$lnf_{max}$ & lnf/$lnf_{next-max}$&lnf/$lnf_{nnmax}$&lnf/$lnf_{nnnmax}$&lnf/$lnf_{nnnnmax}$&lnf/$lnf_{nnnnnmax}$&lnf/$lnf_{nnnnnnmax}$&lnf/$lnf_{nnnnnnnmax}$&lnf/$lnf_{nnnnnnmax}$\\\hline
1 & 0 & 0                 & 824 & 6.71 & 1  & Blank & Blank& Blank& Blank& Blank& Blank& Blank& Blank\\\hline
2 & 0.69 & 0.235          & 656 & 6.49 & 0.967 & 1  & Blank& Blank& Blank& Blank& Blank& Blank& Blank\\\hline
3 & 1.10 & 0.374          & 650 & 6.48 & 0.966 & 0.998& 1& Blank& Blank& Blank& Blank& Blank& Blank\\\hline
4 & 1.39 & 0.473          & 585 & 6.37 & 0.949 & 0.982 & 0.983& 1& Blank& Blank& Blank& Blank& Blank\\\hline
5 & 1.61 & 0.548          & 568 & 6.34 & 0.945 & 0.977& 0.978& 0.995& 1& Blank& Blank& Blank& Blank\\\hline
6 & 1.79 & 0.609          & 552 & 6.31 & 0.940 & 0.972& 0.974& 0.991& 0.995& 1& Blank& Blank& Blank\\\hline
7 & 1.95 & 0.663          & 471 & 6.15 & 0.917 & 0.948& 0.949& 0.965& 0.970& 0.975& 1& Blank& Blank\\
\hline
\end{tabular}
}
\end{center}
\caption{Abor-Miri words: ranking,natural logarithm, normalisations}
\end{table}
\clearpage
\begin{table}
\resizebox{12cm}{.5cm}{
\begin{tabular}{|l|l|l|l|l|l|l|l|l|l|l|l|l|l|l|l|l|l|l|l|l|l|l|l|l|l|l|l|l|l|l|l|l|l|l|l|l|}\hline
 1&2&3&4&5&6&7&8&9&10&11&12&13&14&15&16&17&18&19&20&21&22&23&24&25&26&vao&28 \\ \hline 
 525&1080&300&270&960&1380&1200&915&255&1440&660&1320&1150&760&345&600&915&1970&795&1210&1680&1005&945&1140&2050&1230&1330&180\\ \hline
\end{tabular}
}
\caption{Arabian words: the first row represents letters of the Arabian alphabet in the serial order}
\end{table}
\begin{table}
\begin{center}{}
\resizebox{14cm}{4.5cm}{
\begin{tabular}{|l|l|l|l|l|l|l|l|l|l|l|}\hline
k&lnk&lnk/$lnk_{lim}$&f&lnf&lnf/$lnf_{max}$&lnf/$lnf_{nmax}$&lnf/$lnf_{nnmax}$&lnf/$lnf_{nnnmax}$&lnf/$lnf_{nnnnmax}$&lnf/$lnf_{nnnnnmax}$\\\hline
1 & 0    & 0             & 2050 & 7.63 & 1 & Blank    & Blank& Blank& Blank& Blank\\\hline
2 & 0.69 & 0.207         & 1970& 7.59 & 0.995 & 1     & Blank& Blank& Blank& Blank\\\hline
3 & 1.10 & 0.330         & 1680& 7.43 & 0.974 & 0.979 & 1    &Blank & Blank   & Blank             \\\hline
4 & 1.39 & 0.417         & 1440 & 7.27 & 0.953 & 0.958&0.978 & 1    & Blank   & Blank          \\\hline
5 & 1.61 & 0.483         & 1380 & 7.23 & 0.948 & 0.953&0.973 &0.994 & 1       & Blank         \\\hline
6 & 1.79 & 0.538         & 1330 & 7.19 & 0.942 & 0.947&0.968 &0.989 &0.994   & 1\\\hline
7 & 1.95 & 0.586         & 1320& 7.19 & 0.942 & 0.947 &0.968 &0.989 &0.994   & 1     \\
\hline
\end{tabular}
}
\end{center}
\caption{Arabian words: ranking,natural logarithm, normalisations}
\end{table}
\clearpage
\begin{table}
\resizebox{14cm}{0.5cm}{
\begin{tabular}{|l|l|l|l|l|l|l|l|l|l|l|l|l|l|l|l|l|l|l|l|l|l|l|l|l|l|}\hline
 A&B&C&E&G&H&J&K&L&M&N&O&P&Q&R&S&T&\"{U}&U&V&W&Y&Z \\ \hline 
 24&8&74&753&7&102&14&315&134&230&524&251&185&8&153&219&377&29&12&42&47&162&99 \\ \hline
\end{tabular}
}
\caption{Lotha words: the first row represents letters of the Lotha alphabet in the serial order}
\end{table}
\begin{table}
\begin{center}
\resizebox{14cm}{4.5cm}{
\begin{tabular}{|l|l|l|l|l|l|l|l|l|l|}\hline
k & lnk & lnk/$lnk_{lim}$ & f & lnf & lnf/$lnf_{max}$ & lnf/$lnf_{next-max}$&lnf/$lnf_{nnmax}$&lnf/$lnf_{nnnmax}$\\\hline
1 & 0 & 0        & 753 & 6.62 & 1  & Blank&Blank&Blank\\\hline
2 & 0.69 & 0.220 & 524 & 6.26 & 0.946 & 1&Blank&Blank\\\hline
3 & 1.10 & 0.350 & 377 & 5.93 & 0.896 & 0.947&1&Blank\\\hline
4 & 1.39 & 0.443 & 315 & 5.75 & 0.869 & 0.919&.970&1\\\hline
5 & 1.61 & 0.513 & 251 & 5.53 & 0.835 &0.883&.933&.962\\\hline
6 & 1.79 & 0.570 & 230 & 5.44 & 0.822 &0.869&.917&.946\\\hline
7 & 1.95 & 0.621 & 219 & 5.39 & 0.814 &0.861&.909&.937\\\hline
8 & 2.08 & 0.662 & 185 & 5.22 & 0.789 &0.834&.880&.908\\\hline
9 & 2.20 & 0.701 & 162 & 5.09 & 0.769 &0.813&.858&.885\\\hline
10 & 2.30 & 0.732 & 153 & 5.03 & 0.760&0.804&.848&.875\\\hline
11 & 2.40 & 0.764 & 134 & 4.90 & 0.740&0.783&.826&.852\\\hline
12 & 2.48 & 0.790 & 102 & 4.62 & 0.698&0.738&.779&.803\\\hline
13 & 2.56 & 0.815 & 99 & 4.60 & 0.695&0.735&.776&.800\\\hline
14 & 2.64 & 0.841 & 74 & 4.30 & 0.650&0.687&.725&.748\\\hline
15 & 2.71 & 0.863 & 47 & 3.85 & 0.582&0.615&.649&.670\\\hline
16 & 2.77 & 0.882 & 42 & 3.74 & 0.565&0.597&.631&.650\\\hline
17 & 2.83 & 0.901 & 29 & 3.37 & 0.509&0.538&.568&.586\\\hline
18 & 2.89 & 0.920 & 24 & 3.18 & 0.480&0.510&.536&.553 \\\hline
19&2.94 &0.936    &14   &2.64&0.399   &0.422&.445&.459\\\hline
20&3.00 & 0.955   &12   &2.48&0.375   &0.396&.418&.431 \\\hline
21&3.04 & 0.968  &8    &2.08 &0.314  &0.332&.351 &.362\\\hline
22&3.09 & 0.984  &7    &1.95 &0.295 &0.312&.329 &.339\\\hline
23&3.14 & 1      &1    &0 &0  &0 &0&0\\\hline
\end{tabular}
}
\end{center}
\caption{Lotha words: ranking, natural logarithm, normalisations}
\end{table}
\clearpage
\begin{table}
\resizebox{14cm}{0.5cm}{
\begin{tabular}{|l|l|l|l|l|l|l|l|l|l|l|l|l|l|l|l|l|l|l|l|l|l|l|l|l|l|}\hline
 A&B&C&D&E&F&G&H&I&J&K&L&M&N&O&P&Q&S&T&U&V&W&X&Y&Z \\ \hline 
 986&20&34&15&1&4&80&61&134&46&348&98&220&70&1&242&44&183&216&5&31&12&26&24&68 \\ \hline
\end{tabular}
}
\caption{Sema words: the first row represents letters of the Sema alphabet in the serial order}
\end{table}
\begin{table}
\begin{center}
\resizebox{14cm}{4.5cm}{
\begin{tabular}{|l|l|l|l|l|l|l|l|l|l|}\hline
k & lnk & lnk/$lnk_{lim}$ & f & lnf & lnf/$lnf_{max}$ &lnf/$lnf_{next-max}$&lnf/$lnf_{nnmax}$&lnf/$lnf_{nnnmax}$\\\hline
1 & 0    & 0     & 986 & 6.89 & 1 & Blank&Blank&Blank\\\hline
2 & 0.69 & 0.217 & 348 & 5.85 & 0.849& 1 &Blank&Blank\\\hline
3 & 1.10 & 0.346 & 242 & 5.49 & 0.797& 0.938&1 &Blank\\\hline
4 & 1.39 & 0.437 &220  & 5.39 & 0.782& 0.921&.982 &1\\\hline
5 & 1.61 & 0.506 & 216 & 5.38 & 0.781& 0.920&.980&.998 \\\hline
6 & 1.79 & 0.563 & 183 & 5.21& 0.756& 0.891 &.949&.967\\\hline
7 & 1.95 & 0.613 & 134 & 4.90 & 0.711& 0.838&.893&.909 \\\hline
8 & 2.08 & 0.654 & 98 & 4.58 & 0.665& 0.783 &.834&.850\\\hline
9 & 2.20 & 0.692 & 80 & 4.38 & 0.636& 0.749 &.798&.813\\\hline
10 & 2.30 & 0.723 & 70 & 4.25 & 0.617& 0.726&.774&.788 \\\hline
11 & 2.40 & 0.755 & 68 & 4.22 & 0.612& 0.721&.769&.783 \\\hline
12 & 2.48 & 0.780 & 61 & 4.11 & 0.597& 0.703&.749&.763 \\\hline
13 & 2.56 & 0.805 & 46 & 3.83 & 0.556& 0.655&.698&.711 \\\hline
14 & 2.64 & 0.830 & 44 & 3.78 & 0.549& 0.646&.689&.701 \\\hline
15 & 2.71 & 0.852 & 34 & 3.53 & 0.512& 0.603&.643&.655 \\\hline
16 & 2.77 & 0.871 & 31 & 3.43 & 0.498& 0.586&.625&.636 \\\hline
17 & 2.83 & 0.890 & 26 & 3.26 & 0.473& 0.557&.594&.605 \\\hline
18 & 2.89 & 0.909 & 24 & 3.18 & 0.462& 0.544&.579&.590 \\\hline
19 & 2.94 & 0.925 & 20 & 3.00 & 0.435& 0.513&.546&.557 \\\hline
20 & 3.00 & 0.943 & 15 & 2.71 & 0.393& 0.463&.494&.503 \\\hline
21 & 3.04 & 0.956 & 12 & 2.48 & 0.360& 0.424&.452&.460 \\\hline
22 & 3.09 & 0.972 & 5 & 1.61 & 0.234& 0.275 &.293&.299\\\hline
23 & 3.14 & 0.987 & 4  & 1.39 & 0.202& 0.238&.253&.258 \\\hline
24 & 3.18 & 1 & 1  & 0   & 0& 0 &0&0   \\\hline
\end{tabular}
}
\end{center}
\caption{Sema words: ranking, natural logarithm, normalisations}
\end{table}
\end{section}
\clearpage

\begin{center}
\large\bf{Verbs and Graphical law beneath a written natural language}
\end{center}
\begin{abstract}
We study more than five written natural languages. We draw in the log scale, 
number of verbs starting with a letter vs rank of the letter, both normalised.
We find that all the graphs are closer to the curves of reduced 
magnetisation vs reduced temperature for various approximations of Ising model. 
We make a weak conjecture that a curve of magnetisation underlies 
a written natural language, even if we consider verbs exclusively.
\end{abstract}

\begin{section}
\noindent
In this module, we take a smaller set of 
languages, six in number, and study verbs in place of words. The languages 
are German, French, Abor-Miri, Khasi, Garo and Hindi respectively. 
The organisation of this module is as follows:

\noindent
We explain our method of study in the section XII.
In the ensuing section, section XIII, we narrate our graphical results. 
Then we conclude with a conjecture about the graphical law, 
in the section XIV. 
In the section XV, we adjoin tables related to the plots of this module.
\end{section}

\begin{section}{Method of study}
We take bilingual dictionaries of this six languages 
say French to 
English, Khasi to English etc. Then we count the verbs,  
one by one from the beginning to the end, starting with different letters. 
For the French language, \cite{French}, we have taken verbs marked vt, vi; did 
not count verbs marked v.r. i.e. starting with se. For the Garo language, 
\cite{Garo}, we have taken simple verbs only into counting, avoided counting verbs 
like Miksi-mikat daka. Moreover, we have counted a verb 
with different meanings, but the same spelling, only once. For 
the German language, \cite{German}, we have taken only one entry for a
verb. For the Abor-Miri language, \cite{Abor}, we have counted a verb 
with various meanings but the same spelling only once.
\noindent
For each language, we assort the letters according to their rankings. 
We take natural logarithm of both number of verbs, denoted by $f$ 
and the respective rank, denoted by $k$. $k$ is a positive 
integer starting from one. Since each language has a 
letter, number of verbs initiating with it being very close to one or, one, we attach 
a limiting rank, $k_{lim}$ or,$k_{d}$, and a limiting number of verb to each language. 
The limiting rank is just maximum rank 
(maximum rank plus one) if it is one (close to one) and 
the limiting number of verb is one. 
As a result both $\frac{lnf}{lnf_{max}}$ and $\frac{lnk}{lnk_{lim}}$ varies from
zero to one. Then we plot $\frac{lnf}{lnf_{max}}$ against $\frac{lnk}{lnk_{lim}}$.
We note that the ranking of the letters in a language for the 
verbs is independent of the ranking for the words used in the first module.
\end{section}

\begin{section}{Results}
We describe our results, here, in three consecutive subsections.
\begin{subsection}{Verbs}
In this first subsection, 
we plot $\frac{lnf}{lnf_{max}}$ vs $\frac{lnk}{lnk_{lim}}$ 
for the six languages(\cite{French}-\cite{Abor}) . 
On each plot we superimpose a curve of magnetisation.
For German and Khasi languages, it is Bragg-Williams line, BW(c=0), as a comparator. 
For Garo, Bethe-Peierls curve for four neighbour, BP(4,$\beta H=0$) is used as fit curve. 
Bethe-Peierls curve for four neighbours and in presence 
of little magnetic field, $\beta H=0.04$, BP(4,$\beta H=0.04$)
is used for Abor-Miri.
Bragg-Williams curve in presence of little magnetic field, $c=0.01$, BW(c=0.01), is 
utilised for matching with French and Hindi languages.
\begin{figure}
\centering
\includegraphics[width=13cm,height=13cm]{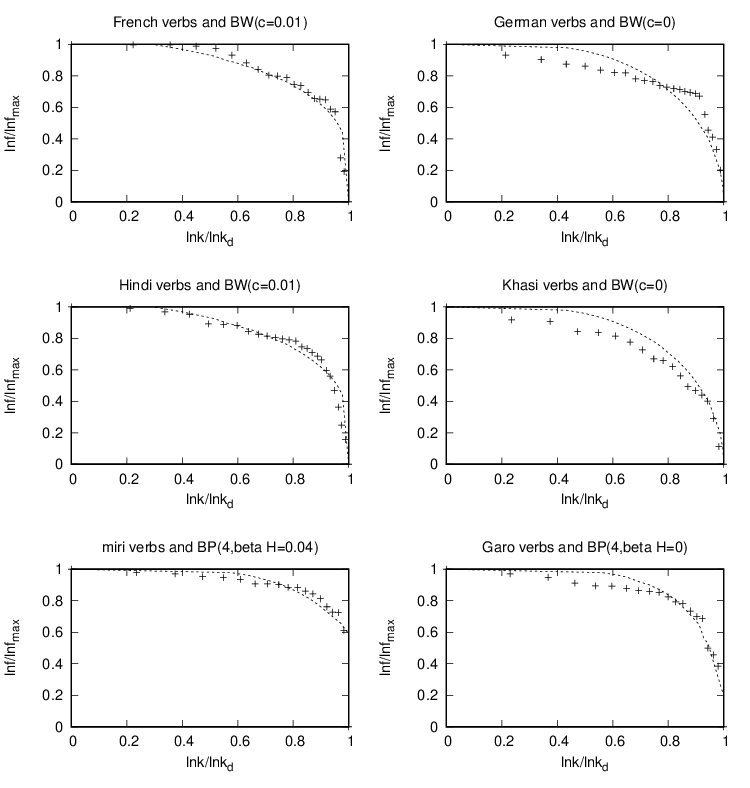}
\caption{Vertical axis is $\frac{lnf}{lnf_{max}}$ and horizontal 
axis is $\frac{lnk}{lnk_{lim}}$. The $+$ points represent the verbs of the languages 
in the titles. The fit curve is different 
for the verbs of the different languages. For German and Khasi it is Bragg-Williams line.
For French and Hindi 
it is the Bragg-Williams line in presence of little external magnetic field.
For the verbs of the Garo language, 
the fit curve is Bethe-Peierls curve in presence of four nearest neighbours with no external 
magnetic field, $\beta H=0$.
For the verbs of the Abor-Miri language, 
the fit curve is the Bethe-Peierls line in presence of four nearest neighbours and little 
external magnetic field, $\beta H=0.04$.
}
\label{Verb1}
\end{figure}
\end{subsection}
\newpage
\begin{subsection}{Khasi language with next-to-maximum}
We observe that the Khasi language is not matched with Bragg-Williams line 
fully in fig.\ref{Verb1}. We then ignore the letter with 
the highest number of verbs and redo the plot, normalising the $lnf$s with 
next-to-maximum $lnf$, and starting from $k=2$. We see, to our surprise, that 
the resulting points almost fall on the Bragg-Williams line in the absence of 
magnetic field.
\begin{figure}
\centering
\includegraphics[width=10cm,height=4cm]{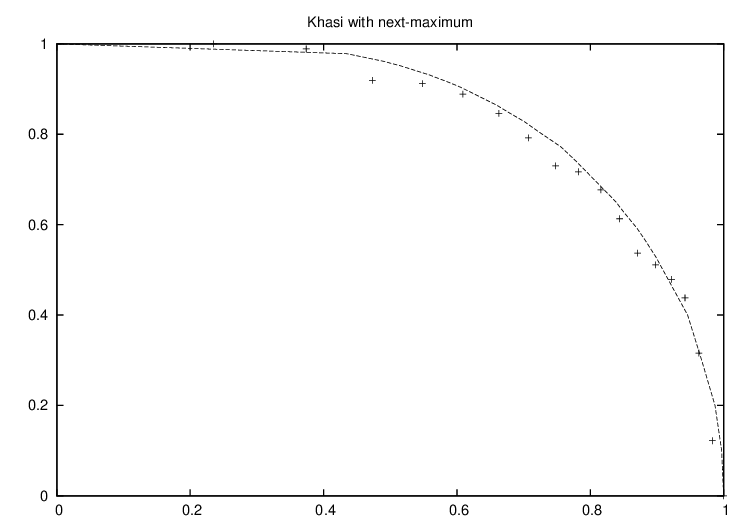}
\caption{The $+$ points represent the Khasi language.
Vertical axis is $\frac{lnf}{lnf_{nextmax}}$ 
and horizontal axis is $\frac{lnk}{lnk_{lim}}$. The fit curve is 
the Bragg-Williams line.}
\label{Verb2}
\end{figure}
\end{subsection}
\noindent
Hence, we can put the six languages in the following classification:
\begin{table}
\begin{center}
\resizebox{10cm}{.5cm}{
\begin{tabular}{|l|l|l|l|l|l|}\hline
French & Garo & Abor-Miri & Hindi & Khasi & German\\\hline
BW(c=0.01) & BP(4,$\beta H=0$) & BP(4,$\beta H=0.04$)& BW(c=0.01) & BW(c=0) & BW(c=0)\\\hline      
\end{tabular}
}
\end{center}
\caption{classification of verbs of six languages according to the underlying magnetisation curves}
\end{table}
\newpage
\begin{subsection}{Spin-Glass}
\begin{figure}
\centering
\includegraphics[width=10cm,height=4cm]{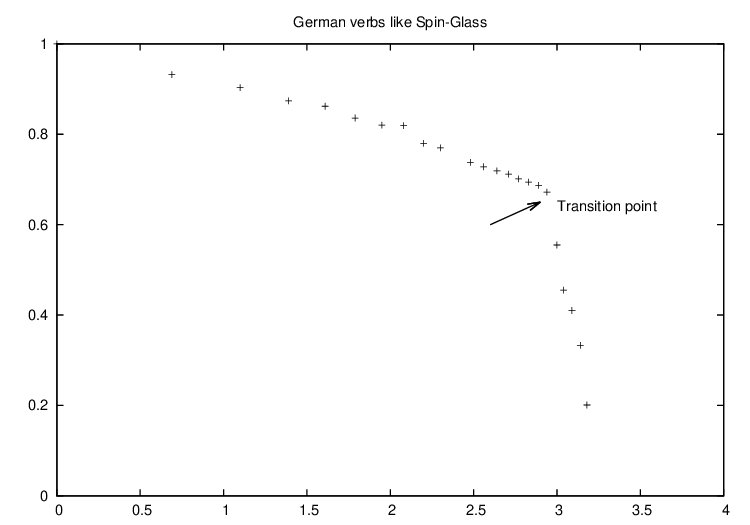}
\caption{Vertical axis is $\frac{lnf}{lnf_{max}}$ and horizontal 
axis is $lnk$. The $+$ points represent the German language. }
\label{Verb3}
\end{figure}
\end{subsection}
\end{section}
\noindent
We note that the pointsline in the figure fig.\ref{Verb3}
has a clear-cut transition point. Hence, it
seems that the German language is better suited to be 
described by a Spin-Glass magnetisation curve, in presence of 
magnetic field. The same feature we have observed for this language, 
in the previous module, in relation to words.
\begin{section}{Conclusion}
From the figures (fig.\ref{Verb1}-fig.\ref{Verb3}), 
hence, we tend to conjecture, behind each written language 
there is a curve of magnetisation, even if verbs only are considered. 

\noindent
Moreover, for the languages we 
have studied here, excepting Khasi, the following correspondance works, 
\begin{eqnarray}
\frac{lnf}{lnf_{max}} \longleftrightarrow \frac{M}{M_{max}},\nonumber\\
lnk\longleftrightarrow T.\nonumber
\end{eqnarray}
For the Khasi language, the correspondance is similar with $lnf_{next-to-maximum}$ 
coming in place of $lnf_{max}$.
\end{section}

\begin{section}{appendix}
\clearpage
\begin{table}
\resizebox{14cm}{.5cm}{
\begin{tabular}{|l|l|l|l|l|l|l|l|l|l|l|l|l|l|l|l|l|l|l|l|l|}\hline
 A&B&K&D&E&G&NG&H&I&J&L&M&N&O&P&R&S&T&U&W&Y \\ \hline 
 14&50&96&29&4&0&5&4&29&48&49&19&6&0&41&17&111&54&4&1&0 \\ \hline
\end{tabular}
}
\caption{Khasi verbs: the first row represents letters of the khasi alphabet in
the serial order}
\end{table}
\begin{table}
\begin{center}
\begin{tabular}{|l|l|l|l|l|l|l|}\hline
k & lnk  & lnk/$lnk_{lim}$ & f & lnf & lnf/$lnf_{max}$ & lnf/$lnf_{next-next-max}$\\\hline
1 & 0    & 0     & 111 & 4.71 & 1 & Blank\\\hline
2 & 0.69 & 0.255 & 96 & 4.56 & 0.968 & Blank\\\hline
3 & 1.10 & 0.406 & 54 & 3.99 & 0.847 & 1\\\hline
4 & 1.39 & 0.513 & 50 & 3.91 & 0.830 & 0.980\\\hline
5 & 1.61 & 0.594 & 49 & 3.89 & 0.826 & 0.975\\\hline
6 & 1.79 & 0.661 & 48 & 3.87 & 0.822 & 0.970\\\hline
7 & 1.95 & 0.720 & 41 & 3.71 & 0.788 & 0.930\\\hline
8 & 2.08 & 0.768 & 29 & 3.37 & 0.715 & 0.845\\\hline
9 & 2.20 & 0.812 & 19 & 2.94 & 0.624 & 0.737\\\hline
10 & 2.30 & 0.849 & 17 & 2.83 & 0.601 & 0.709\\\hline
11 & 2.40 & 0.886 & 14 & 2.64 & 0.561 & 0.662\\\hline
12 & 2.48 & 0.915 & 6 & 1.79 & 0.380 & 0.449\\\hline
13 & 2.56 & 0.945 & 5 & 1.61 & 0.342 & 0.404\\\hline
14 & 2.64 & 0.974 & 4 & 1.39 & 0.295 & 0.348\\\hline
15 & 2.71 & 1     & 1 & 0 & 0 & 0\\\hline
\end{tabular}
\end{center}
\caption{Khasi verbs: ranking, natural logarithm, normalisations}
\end{table}
\clearpage
\begin{table}
\resizebox{14cm}{0.5cm}{
\begin{tabular}{|l|l|l|l|l|l|l|l|l|l|l|l|l|l|l|l|l|l|l|l|l|l|l|l|l|l|}\hline
 A&B&C&D&E&F&G&H&I&J&K&L&M&N&O&P&Q&R&S&T&U&V&W&X&Y&Z \\ \hline 
 183&84&231&126&140&114&57&49&279&21&2&59&179&63&64&230&14&135&212&95&16&99&1&0&1&4 \\ \hline
\end{tabular}
}
\caption{French verbs: the first row represents letters of the French alphabet in
the serial order}
\end{table}

\begin{table}
\begin{center}
\resizebox{14cm}{4.5cm}{
\begin{tabular}{|l|l|l|l|l|l|l|}\hline
k & lnk & lnk/$lnk_{lim}$ & f & lnf & lnf/$lnf_{max}$ \\\hline
1 & 0    & 0     & 279 & 5.63 & 1 \\\hline
2 & 0.69 & 0.217 & 231 & 5.44 & 0.966 \\\hline
3 & 1.10 & 0.346 & 230 & 5.44 & 0.966 \\\hline
4 & 1.39 & 0.437 & 212 & 5.36 & 0.952 \\\hline
5 & 1.61 & 0.506 & 183 & 5.21 & 0.925 \\\hline
6 & 1.79 & 0.563 & 179 & 5.19& 0.922 \\\hline
7 & 1.95 & 0.613 & 140 & 4.94 & 0.877 \\\hline
8 & 2.08 & 0.654 & 135 & 4.91 & 0.872 \\\hline
9 & 2.20 & 0.692 & 126 & 4.84 & 0.860 \\\hline
10 & 2.30 & 0.723 & 114 & 4.74 & 0.842 \\\hline
11 & 2.40 & 0.755 & 99 & 4.60 & 0.817 \\\hline
12 & 2.48 & 0.780 & 95 & 4.55 & 0.808 \\\hline
13 & 2.56 & 0.805 & 84 & 4.43 & 0.787 \\\hline
14 & 2.64 & 0.830 & 64 & 4.16 & 0.739 \\\hline
15 & 2.71 & 0.852 & 63 & 4.14 & 0.735 \\\hline
16 & 2.77 & 0.871 & 59 & 4.08 & 0.725 \\\hline
17 & 2.83 & 0.890 & 57 & 4.04 & 0.718 \\\hline
18 & 2.89 & 0.909 & 49 & 3.89 & 0.691 \\\hline
19 & 2.94 & 0.925 & 21 & 3.04 & 0.540 \\\hline
20 & 3.00 & 0.943 & 16 & 2.77 & 0.492 \\\hline
21 & 3.04 & 0.956 & 14 & 2.64 & 0.469 \\\hline
22 & 3.09 & 0.972 & 4  & 1.39 & 0.247 \\\hline
23 & 3.14 & 0.987 & 2  & .693 & 0.123 \\\hline
24 & 3.18 & 1     & 1  & 0    & 0     \\\hline
\end{tabular}
}
\end{center}
\caption{French verbs: ranking, natural logarithm, normalisations}
\end{table}
\clearpage
\begin{table}
\resizebox{14cm}{.5cm}{
\begin{tabular}{|l|l|l|l|l|l|l|l|l|l|l|l|l|l|l|l|l|l|l|l|l|}\hline
 A&B&D&E&G&I&J&K&L&M&N&O&P&R&S&T&U&Y \\ \hline 
 28&10&6&3&4&3&3&15&4&8&8&6&9&3&26&10&4&8 \\ \hline
\end{tabular}
}
\caption{Abor-Miri verbs: the first row represents letters of the  Abor-Miri alphabet in
the serial order}
\end{table}
\begin{table}
\begin{center}
\begin{tabular}{|l|l|l|l|l|l|l|}\hline
k & lnk & lnk/$lnk_{lim}$ & f & lnf & lnf/$lnf_{max}$ & lnf/$lnf_{next-next-max}$\\\hline
1 & 0 & 0                 & 28 & 3.33 & 1  & Blank \\\hline
2 & 0.69 & 0.3            & 26 & 3.26 & 0.979 & Blank \\\hline
3 & 1.10 & 0.478          & 15 & 2.71 & 0.814 & 1\\\hline
4 & 1.39 & 0.604          & 10 & 2.30 & 0.691 & 0.849 \\\hline
5 & 1.61 & 0.7            & 9 & 2.20 & 0.661 & 0.812\\\hline
6 & 1.79 & 0.778          & 8 & 2.08 & 0.625 & 0.768\\\hline
7 & 1.95 & 0.848          & 6 & 1.79 & 0.538 & 0.661\\\hline
8 & 2.08 & 0.904          & 4 & 1.39 & 0.417 & 0.512\\\hline
9 & 2.20 & 0.957          & 3 & 1.10 & 0.330 &0.405\\\hline
10 & 2.30 & 1             & 1 & 0 & 0 & 0 \\\hline
\end{tabular}
\end{center}
\caption{Abor-Miri verbs: ranking, natural logarithm, normalisations}
\end{table}
\clearpage
\begin{table}
\resizebox{14cm}{.5cm}{
\begin{tabular}{|l|l|l|l|l|l|l|l|l|l|l|l|l|l|l|l|l|l|l|l|l|}\hline
 A&B&C&D&E&G&H&I&J&K&L&M&N&O&P&R&S&T&U&W \\ \hline 
 104&213&94&64&15&190&9&8&38&108&3&153&56&24&96&189&133&55&9&18 \\ \hline
\end{tabular}
}
\caption{Garo verbs: the first row represents letters of the  Garo alphabet in
the serial order}
\end{table}
\begin{table}
\begin{center}
\begin{tabular}{|l|l|l|l|l|l|l|}\hline
k & lnk & lnk/$lnk_{lim}$ & f & lnf & lnf/$lnf_{max}$ \\\hline
1 & 0 & 0                 & 213 & 5.36 & 1 \\\hline
2 & 0.69 & 0.230          & 190 & 5.25 & 0.979 \\\hline
3 & 1.10 & 0.367          & 189 & 5.24 & 0.978 \\\hline
4 & 1.39 & 0.463          & 153 & 5.03 & 0.938 \\\hline
5 & 1.61 & 0.537          & 133 & 4.89 & 0.912 \\\hline
6 & 1.79 & 0.597          & 108 & 4.68 & 0.873 \\\hline
7 & 1.95 & 0.650          & 104 & 4.64 & 0.866 \\\hline
8 & 2.08 & 0.693          & 96 & 4.56 & 0.851 \\\hline
9 & 2.20 & 0.733          & 94 & 4.54 & 0.847 \\\hline
10 & 2.30 & 0.767         & 64 & 4.16 & 0.776 \\\hline
11 & 2.40 & 0.800         & 56 & 4.03 & 0.752 \\\hline
12 & 2.48 & 0.827         & 55 & 4.01 & 0.748 \\\hline
13 & 2.56 & 0.853         & 38 & 3.64 & 0.679 \\\hline
14 & 2.64 & 0.880         & 24 & 3.18 & 0.593 \\\hline
15 & 2.71 & 0.903         & 18 & 2.89  & 0.539 \\\hline
16 & 2.77 & 0.923         & 15 & 2.71 & 0.506 \\\hline
17 & 2.83 & 0.943         & 9 & 2.20 & 0.410 \\\hline
18 & 2.89 & 0.963         & 8 & 2.08 & 0.388  \\\hline
19 & 2.94 & 0.980         & 3 & 1.10 & 0.205  \\\hline
20 & 3.00 & 1             & 1 & 0    & 0      \\\hline
\end{tabular}
\end{center}
\caption{Garo verbs: ranking, natural logarithm, normalisations}
\end{table}
\clearpage
\begin{table}
\resizebox{14cm}{.5cm}{
\begin{tabular}{|l|l|l|l|l|l|l|l|l|l|l|l|l|l|l|l|l|l|l|l|l|l|l|l|l|l|l|}\hline
 A&B&C&D&E&F&G&H&I&J&K&L&M&N&O&P&Q&R&S&T&U&V&W&X&Y&Z \\ \hline 
 346&282&15&138&257&200&318&205&284&24&219&81&307&125&67&76&10&163&492&122&651&319&210&4&0&137 \\ \hline
\end{tabular}
}
\caption{German verbs: the first row represents letters of the  German alphabet in
the serial order}
\end{table}
\begin{table}
\begin{center}
\resizebox{14cm}{4.5cm}{
\begin{tabular}{|l|l|l|l|l|l|l|}\hline
k & lnk & lnk/$lnk_{lim}$ & f & lnf & lnf/$lnf_{max}$ & lnf/$lnf_{next-next-max}$\\\hline
1 & 0 & 0        & 651 & 6.48 & 1  & Blank\\\hline
2 & 0.69 & 0.212 & 492 & 6.20 & 0.957 & Blank\\\hline
3 & 1.10 & 0.337 & 346 & 5.85 & 0.903 & 1\\\hline
4 & 1.39 & 0.426 & 319 & 5.77 & 0.890 & 0.986\\\hline
5 & 1.61 & 0.494 & 318 & 5.76 & 0.889 &0.985\\\hline
6 & 1.79 & 0.549 & 307 & 5.73 & 0.884 &0.979\\\hline
7 & 1.95 & 0.598 & 284 & 5.65 & 0.872 &0.966\\\hline
8 & 2.08 & 0.638 & 282 & 5.64 & 0.870 &0.964\\\hline
9 & 2.20 & 0.675 & 257 & 5.55 & 0.856 &0.949\\\hline
10 & 2.30 & 0.706 & 219 & 5.39 & 0.832&0.921\\\hline
11 & 2.40 & 0.736 & 210 & 5.35 & 0.826&0.915\\\hline
12 & 2.48 & 0.761 & 205 & 5.32 & 0.821&0.909\\\hline
13 & 2.56 & 0.785 & 200 & 5.30 & 0.818&0.906\\\hline
14 & 2.64 & 0.810 & 163 & 5.09 & 0.785&0.870\\\hline
15 & 2.71 & 0.831 & 138 & 4.93 & 0.761&0.843\\\hline
16 & 2.77 & 0.850 & 137 & 4.92 & 0.759&0.841\\\hline
17 & 2.83 & 0.868 & 125 & 4.83 & 0.745&0.826\\\hline
18 & 2.89 & 0.887 & 122 & 4.80 & 0.741&0.821 \\\hline
19&2.94 &0.902    &81   &4.39&0.677   &0.750\\\hline
20&3.00 & 0.920   &76   &4.33&0.668   &0.740 \\\hline
21&3.04 & 0.933  &67    &4.20 &0.648  &0.718 \\\hline
22&3.09 & 0.948  &24    &3.18 &0.491  &0.544 \\\hline
23&3.14 & 0.963  &15    &2.71 &0.418  &0.463 \\\hline
24&3.18 & 0.975  &10    &2.30 &0.355  &0.393 \\\hline
25&3.22 & 0.988  &4     &1.39 &0.215  &0.238 \\\hline
26&3.26 & 1      &1     &0    & 0     &0      \\\hline
\end{tabular}
}
\end{center}
\caption{German verbs: ranking, natural logarithm, normalisations}
\end{table}
\clearpage
\begin{table}
\resizebox{14cm}{.5cm}{
\begin{tabular}{|l|l|l|l|l|l|l|l|l|l|l|l|l|l|l|l|l|l|l|l|l|l|l|l|l|l|l|l|l|l|l|l|l|l|l|l|l|l|l|l|l|}\hline
 1&2&3&4&5&6&7&8&9&10&11&12&13&14&15&16&17&18&19&20&21&22&23&24&25&26&27&28&29&30&31&32&33&34&35&36&37&38&39&40&41 \\ \hline 
 288&50&16&2&67&8&1&23&8&4&6&113&53&77&20&118&24&54&10&7&10&10&9&70&8&95&15&158&179&28&160&38&92&8&26&56&97&32&3&238&35 \\ \hline
\end{tabular}
}
\caption{Hindi verbs: the first row represents letters of the Hindi alphabet in
the serial order}
\end{table}
\begin{table}
\begin{center}
\resizebox{14cm}{4.5cm}{
\begin{tabular}{|l|l|l|l|l|l|l|}\hline
k & lnk & lnk/$lnk_{lim}$ & f & lnf & lnf/$lnf_{max}$ & lnf/$lnf_{nextnextmax}$\\\hline
1 & 0    & 0             & 288 & 5.66 & 1 & Blank\\\hline
2 & 0.69 & 0.193         & 238& 5.47 & 0.966 & Blank\\\hline
3 & 1.10 & 0.307         & 179 & 5.19 & 0.917 & 1\\\hline
4 & 1.39 & 0.388         & 160 & 5.08 & 0.898 & 0.979\\\hline
5 & 1.61 & 0.450         & 158 & 5.06 & 0.894 & 0.975\\\hline
6 & 1.79 & 0.500         & 118 & 4.77 & 0.843 & 0.919\\\hline
7 & 1.95 & 0.545         & 113& 4.73 & 0.836 & 0.912\\\hline
8 & 2.08 & 0.581         & 97 &  4.57 & 0.807 & 0.880\\\hline
9 & 2.20 & 0.615         & 95 & 4.55 & 0.804 & 0.877\\\hline
10 & 2.30 & 0.642        & 92 & 4.52 & 0.799 & 0.871\\\hline
11 & 2.40 & 0.670        & 77 & 4.34 & 0.767 & 0.836\\\hline
12 & 2.48 & 0.693        & 70 & 4.25 & 0.751 & 0.819\\\hline
13 & 2.56 & 0.715        & 67 & 4.20 & 0.742 & 0.809\\\hline
14&2.64   & 0.737        &56  & 4.03& 0.712&0.776\\\hline
15&2.71   &0.757         &54  &3.99 &0.705 &0.769\\\hline
16&2.77   &0.774         &53  &3.97 &0.701 &0.764\\\hline
17&2.83   &0.791         &50  &3.91 &0.691 &0.754\\\hline
18&2.89   &0.807         &38  &3.64 &0.643 &0.701\\\hline
19&2.94   &0.821         &35  &3.56 &0.629 &0.686\\\hline
20&3.00   &0.838         &32  &3.47 &0.613 &0.668\\\hline
21&3.04   &0.849         &28  &3.33 &0.588 &0.641\\\hline
22&3.09   &0.863         &26  &3.26 &0.576 &0.628\\\hline
23&3.14   &0.877         &24  &3.18 &0.562 &0.613\\\hline
24&3.18   &0.888         &23  &3.14 &0.555 &0.605\\\hline
25&3.22   &0.899         &20  &3.00 &0.530 &0.578\\\hline
26&3.26   &0.911         &16  &2.77 &0.489 &0.533\\\hline
27&3.30   &0.922         &15  &2.71 &0.479 &0.522\\\hline
28&3.33   &0.930         &10  &2.30 &0.406 &0.443\\\hline
29&3.37   &0.941         &9   &2.20 &0.389 &0.424\\\hline
30&3.40   &0.950         &8   &2.08 &0.367 &0.400\\\hline
31&3.43   &0.958         &7   &1.95 &0.345 &0.376\\\hline
32&3.47   &0.969         &6   &1.79 &0.316 &0.345\\\hline
33&3.50   &0.978         &4   &1.39 &0.246 &0.268\\\hline
34&3.53   &0.986         &3   &1.10 &0.194 &0.212\\\hline
35&3.56   &0.994         &2   &0.693&0.122 &0.133\\\hline
36&3.58   &1             &1   &0    &0     &0    \\\hline
\end{tabular}
}
\end{center}
\caption{Hindi verbs: ranking, natural logarithm, normalisations}
\end{table}
\end{section}
\clearpage

\begin{center}
\large\bf{Adverbs and Graphical law beneath a written natural language}
\end{center}
\begin{abstract}
We study more than five written natural languages. We draw in the log scale, 
number of adverbs starting with a letter vs rank of the letter, both normalised.
We find that all the graphs are closer to the curves of reduced 
magnetisation vs reduced temperature for various approximations of Ising model. 
We make a weak conjecture that a curve of magnetisation underlies 
a written natural language, even if we consider adverbs exclusively.
\end{abstract}

\begin{section}
\noindent
In this module, we take the same set of six
languages and study adverbs in place of verbs. 
The organisation of this module is as follows:

\noindent
We explain our method of study in the section XVII.
In the ensuing section, section XVIII, we narrate our graphical results. 
Then we conclude with a conjecture about the graphical law, 
in the section XIX. 
In an 
adjoining appendix, section XX, we give the adverb datas for the six languages.
\end{section}

\begin{section}{Method of study}
We take bilingual dictionaries of six different languages 
(\cite{French}-\cite{Abor}), say French to 
English, Khasi to English etc. Then we count the adverbs,  
one by one from the beginning to the end, starting with different letters. 

\noindent
For each language, we assort the letters according to their rankings. 
We take natural logarithm of both number of adverbs, denoted by $f$ 
and the respective rank, denoted by $k$. $k$ is a positive 
integer starting from one. Since each language has a 
letter, number of adverbs initiating with it being very close to one or, one, we attach 
a limiting rank, $k_{lim}$, and a limiting number of adverb to each language. 
The limiting rank is just maximum rank 
(maximum rank plus one) if it is one (close to one) and 
the limiting number of adverb is one. 
As a result both $\frac{lnf}{lnf_{max}}$ and $\frac{lnk}{lnk_{lim}}$ varies from
zero to one. Then we plot $\frac{lnf}{lnf_{max}}$ against $\frac{lnk}{lnk_{lim}}$.
We note that the ranking of the letters in a language for the 
adverbs is independent of the ranking for the words and verbs used in 
the previous two modules.
\end{section}

\newpage

\begin{section}{Results}
We describe our results, here, in four consecutive subsections.
\begin{subsection}{Adverbs}
In this first subsection, 
we plot $\frac{lnf}{lnf_{max}}$ vs $\frac{lnk}{lnk_{lim}}$ 
for the languages(\cite{French}-\cite{Abor}) . 
On each plot we superimpose a curve of magnetisation.
For adverbs of French, Hindi and Khasi languages, it is Bragg-Williams line 
is placed as a comparator. 
For adverbs of Garo language
, Bethe-Peierls curve for four neighbour 
in presence of little magnetic field, BP(4, $\beta H=0.01$) 
is used as fit curve. 
Bragg-Williams curve in presence of little magnetic field, $c=0.01$, BW(c=0.01) is 
utilised for matching with the adverbs of German and Abor-Miri languages.
\begin{figure}
\centering
\includegraphics[width=13cm,height=13cm]{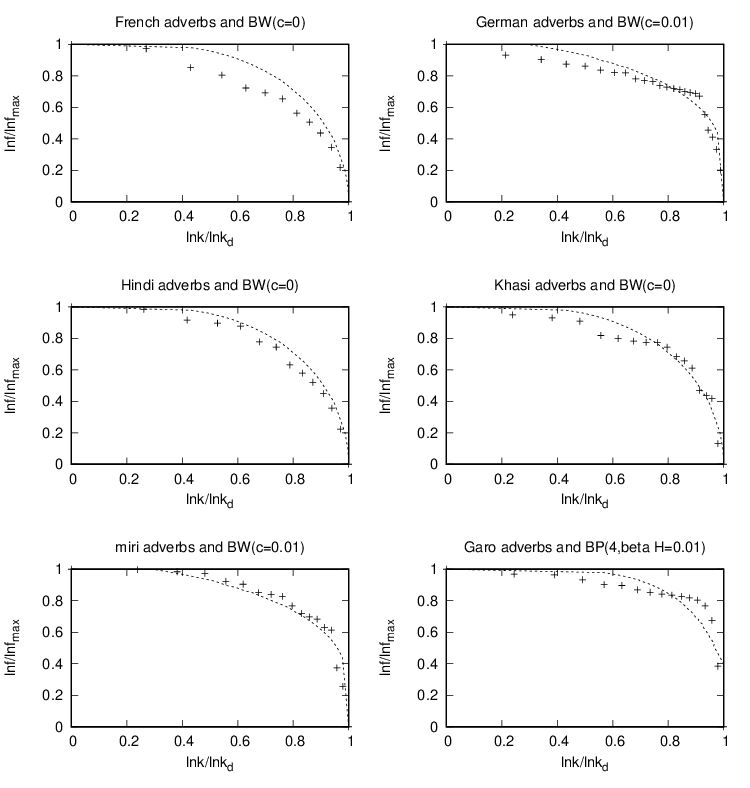}
\caption{Vertical axis is $\frac{lnf}{lnf_{max}}$ and horizontal 
axis is $\frac{lnk}{lnk_{lim}}$. The $+$ points represent the adverbs of the languages 
in the titles,(\cite{French}-\cite{Abor}). The fit curve is different 
for adverbs of different languages. For French, Hindi and Khasi it is Bragg-Williams; 
for Abor-Miri, German it is the Bragg-Williams line in presence of little magnetic field.
For Garo the fit curve is Bethe-Peierls line in presence of four nearest neighbours and little 
magnetic field, $\beta H=0.01$.}
\label{adverb1}
\end{figure}
\end{subsection}
\newpage
\begin{subsection}{Khasi language with next-to-maximum}
We observe that the Khasi language is not matched with Bragg-Williams line 
fully in fig.\ref{adverb1}. We then ignore the letter with 
the highest number of adverbs and redo the plot, normalising the $lnf$s with 
next-to-maximum $lnf$, and starting from $k=2$. We see that the best fit curve for 
the resulting points is the Bragg-Williams line in the presence of 
little magnetic field, $c=0.01$, BW(c=0.01).
\begin{figure}
\centering
\includegraphics[width=10cm,height=3cm]{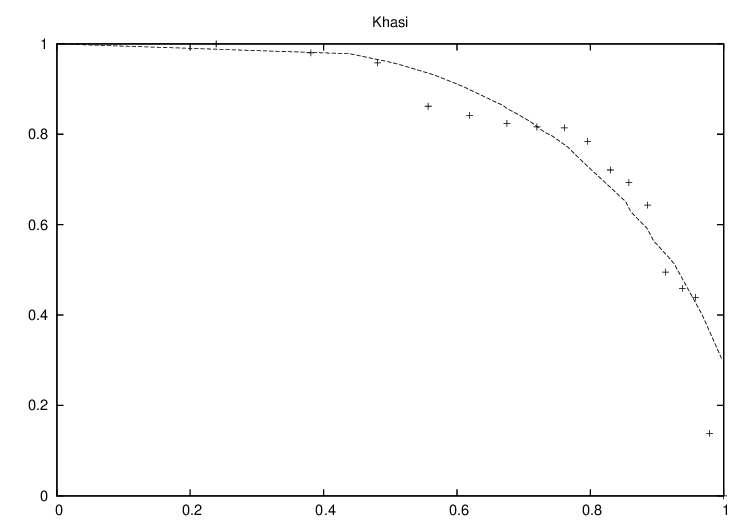}
\caption{The $+$ points represent the Khasi language.
Vertical axis is $\frac{lnf}{lnf_{nextmax}}$ 
and horizontal axis is $\frac{lnk}{lnk_{lim}}$. The fit curve is 
the Bragg-Williams line with little magnetic field.}
\label{adverb2}
\end{figure}
\end{subsection}
\begin{subsection}{French and Hindi with next-to-next-to-maximum }
\begin{figure}
\centering
\includegraphics[width=13cm,height=3cm]{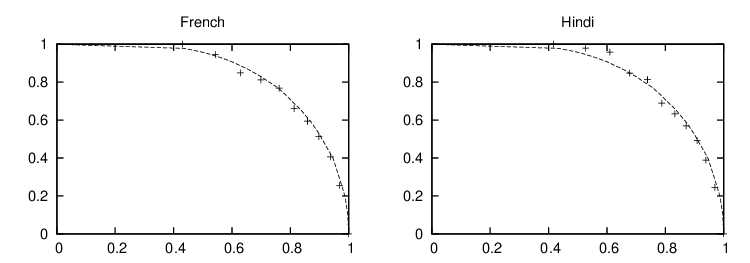}
\caption{The $+$ points represent the languages in the title.
Vertical axis is $\frac{lnf}{lnf_{nextnextmax}}$ 
and horizontal axis is $\frac{lnk}{lnk_{lim}}$. The fit curve is 
the Bragg-Williams line.}
\label{adverb3}
\end{figure}
We observe that the French and Hindi languages are not matched 
with Bragg-Williams line 
fully in fig.\ref{adverb1}. We then ignore the letters with 
the highest and next highest number of adverbs and redo the plot, 
normalising the $lnf$s with 
next-to-next-to-maximum $lnf_{nextnextmax}$, and starting from $k=3$. 
We see, to our surprise, that 
the resulting points almost fall on the Bragg-Williams line.
\noindent
Hence, we can put the six languages in the following classification
\begin{table}
\begin{center}
\resizebox{15cm}{1cm}{
\begin{tabular}{|l|l|l|l|l|l|}\hline
French & Garo & Abor-Miri & Hindi & Khasi & German\\\hline
BW(c=0) & BP(4, $\beta H=0.01$) & BW(c=0.01) & BW(c=0) & BW(c=0.01) &BW(c=0.01) \\\hline      
\end{tabular}
}
\end{center}
\caption{classification of adverbs of six languages according to the underlying magnetisation curves}
\end{table}
\end{subsection}
\newpage
\begin{subsection}{Spin-Glass}
\begin{figure}
\centering
\includegraphics[width=15cm,height=3cm]{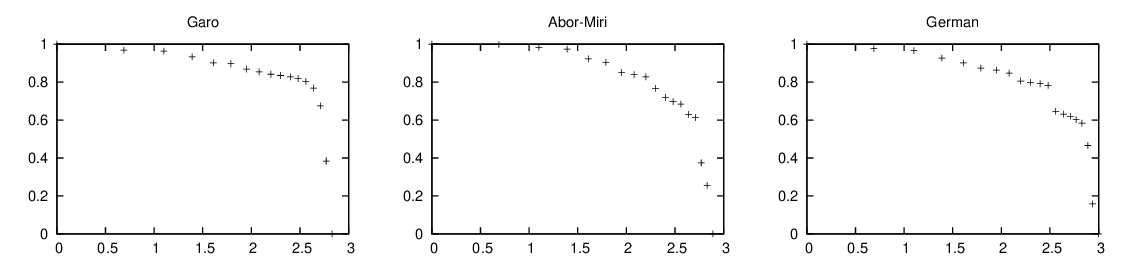}
\caption{Vertical axis is $\frac{lnf}{lnf_{max}}$ and horizontal 
axis is $lnk$. The $+$ points represent the languages in the title. }
\label{adverb4}
\end{figure}
\end{subsection}
\end{section}
\noindent
We note that the pointslines in the fig.\ref{adverb4}
has a clear-cut transition point, at least  for Garo. Hence, it
seems that at least the Garo language is better suited to be 
described by a Spin-Glass magnetisation curve, 
\cite{SpinGlass1, SpinGlass7, SpinGlass2, SpinGlass3, SpinGlass4, SpinGlass5, SpinGlass6}, in presence of 
magnetic field.

\begin{section}{Conclusion}
From the figures (fig.\ref{adverb1}-fig.\ref{adverb4}), 
hence, we tend to conjecture, behind each written language 
there is a curve of magnetisation, even if adverbs only are considered. 

\noindent
Moreover, for the languages we 
have studied here, excepting Khasi, Hindi and French,
the following correspondance works, 
\begin{eqnarray}
\frac{lnf}{lnf_{max}} \longleftrightarrow \frac{M}{M_{max}},\nonumber\\
lnk\longleftrightarrow T.\nonumber
\end{eqnarray}
For the Khasi language, the correspondance is similar with 
$lnf_{next-to-maximum}$  coming in place of $lnf_{max}$.
For French and Hindi languages, the correspondance is also similar with 
$lnf_{next-to-next-to-maximum}$  coming in place of $lnf_{max}$.

\end{section}

\begin{section}{appendix}
\newpage
\begin{table}
\resizebox{14cm}{.5cm}{
\begin{tabular}{|l|l|l|l|l|l|l|l|l|l|l|l|l|l|l|l|l|l|l|l|l|}\hline
 A&B&K&D&E&G&NG&H&I&J&L&M&N&O&P&R&S&T&U&W&Y \\ \hline 
 12&62&196&37&2&0&25&59&10&75&150&60&32&1&68&51&135&122&1&9&2 \\ \hline
\end{tabular}
}
\caption{Khasi adverbs: the first row represents letters of the khasi alphabet in
the serial order}
\end{table}
\begin{table}
\begin{center}
\begin{tabular}{|l|l|l|l|l|l|l|}\hline
k & lnk & lnk/$lnk_{lim}$ & f & lnf & lnf/$lnf_{max}$ & lnf/$lnf_{next-max}$\\\hline
1 & 0 & 0 & 196 & 5.28 & 1 & Blank\\\hline
2 & 0.69 & 0.239 & 150 & 5.01 & 0.949 & 1\\\hline
3 & 1.10 & 0.381 & 135 & 4.91 & 0.930 & 0.98\\\hline
4 & 1.39 & 0.481 & 122 & 4.80 & 0.909 & 0.958\\\hline
5 & 1.61 & 0.557 & 75 & 4.32 & 0.818 & 0.862\\\hline
6 & 1.79 & 0.619 & 68 & 4.22 & 0.799 & 0.842\\\hline
7 & 1.95 & 0.675 & 62 & 4.13 & 0.782 & 0.824\\\hline
8 & 2.08 & 0.720 & 60 & 4.09 & 0.775 & 0.816\\\hline
9 & 2.20 & 0.761 & 59 & 4.08 & 0.773 & 0.814\\\hline
10 & 2.30 & 0.796 & 51 & 3.93 & 0.744 & 0.784\\\hline
11 & 2.40 & 0.830 & 37 & 3.61 & 0.684 & 0.721\\\hline
12 & 2.48 & 0.858 & 32 & 3.47 & 0.657 & 0.693\\\hline
13 & 2.56 & 0.886 & 25 & 3.22 & 0.610 & 0.643\\\hline
14 & 2.64 & 0.913 & 12 & 2.48 & 0.470 & 0.495\\\hline
15 & 2.71 & 0.938 & 10 & 2.30 & 0.436 & 0.459\\\hline
16 & 2.77 & 0.958 & 9 & 2.20 & 0.417 & 0.439\\\hline
17 & 2.83 & 0.979 & 2 & .693 & 0.131 & 0.138\\\hline
18 & 2.89 & 1 & 1 & 0 & 0 & 0\\\hline
\end{tabular}
\end{center}
\caption{Khasi adverbs: ranking, natural logarithm, normalisations}
\end{table}
\clearpage
\begin{table}
\resizebox{14cm}{.5cm}{
\begin{tabular}{|l|l|l|l|l|l|l|l|l|l|l|l|l|l|l|l|l|l|l|l|l|l|l|l|l|l|}\hline
 A&B&C&D&E&F&G&H&I&J&K&L&M&N&O&P&Q&R&S&T&U&V&W&X&Y&Z \\ \hline 
 24&4&13&15&10&5&5&8&9&4&0&9&13&8&3&22&3&2&15&10&0&6&0&0&1&0 \\ \hline
\end{tabular}
}
\caption{French adverbs: the first row represents letters of the French alphabet in
the serial order}
\end{table}
\begin{table}
\begin{center}
\begin{tabular}{|l|l|l|l|l|l|l|}\hline
k & lnk & lnk/$lnk_{lim}$ & f & lnf & lnf/$lnf_{max}$ & lnf/$lnf_{nextnextmax}$\\\hline
1 & 0    & 0     & 24 & 3.18 & 1 & Blank\\\hline
2 & 0.69 & 0.27 & 22 & 3.09 & 0.972 & Blank\\\hline
3 & 1.10 & 0.430 & 15 & 2.71 & 0.852 & 1\\\hline
4 & 1.39 & 0.543 & 13 & 2.56 & 0.805 & 0.945\\\hline
5 & 1.61 & 0.629 & 10 & 2.30 & 0.723 & 0.849\\\hline
6 & 1.79 & 0.699 & 9 & 2.20& 0.692 & 0.812\\\hline
7 & 1.95 & 0.762 & 8 & 2.08 & 0.654 & 0.768\\\hline
8 & 2.08 & 0.813 & 6 & 1.79 & 0.563 & 0.661\\\hline
9 & 2.20 & 0.859 & 5 & 1.61 & 0.506 & 0.594\\\hline
10 & 2.30 & 0.898 & 4 & 1.39 & 0.437 & 0.513\\\hline
11 & 2.40 & 0.838 & 3 & 1.10 & 0.346 & 0.406\\\hline
12 & 2.48 & 0.969 & 2 & .693 & 0.218 & 0.256\\\hline
13 & 2.56 & 1     & 1 & 0 & 0 & 0\\\hline
\end{tabular}
\end{center}
\caption{French adverbs: ranking, natural logarithm, normalisations}
\end{table}
\clearpage
\begin{table}
\resizebox{14cm}{.5cm}{
\begin{tabular}{|l|l|l|l|l|l|l|l|l|l|l|l|l|l|l|l|l|l|l|l|l|}\hline
 A&B&D&E&G&I&J&K&L&M&N&O&P&R&S&T&U&Y \\ \hline 
 69&27&39&22&15&14&5&66&73&53&19&20&35&27&74&49&3&37 \\ \hline
\end{tabular}
}
\caption{Abor-Miri adverbs: the first row represents letters of the Abor-Miri alphabet in
the serial order}
\end{table}
\begin{table}
\begin{center}
\begin{tabular}{|l|l|l|l|l|l|l|}\hline
k & lnk & lnk/$lnk_{lim}$ & f & lnf & lnf/$lnf_{max}$ \\\hline
1 & 0 & 0                 & 74 & 4.30 & 1 \\\hline
2 & 0.69 & 0.239          & 73 & 4.29 & 0.998 \\\hline
3 & 1.10 & 0.381          & 69 & 4.23 & 0.984 \\\hline
4 & 1.39 & 0.481          & 66 & 4.19 & 0.974 \\\hline
5 & 1.61 & 0.557          & 53 & 3.97 & 0.923 \\\hline
6 & 1.79 & 0.619          & 49 & 3.89 & 0.905 \\\hline
7 & 1.95 & 0.675          & 39 & 3.66 & 0.851 \\\hline
8 & 2.08 & 0.720          & 37 & 3.61 & 0.840 \\\hline
9 & 2.20 & 0.761          & 35 & 3.56 & 0.828 \\\hline
10 & 2.30 & 0.796         & 27 & 3.30 & 0.767 \\\hline
11 & 2.40 & 0.830         & 22 & 3.09 & 0.719 \\\hline
12 & 2.48 & 0.858         & 20 & 3.00 & 0.698 \\\hline
13 & 2.56 & 0.886         & 19 & 2.94 & 0.684 \\\hline
14 & 2.64 & 0.913         & 15 & 2.71 & 0.630 \\\hline
15 & 2.71 & 0.938         & 14& 2.64 & 0.614 \\\hline
16 & 2.77 & 0.958         & 5 & 1.61 & 0.374 \\\hline
17 & 2.83 & 0.979         & 3 & 1.10 & 0.256 \\\hline
18 & 2.89 & 1             & 1 & 0 & 0 \\\hline
\end{tabular}
\end{center}
\caption{Abor-Miri adverbs: ranking, natural logarithm, normalisations}
\end{table}
\clearpage
\begin{table}
\resizebox{14cm}{.5cm}{
\begin{tabular}{|l|l|l|l|l|l|l|l|l|l|l|l|l|l|l|l|l|l|l|l|l|}\hline
 A&B&C&D&E&G&H&I&J&K&L&M&N&O&P&R&S&T&U&W \\ \hline 
 34&58&32&50&0&31&5&33&57&43&0&44&29&17&31&38&66&25&43&36 \\ \hline
\end{tabular}
}
\caption{Garo adverbs: the first row represents letters of the Garo alphabet in
the serial order}
\end{table}
\begin{table}
\begin{center}
\begin{tabular}{|l|l|l|l|l|l|l|}\hline
k & lnk & lnk/$lnk_{lim}$ & f & lnf & lnf/$lnf_{max}$ \\\hline
1 & 0 & 0                 & 66 & 4.19 & 1 \\\hline
2 & 0.69 & 0.244          & 58 & 4.06 & 0.969 \\\hline
3 & 1.10 & 0.389          & 57 & 4.04 & 0.964 \\\hline
4 & 1.39 & 0.491          & 50 & 3.91 & 0.933 \\\hline
5 & 1.61 & 0.569          & 44 & 3.78 & 0.902 \\\hline
6 & 1.79 & 0.633          & 43 & 3.76 & 0.897 \\\hline
7 & 1.95 & 0.689          & 38 & 3.64 & 0.869 \\\hline
8 & 2.08 & 0.735          & 36 & 3.58 & 0.854 \\\hline
9 & 2.20 & 0.777          & 34 & 3.53 & 0.842 \\\hline
10 & 2.30 & 0.813         & 33 & 3.50 & 0.835 \\\hline
11 & 2.40 & 0.848         & 32 & 3.47 & 0.828 \\\hline
12 & 2.48 & 0.876         & 31 & 3.43 & 0.819 \\\hline
13 & 2.56 & 0.905         & 29 & 3.37 & 0.804 \\\hline
14 & 2.64 & 0.933         & 25 & 3.22 & 0.768 \\\hline
15 & 2.71 & 0.958         & 17& 2.83& 0.675 \\\hline
16 & 2.77 & 0.979         & 5 & 1.61 & 0.384 \\\hline
17 & 2.83 & 1             & 1 & 0 & 0 \\\hline
\end{tabular}
\end{center}
\caption{Garo adverbs: ranking, natural logarithm, normalisations}
\end{table}
\clearpage
\begin{table}
\resizebox{14cm}{.5cm}{
\begin{tabular}{|l|l|l|l|l|l|l|l|l|l|l|l|l|l|l|l|l|l|l|l|l|l|l|l|l|l|l|}\hline
 A&B&C&D&E&F&G&H&I&J&K&L&M&N&O&P&Q&R&S&T&U&V&W&X&Y&Z \\ \hline 
 52&32&2&80&33&17&31&69&34&15&14&16&32&44&13&12&2&12&72&15&41&52&58&1&0&46 \\ \hline
\end{tabular}
}
\caption{German adverbs: the first row represents letters of the German alphabet in
the serial order}
\end{table}
\begin{table}
\begin{center}
\begin{tabular}{|l|l|l|l|l|l|}\hline
k & lnk & lnk/$lnk_{lim}$ & f & lnf & lnf/$lnf_{max}$ \\\hline
1 & 0 & 0        & 80 & 4.38 & 1 \\\hline
2 & 0.69 & 0.230 & 72 & 4.28 & 0.977 \\\hline
3 & 1.10 & 0.367 & 69 & 4.23 & 0.966 \\\hline
4 & 1.39 & 0.463 & 58 & 4.06 & 0.927 \\\hline
5 & 1.61 & 0.537 & 52 & 3.95 & 0.902 \\\hline
6 & 1.79 & 0.597 & 46 & 3.83 & 0.874 \\\hline
7 & 1.95 & 0.650 & 44 & 3.78 & 0.863 \\\hline
8 & 2.08 & 0.693 & 41 & 3.71 & 0.847 \\\hline
9 & 2.20 & 0.733 & 34 & 3.53 & 0.806 \\\hline
10 & 2.30 & 0.767 & 33 & 3.50 & 0.799\\\hline
11 & 2.40 & 0.800 & 32 & 3.47 & 0.792\\\hline
12 & 2.48 & 0.827 & 31 & 3.43 & 0.783\\\hline
13 & 2.56 & 0.853 & 17 & 2.83 & 0.646\\\hline
14 & 2.64 & 0.880 & 16 & 2.77 & 0.632\\\hline
15 & 2.71 & 0.903 & 15 & 2.71 & 0.619\\\hline
16 & 2.77 & 0.923 & 14 & 2.64 & 0.603\\\hline
17 & 2.83 & 0.943 & 13 & 2.56 & 0.584\\\hline
18 & 2.89 & 0.963 & 12 & 2.48 & 0.566 \\\hline
19&2.94 &0.980&2&.693&0.158\\\hline
20&3.00 & 1&1&0&0\\\hline
\end{tabular}
\end{center}
\caption{German adverbs: ranking, natural logarithm, normalisations}
\end{table}
\clearpage
\begin{table}
\resizebox{14cm}{.5cm}{
\begin{tabular}{|l|l|l|l|l|l|l|l|l|l|l|l|l|l|l|l|l|l|l|l|l|l|l|l|l|l|l|l|l|l|l|l|l|l|l|l|l|l|l|l|l|}\hline
 1&2&3&4&5&6&7&8&9&10&11&12&13&14&15&16&17&18&19&20&21&22&23&24&25&26&27&28&29&30&31&32&33&34&35&36&37&38&39&40&41 \\ \hline 
 22&11&3&0&2&1&0&3&1&0&2&21&5&2&0&7&0&10&2&2&1&0&0&10&0&11&5&11&15&4&17&4&6&6&2&2&6&4&0&16&5 \\ \hline
\end{tabular}
}
\caption{Hindi adverbs: the first row represents letters of the Hindi alphabet in
the serial order}
\end{table}
\begin{table}
\begin{center}
\begin{tabular}{|l|l|l|l|l|l|l|}\hline
k & lnk & lnk/$lnk_{lim}$ & f & lnf & lnf/$lnf_{max}$ & lnf/$lnf_{nextnextmax}$\\\hline
1 & 0    & 0             & 22 & 3.09 & 1 & Blank\\\hline
2 & 0.69 & 0.261         & 21 & 3.04 & 0.984 & Blank\\\hline
3 & 1.10 & 0.417         & 17 & 2.83 & 0.916 & 1\\\hline
4 & 1.39 & 0.527         & 16 & 2.77 & 0.896 & 0.979\\\hline
5 & 1.61 & 0.610         & 15 & 2.71 & 0.877 & 0.958\\\hline
6 & 1.79 & 0.678         & 11 & 2.40 & 0.777 & 0.848\\\hline
7 & 1.95 & 0.739         & 10& 2.30 & 0.744 & 0.813\\\hline
8 & 2.08 & 0.788         & 7&  1.95 & 0.631 & 0.689\\\hline
9 & 2.20 & 0.833         & 6 & 1.79 & 0.579 & 0.633\\\hline
10 & 2.30 & 0.871        & 5 & 1.61 & 0.521 & 0.569\\\hline
11 & 2.40 & 0.909        & 4 & 1.39 & 0.449 & 0.491\\\hline
12 & 2.48 & 0.939        & 3 & 1.10 & 0.356 & 0.389\\\hline
13 & 2.56 & 0.970        & 2 & 0.693 & 0.224 & 0.245\\\hline
14&2.64& 1               &1  & 0& 0&0\\\hline
\end{tabular}
\end{center}
\caption{Hindi adverbs: ranking, natural logarithm, normalisations}
\end{table}
\end{section}

\clearpage
\begin{center}
\large\bf{Adjectives and Graphical law beneath a written natural language}
\end{center}
\begin{abstract}
\noindent
We study more than five written natural languages. We draw in the log scale, 
number of adjectives starting with a letter vs rank of the letter, both normalised.
We find that all the graphs are closer to the curves of reduced 
magnetisation vs reduced temperature for various approximations of Ising model. 
We make a weak conjecture that a curve of magnetisation underlies 
a written natural language, even if we consider adjectives exclusively.
\end{abstract}

\begin{section}
\noindent
In this module, we study adjectives 
for German, French, Abor-Miri, Khasi, Garo and Hindi respectively. 
The organisation of this module is as follows:

\noindent
We explain our method of study in the section XXII.
In the ensuing section, section XXIII, we narrate our graphical results. 
Then we conclude with a conjecture about the graphical law, 
in the section XXIV. 
In an 
adjoining appendix, section XXV, we give the adjective datas for the six languages.
\end{section}

\begin{section}{Method of study}
We take bilingual dictionaries of six different languages 
(\cite{French}-\cite{Abor}), say French to 
English, Khasi to English etc. Then we count the adjectives,  
one by one from the beginning to the end, starting with different letters. 

\noindent
For each language, we assort the letters according to their rankings. 
We take natural logarithm of both number of adjectives, denoted by $f$ 
and the respective rank, denoted by $k$. $k$ is a positive 
integer starting from one. Since each language has a 
letter, number of adjectives initiating with it being very close to one or, 
one, we attach a limiting rank, $k_{lim}$, and a limiting number of 
adjective to each language. 
The limiting rank is just maximum rank 
(maximum rank plus one) if it is one (close to one) and 
the limiting number of adjective is one. 
As a result both $\frac{lnf}{lnf_{max}}$ and $\frac{lnk}{lnk_{lim}}$ varies from
zero to one. Then we plot $\frac{lnf}{lnf_{max}}$ against $\frac{lnk}{lnk_{lim}}$.
We note that the ranking of the letters in a language for the 
adjectives is independent of the ranking for that of the words, verbs and adverbs.
\end{section}
\begin{section}{Results}
\noindent
We plot $\frac{lnf}{lnf_{max}}$ vs $\frac{lnk}{lnk_{lim}}$ 
for the languages(\cite{French}-\cite{Abor}) . 
On each plot we superimpose a curve of magnetisation.
For French, German and Garo languages, 
Bragg-Williams curve in presence of little magnetic field, $c=0.01$,
is placed as a comparator.
For Hindi, Khasi, Abor-Miri 
Bragg-Williams line is used as a fit.
\begin{figure}
\centering
\includegraphics[width=13cm,height=10cm]{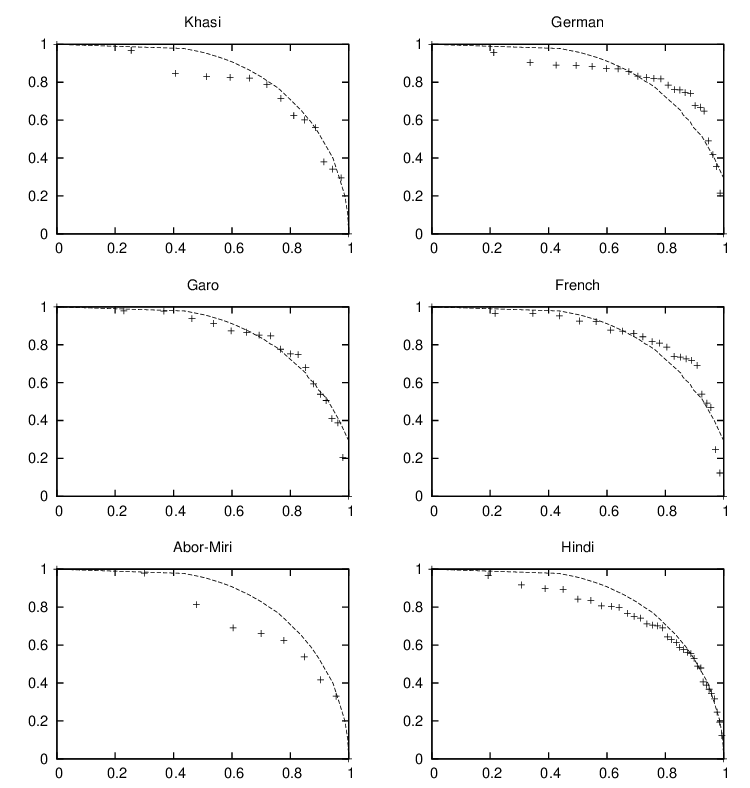}
\caption{Vertical axis is $\frac{lnf}{lnf_{max}}$ and horizontal 
axis is $\frac{lnk}{lnk_{lim}}$. The $+$ points represent the languages 
in the titles. The fit curve is different 
for different languages. For Abor-Miri, Hindi and Khasi it is Bragg-Williams; 
For Garo, French and German it is the Bragg-Williams line in 
presence of little magnetic field.}
\label{adjective1}
\end{figure}
\newpage
\begin{figure}
\centering
\includegraphics[width=13cm,height=7cm]{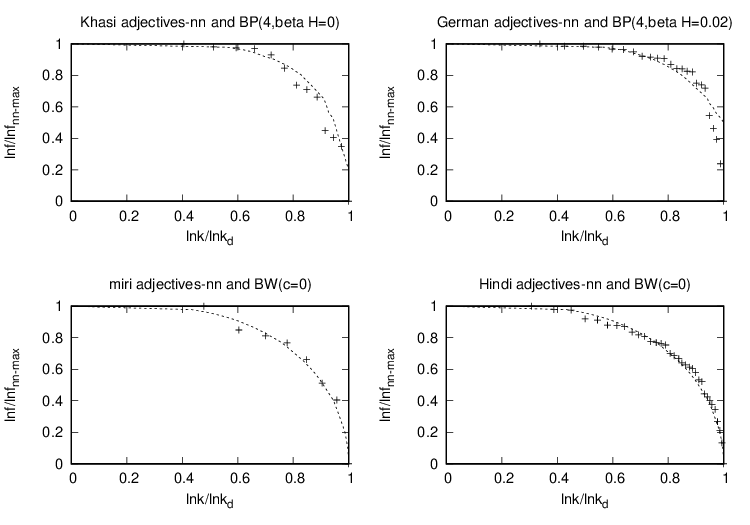}
\caption{The $+$ points represent adjectives of the languages in the title.
Vertical axis is $\frac{lnf}{lnf_{nextnextmax}}$ 
and horizontal axis is $\frac{lnk}{lnk_{lim}}$. The fit curve is 
the Bragg-Williams line for Hindi and Abor-Miri. For Khasi the fit 
curves is Bethe-Peierls curve in presence of four nearest neighbours.
For German the fit curve is Bethe-Peierls curve in presence of four 
nearest neighbours and little 
magnetic field, $\beta H=0.02$.}
\label{adjective2}
\end{figure}
\noindent
We observe that the German, Khasi, Hindi and Abor-Miri are  
not matched 
with Bragg-Williams line, in presence or, absence of magnetic field,  
fully in fig.\ref{adjective1}. We then ignore the letters with 
the highest and next highest number of adjectives and redo the plot, 
normalising the $lnf$s with 
next-to-next-to-maximum $lnf_{nextnextmax}$, and starting from $k=3$. 
We see, to our surprise, that 
the resulting points almost fall on the Bragg-Williams line for Hindi 
, Abor-Miri languages; on BP(4,$\beta H=0$) and BP(4,$\beta H=0.02$) 
for Khasi and German languages respectively.

\noindent
Hence, we can put the six languages in the following classification:
\begin{table}
\begin{center}
\resizebox{12cm}{1cm}{
\begin{tabular}{|l|l|l|l|l|l|}\hline
French & Garo & Abor-Miri & Hindi & Khasi & German\\\hline
BW(c=0.01) & BW(c=0.01) & BW(c=0) & BW(c=0) & BP(4,$\beta H=0$) & BP(4,$\beta H=0.02$)\\\hline      
\end{tabular}
}
\end{center}
\caption{classification of adjectives of six languages according to the underlying magnetisation curves}
\end{table}
\newpage
\noindent
We also observe that the pointslines in the fig.\ref{Figure3}
for German and Khasi appear like 
Spin-Glass magnetisation curve, 
\cite{SpinGlass1, SpinGlass7, SpinGlass2, SpinGlass3, SpinGlass4, SpinGlass5, SpinGlass6}, in presence of 
magnetic field, if we leave the maximum and the next-to-maximum points.
\begin{figure}
\centering
\includegraphics[width=15cm,height=3.5cm]{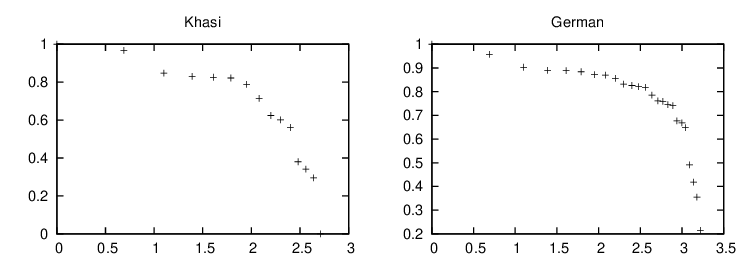}
\caption{Vertical axis is $\frac{lnf}{lnf_{max}}$ and horizontal 
axis is $lnk$. The $+$ points represent the languages in the title. }
\label{adjective3}
\end{figure}
\end{section}

\begin{section}{Conclusion}
From the figures (fig.\ref{adjective1}-fig.\ref{adjective3}), 
hence, we tend to conjecture, behind each written language 
there is a curve of magnetisation, even if adjectives only are considered. 

\noindent
Moreover, for the languages we 
have studied here
the following correspondance works, 
\begin{equation}
\frac{lnf}{lnf_{max}} or, \frac{lnf}{lnf_{next-to-next-to-maximum}}\longleftrightarrow \frac{M}{M_{max}},\nonumber\\
\end{equation}
\begin{equation}
lnk\longleftrightarrow T.\nonumber
\end{equation}

\end{section}

\begin{section}{appendix}
\newpage
\begin{table}
\resizebox{14cm}{.5cm}{
\begin{tabular}{|l|l|l|l|l|l|l|l|l|l|l|l|l|l|l|l|l|l|l|l|l|}\hline
 A&B&K&D&E&G&NG&H&I&J&L&M&N&O&P&R&S&T&U&W&Y \\ \hline 
 14&50&96&29&4&0&5&4&29&48&49&19&6&0&41&17&111&54&4&1&0 \\ \hline
\end{tabular}
}
\caption{Khasi adjectives: the first row represents letters of the khasi alphabet in
the serial order}
\end{table}
\begin{table}
\begin{center}
\begin{tabular}{|l|l|l|l|l|l|l|}\hline
k & lnk  & lnk/$lnk_{lim}$ & f & lnf & lnf/$lnf_{max}$ & lnf/$lnf_{next-next-max}$\\\hline
1 & 0    & 0     & 111 & 4.71 & 1 & Blank\\\hline
2 & 0.69 & 0.255 & 96 & 4.56 & 0.968 & Blank\\\hline
3 & 1.10 & 0.406 & 54 & 3.99 & 0.847 & 1\\\hline
4 & 1.39 & 0.513 & 50 & 3.91 & 0.830 & 0.980\\\hline
5 & 1.61 & 0.594 & 49 & 3.89 & 0.826 & 0.975\\\hline
6 & 1.79 & 0.661 & 48 & 3.87 & 0.822 & 0.970\\\hline
7 & 1.95 & 0.720 & 41 & 3.71 & 0.788 & 0.930\\\hline
8 & 2.08 & 0.768 & 29 & 3.37 & 0.715 & 0.845\\\hline
9 & 2.20 & 0.812 & 19 & 2.94 & 0.624 & 0.737\\\hline
10 & 2.30 & 0.849 & 17 & 2.83 & 0.601 & 0.709\\\hline
11 & 2.40 & 0.886 & 14 & 2.64 & 0.561 & 0.662\\\hline
12 & 2.48 & 0.915 & 6 & 1.79 & 0.380 & 0.449\\\hline
13 & 2.56 & 0.945 & 5 & 1.61 & 0.342 & 0.404\\\hline
14 & 2.64 & 0.974 & 4 & 1.39 & 0.295 & 0.348\\\hline
15 & 2.71 & 1     & 1 & 0 & 0 & 0\\\hline
\end{tabular}
\end{center}
\caption{Khasi adjectives: ranking, natural logarithm, normalisations}
\end{table}
\clearpage
\begin{table}
\resizebox{14cm}{0.5cm}{
\begin{tabular}{|l|l|l|l|l|l|l|l|l|l|l|l|l|l|l|l|l|l|l|l|l|l|l|l|l|l|}\hline
 A&B&C&D&E&F&G&H&I&J&K&L&M&N&O&P&Q&R&S&T&U&V&W&X&Y&Z \\ \hline 
 183&84&231&126&140&114&57&49&279&21&2&59&179&63&64&230&14&135&212&95&16&99&1&0&1&4 \\ \hline
\end{tabular}
}
\caption{French adjectives: the first row represents letters of the French alphabet in
the serial order}
\end{table}
\begin{table}
\begin{center}
\resizebox{14cm}{4.5cm}{
\begin{tabular}{|l|l|l|l|l|l|l|}\hline
k & lnk & lnk/$lnk_{lim}$ & f & lnf & lnf/$lnf_{max}$ \\\hline
1 & 0    & 0     & 279 & 5.63 & 1 \\\hline
2 & 0.69 & 0.217 & 231 & 5.44 & 0.966 \\\hline
3 & 1.10 & 0.346 & 230 & 5.44 & 0.966 \\\hline
4 & 1.39 & 0.437 & 212 & 5.36 & 0.952 \\\hline
5 & 1.61 & 0.506 & 183 & 5.21 & 0.925 \\\hline
6 & 1.79 & 0.563 & 179 & 5.19& 0.922 \\\hline
7 & 1.95 & 0.613 & 140 & 4.94 & 0.877 \\\hline
8 & 2.08 & 0.654 & 135 & 4.91 & 0.872 \\\hline
9 & 2.20 & 0.692 & 126 & 4.84 & 0.860 \\\hline
10 & 2.30 & 0.723 & 114 & 4.74 & 0.842 \\\hline
11 & 2.40 & 0.755 & 99 & 4.60 & 0.817 \\\hline
12 & 2.48 & 0.780 & 95 & 4.55 & 0.808 \\\hline
13 & 2.56 & 0.805 & 84 & 4.43 & 0.787 \\\hline
14 & 2.64 & 0.830 & 64 & 4.16 & 0.739 \\\hline
15 & 2.71 & 0.852 & 63 & 4.14 & 0.735 \\\hline
16 & 2.77 & 0.871 & 59 & 4.08 & 0.725 \\\hline
17 & 2.83 & 0.890 & 57 & 4.04 & 0.718 \\\hline
18 & 2.89 & 0.909 & 49 & 3.89 & 0.691 \\\hline
19 & 2.94 & 0.925 & 21 & 3.04 & 0.540 \\\hline
20 & 3.00 & 0.943 & 16 & 2.77 & 0.492 \\\hline
21 & 3.04 & 0.956 & 14 & 2.64 & 0.469 \\\hline
22 & 3.09 & 0.972 & 4  & 1.39 & 0.247 \\\hline
23 & 3.14 & 0.987 & 2  & .693 & 0.123 \\\hline
24 & 3.18 & 1     & 1  & 0    & 0     \\\hline
\end{tabular}
}
\end{center}
\caption{French adjectives: ranking, natural logarithm, normalisations}
\end{table}
\clearpage
\begin{table}
\resizebox{14cm}{.5cm}{
\begin{tabular}{|l|l|l|l|l|l|l|l|l|l|l|l|l|l|l|l|l|l|l|l|l|}\hline
 A&B&D&E&G&I&J&K&L&M&N&O&P&R&S&T&U&Y \\ \hline 
 28&10&6&3&4&3&3&15&4&8&8&6&9&3&26&10&4&8 \\ \hline
\end{tabular}
}
\caption{Abor-Miri adjectives: the first row represents letters of the Abor-Miri alphabet in
the serial order}
\end{table}
\begin{table}
\begin{center}
\begin{tabular}{|l|l|l|l|l|l|l|}\hline
k & lnk & lnk/$lnk_{lim}$ & f & lnf & lnf/$lnf_{max}$ & lnf/$lnf_{next-next-max}$\\\hline
1 & 0 & 0                 & 28 & 3.33 & 1  & Blank \\\hline
2 & 0.69 & 0.3            & 26 & 3.26 & 0.979 & Blank \\\hline
3 & 1.10 & 0.478          & 15 & 2.71 & 0.814 & 1\\\hline
4 & 1.39 & 0.604          & 10 & 2.30 & 0.691 & 0.849 \\\hline
5 & 1.61 & 0.7            & 9 & 2.20 & 0.661 & 0.812\\\hline
6 & 1.79 & 0.778          & 8 & 2.08 & 0.625 & 0.768\\\hline
7 & 1.95 & 0.848          & 6 & 1.79 & 0.538 & 0.661\\\hline
8 & 2.08 & 0.904          & 4 & 1.39 & 0.417 & 0.512\\\hline
9 & 2.20 & 0.957          & 3 & 1.10 & 0.330 &0.405\\\hline
10 & 2.30 & 1             & 1 & 0 & 0 & 0 \\\hline
\end{tabular}
\end{center}
\caption{Abor-Miri adjectives: ranking, natural logarithm, normalisations}
\end{table}
\clearpage
\begin{table}
\resizebox{14cm}{.5cm}{
\begin{tabular}{|l|l|l|l|l|l|l|l|l|l|l|l|l|l|l|l|l|l|l|l|l|}\hline
 A&B&C&D&E&G&H&I&J&K&L&M&N&O&P&R&S&T&U&W \\ \hline 
 104&213&94&64&15&190&9&8&38&108&3&153&56&24&96&189&133&55&9&18 \\ \hline
\end{tabular}
}
\caption{Garo adjectives: the first row represents letters of the Garo alphabet in
the serial order}
\end{table}
\begin{table}
\begin{center}
\begin{tabular}{|l|l|l|l|l|l|l|}\hline
k & lnk & lnk/$lnk_{lim}$ & f & lnf & lnf/$lnf_{max}$ \\\hline
1 & 0 & 0                 & 213 & 5.36 & 1 \\\hline
2 & 0.69 & 0.230          & 190 & 5.25 & 0.979 \\\hline
3 & 1.10 & 0.367          & 189 & 5.24 & 0.978 \\\hline
4 & 1.39 & 0.463          & 153 & 5.03 & 0.938 \\\hline
5 & 1.61 & 0.537          & 133 & 4.89 & 0.912 \\\hline
6 & 1.79 & 0.597          & 108 & 4.68 & 0.873 \\\hline
7 & 1.95 & 0.650          & 104 & 4.64 & 0.866 \\\hline
8 & 2.08 & 0.693          & 96 & 4.56 & 0.851 \\\hline
9 & 2.20 & 0.733          & 94 & 4.54 & 0.847 \\\hline
10 & 2.30 & 0.767         & 64 & 4.16 & 0.776 \\\hline
11 & 2.40 & 0.800         & 56 & 4.03 & 0.752 \\\hline
12 & 2.48 & 0.827         & 55 & 4.01 & 0.748 \\\hline
13 & 2.56 & 0.853         & 38 & 3.64 & 0.679 \\\hline
14 & 2.64 & 0.880         & 24 & 3.18 & 0.593 \\\hline
15 & 2.71 & 0.903         & 18 & 2.89  & 0.539 \\\hline
16 & 2.77 & 0.923         & 15 & 2.71 & 0.506 \\\hline
17 & 2.83 & 0.943         & 9 & 2.20 & 0.410 \\\hline
18 & 2.89 & 0.963         & 8 & 2.08 & 0.388  \\\hline
19 & 2.94 & 0.980         & 3 & 1.10 & 0.205  \\\hline
20 & 3.00 & 1             & 1 & 0    & 0      \\\hline
\end{tabular}
\end{center}
\caption{Garo adjectives: ranking, natural logarithm, normalisations}
\end{table}
\clearpage
\begin{table}
\resizebox{14cm}{.5cm}{
\begin{tabular}{|l|l|l|l|l|l|l|l|l|l|l|l|l|l|l|l|l|l|l|l|l|l|l|l|l|l|l|}\hline
 A&B&C&D&E&F&G&H&I&J&K&L&M&N&O&P&Q&R&S&T&U&V&W&X&Y&Z \\ \hline 
 346&282&15&138&257&200&318&205&284&24&219&81&307&125&67&76&10&163&492&122&651&319&210&4&0&137 \\ \hline
\end{tabular}
}
\caption{German adjectives: the first row represents letters of the German alphabet in
the serial order}
\end{table}
\begin{table}
\begin{center}
\resizebox{14cm}{4.5cm}{
\begin{tabular}{|l|l|l|l|l|l|l|}\hline
k&lnk&lnk/$lnk_{lim}$&f&lnf&lnf/$lnf_{max}$&lnf/$lnf_{nnmax}$\\\hline
1 & 0 & 0        & 651 & 6.48 & 1  & Blank \\\hline
2 & 0.69 & 0.212 & 492 & 6.20 & 0.957 & Blank \\\hline
3 & 1.10 & 0.337 & 346 & 5.85 & 0.903 & 1  \\\hline
4 & 1.39 & 0.426 & 319 & 5.77 & 0.890 & 0.986 \\\hline
5 & 1.61 & 0.494 & 318 & 5.76 & 0.889 &0.985\\\hline
6 & 1.79 & 0.549 & 307 & 5.73 & 0.884 &0.979\\\hline
7 & 1.95 & 0.598 & 284 & 5.65 & 0.872 &0.966\\\hline
8 & 2.08 & 0.638 & 282 & 5.64 & 0.870 &0.964\\\hline
9 & 2.20 & 0.675 & 257 & 5.55 & 0.856 &0.949\\\hline
10 & 2.30 & 0.706 & 219 & 5.39 & 0.832&0.921\\\hline
11 & 2.40 & 0.736 & 210 & 5.35 & 0.826&0.915\\\hline
12 & 2.48 & 0.761 & 205 & 5.32 & 0.821&0.909\\\hline
13 & 2.56 & 0.785 & 200 & 5.30 & 0.818&0.906\\\hline
14 & 2.64 & 0.810 & 163 & 5.09 & 0.785&0.870\\\hline
15 & 2.71 & 0.831 & 138 & 4.93 & 0.761&0.843\\\hline
16 & 2.77 & 0.850 & 137 & 4.92 & 0.759&0.841\\\hline
17 & 2.83 & 0.868 & 125 & 4.83 & 0.745&0.826\\\hline
18 & 2.89 & 0.887 & 122 & 4.80 & 0.741&0.821 \\\hline
19&2.94 &0.902    &81   &4.39&0.677   &0.750\\\hline
20&3.00 & 0.920   &76   &4.33&0.668   &0.740 \\\hline
21&3.04 & 0.933  &67    &4.20 &0.648  &0.718 \\\hline
22&3.09 & 0.948  &24    &3.18 &0.491  &0.544 \\\hline
23&3.14 & 0.963  &15    &2.71 &0.418  &0.463  \\\hline
24&3.18 & 0.975  &10    &2.30 &0.355  &0.393 \\\hline
25&3.22 & 0.988  &4     &1.39 &0.215  &0.238 \\\hline
26&3.26 & 1      &1     &0    & 0     &0      \\\hline
\end{tabular}
}
\end{center}
\caption{German adjectives: ranking, natural logarithm, normalisations}
\end{table}
\clearpage
\begin{table}
\resizebox{14cm}{.5cm}{
\begin{tabular}{|l|l|l|l|l|l|l|l|l|l|l|l|l|l|l|l|l|l|l|l|l|l|l|l|l|l|l|l|l|l|l|l|l|l|l|l|l|l|l|l|l|}\hline
 1&2&3&4&5&6&7&8&9&10&11&12&13&14&15&16&17&18&19&20&21&22&23&24&25&26&27&28&29&30&31&32&33&34&35&36&37&38&39&40&41 \\ \hline 
 288&50&16&2&67&8&1&23&8&4&6&113&53&77&20&118&24&54&10&7&10&10&9&70&8&95&15&158&179&28&160&38&92&8&26&56&97&32&3&238&35 \\ \hline
\end{tabular}
}
\caption{Hindi adjectives: the first row represents letters of the Hindi alphabet in
the serial order}
\end{table}
\begin{table}
\begin{center}
\resizebox{14cm}{4.5cm}{
\begin{tabular}{|l|l|l|l|l|l|l|}\hline
k & lnk & lnk/$lnk_{lim}$ & f & lnf & lnf/$lnf_{max}$ & lnf/$lnf_{nextnextmax}$\\\hline
1 & 0    & 0             & 288 & 5.66 & 1 & Blank\\\hline
2 & 0.69 & 0.193         & 238& 5.47 & 0.966 & Blank\\\hline
3 & 1.10 & 0.307         & 179 & 5.19 & 0.917 & 1\\\hline
4 & 1.39 & 0.388         & 160 & 5.08 & 0.898 & 0.979\\\hline
5 & 1.61 & 0.450         & 158 & 5.06 & 0.894 & 0.975\\\hline
6 & 1.79 & 0.500         & 118 & 4.77 & 0.843 & 0.919\\\hline
7 & 1.95 & 0.545         & 113& 4.73 & 0.836 & 0.912\\\hline
8 & 2.08 & 0.581         & 97 &  4.57 & 0.807 & 0.880\\\hline
9 & 2.20 & 0.615         & 95 & 4.55 & 0.804 & 0.877\\\hline
10 & 2.30 & 0.642        & 92 & 4.52 & 0.799 & 0.871\\\hline
11 & 2.40 & 0.670        & 77 & 4.34 & 0.767 & 0.836\\\hline
12 & 2.48 & 0.693        & 70 & 4.25 & 0.751 & 0.819\\\hline
13 & 2.56 & 0.715        & 67 & 4.20 & 0.742 & 0.809\\\hline
14&2.64   & 0.737        &56  & 4.03& 0.712&0.776\\\hline
15&2.71   &0.757         &54  &3.99 &0.705 &0.769\\\hline
16&2.77   &0.774         &53  &3.97 &0.701 &0.764\\\hline
17&2.83   &0.791         &50  &3.91 &0.691 &0.754\\\hline
18&2.89   &0.807         &38  &3.64 &0.643 &0.701\\\hline
19&2.94   &0.821         &35  &3.56 &0.629 &0.686\\\hline
20&3.00   &0.838         &32  &3.47 &0.613 &0.668\\\hline
21&3.04   &0.849         &28  &3.33 &0.588 &0.641\\\hline
22&3.09   &0.863         &26  &3.26 &0.576 &0.628\\\hline
23&3.14   &0.877         &24  &3.18 &0.562 &0.613\\\hline
24&3.18   &0.888         &23  &3.14 &0.555 &0.605\\\hline
25&3.22   &0.899         &20  &3.00 &0.530 &0.578\\\hline
26&3.26   &0.911         &16  &2.77 &0.489 &0.533\\\hline
27&3.30   &0.922         &15  &2.71 &0.479 &0.522\\\hline
28&3.33   &0.930         &10  &2.30 &0.406 &0.443\\\hline
29&3.37   &0.941         &9   &2.20 &0.389 &0.424\\\hline
30&3.40   &0.950         &8   &2.08 &0.367 &0.400\\\hline
31&3.43   &0.958         &7   &1.95 &0.345 &0.376\\\hline
32&3.47   &0.969         &6   &1.79 &0.316 &0.345\\\hline
33&3.50   &0.978         &4   &1.39 &0.246 &0.268\\\hline
34&3.53   &0.986         &3   &1.10 &0.194 &0.212\\\hline
35&3.56   &0.994         &2   &0.693&0.122 &0.133\\\hline
36&3.58   &1             &1   &0    &0     &0    \\\hline
\end{tabular}
}
\end{center}
\caption{Hindi adjectives: ranking, natural logarithm, normalisations}
\end{table}
\end{section}

\clearpage
\begin{center}
\large\bf{Contemporary Chinese usage and Graphical law}
\end{center}
\begin{abstract}
\noindent
We study Chinese-English dictionary of contemporary usage. We draw in the log scale, 
number of usages starting with a letter vs rank of the letter, both normalised.
We find that the graphs are closer to the curves of reduced 
magnetisation vs reduced temperature for various approximations of Ising model. 
\end{abstract}

\begin{section}
\noindent
In this module, we continue our enquiry into the chinese language. 
Chinese is comparable in terms of natural language users to English. 
Originally, 
chinese language had pictograms, no alphabet. 
Recently adopting Wade-Giles romanization
system a compilation has been done for the contemporary chinese usages, \cite{Chin}.
We take that as a source for written chinese language. We try to see whether 
a graphical law is buried within, through this version, in the chinese language, 
in this module

\noindent
We organise this module as follows. 
We explain our method of study in the section XXVII.
In the ensuing section, section XXVIII, we narrate our graphical results. 
Then we conclude about the existence of the graphical law 
in the section XXIX. 
In an adjoining appendix, section XXX, 
we give the usage datas for the chinese language. 
\end{section}

\begin{section}{Method of study}
We take the Chinese-English dictionary of contemporary usage,\cite{Chin}.
Then we count the usages,  
one by one from the beginning to the end, starting with different letters. 
We assort the letters according to their rankings. 
We take natural logarithm of both number of usages, denoted by $f$ 
and the respective rank, denoted by $k$. $k$ is a positive 
integer starting from one. Moreover,
we attach a limiting rank, $k_{lim}$, and a limiting number of usage. 
The limiting rank is maximum rank plus one and 
the limiting number of usage is one. 
As a result both $\frac{lnf}{lnf_{max}}$ and $\frac{lnk}{lnk_{lim}}$ varies from
zero to one. Then we plot $\frac{lnf}{lnf_{max}}$ against $\frac{lnk}{lnk_{lim}}$.
\end{section}

\begin{section}{Results}
\noindent
We plot $\frac{lnf}{lnf_{max}}$ vs $\frac{lnk}{lnk_{lim}}$ 
for the chinese language(\cite{Chin}). 
On the plot we superimpose a 
Bragg-Williams curve in presence of little magnetic field, $c=0.01$,
as a comparator.
\begin{figure}
\centering
\includegraphics[width=13cm,height=7cm]{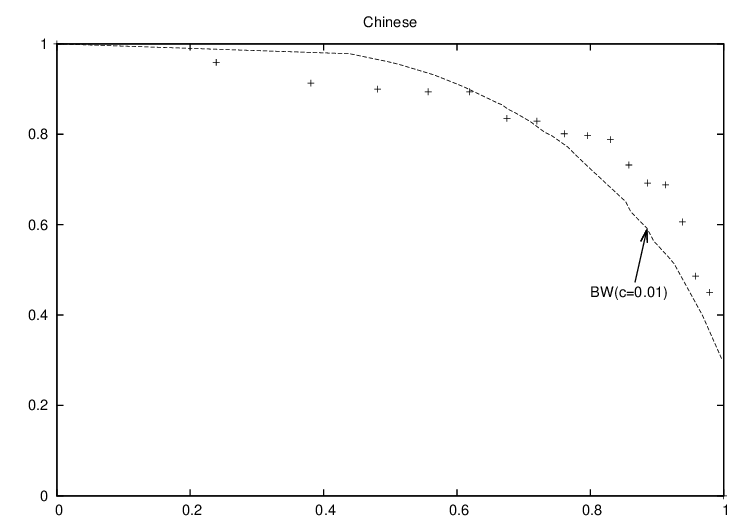}
\caption{Vertical axis is $\frac{lnf}{lnf_{max}}$ and horizontal 
axis is $\frac{lnk}{lnk_{lim}}$. The $+$ points represent the chinese language 
(\cite{Chin}).}
\label{Chin1}
\end{figure}
\begin{figure}
\centering
\includegraphics[width=13cm,height=7cm]{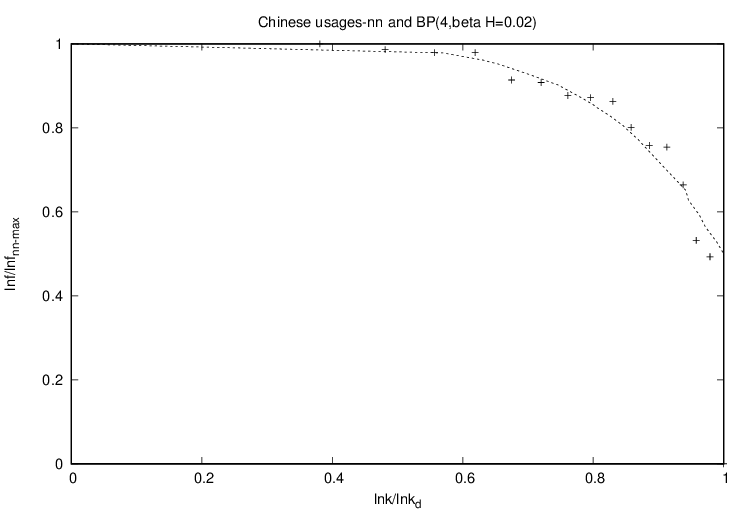}
\caption{The $+$ points represent the chinese usages.
Vertical axis is $\frac{lnf}{lnf_{nextnextmax}}$ 
and horizontal axis is $\frac{lnk}{lnk_{lim}}$. 
Fit curve is BP(4,$\beta H=0.02$) i.e. Bethe-Peierls curve 
in presence of four nearest neighbours and little 
magnetic field, $\beta H=0.02$. }
\label{Chin2}
\end{figure}
\noindent
We observe that the points are  
not matched 
with Bragg-Williams line, in presence of magnetic field,  
fully in fig.\ref{Chin1}. We then ignore the letters with 
the highest and next highest number of usages and redo the plot, 
normalising the $lnf$s with 
next-to-next-to-maximum $lnf_{nextnextmax}$, and starting from $k=3$. 
We see that the best fit curve for 
the resulting points is the Bethe-Peierls curve 
in presence of four nearest neighbours and little 
magnetic field, $\beta H=0.02$ i.e.  BP(4,$\beta H=0.02$) as in fig.\ref{Chin2}. 

\end{section}

\begin{section}{Conclusion}
From the figures (fig.\ref{Chin1}-fig.\ref{Chin2}), 
we observe that there is a curve of magnetisation, behind contemporary chinese usage. This is 
BP(4,$\beta H=0.02$). 
In other words, we conclude that the graphical law exists beneath the contemporary 
chinese usage also.

\noindent
Moreover, the associated correspondance is,
\begin{equation}
\frac{lnf}{lnf_{next-to-next-to-maximum}}\longleftrightarrow \frac{M}{M_{max}},\nonumber\\
\end{equation}
\begin{equation}
lnk\longleftrightarrow T.\nonumber
\end{equation}
\end{section}
\clearpage
\begin{section}{appendix}
\begin{table}
\resizebox{14cm}{.5cm}{
\begin{tabular}{|l|l|l|l|l|l|l|l|l|l|l|l|l|l|l|l|l|l|l|l|l|l|l|l|l|l|l|}\hline
 A&C&E&F&H&I&J&K&L&M&N&O&P&S&T&W&Y \\ \hline 
 155&4058&42&701&1981&439&314&1692&982&751&306&57&1691&1775&2893&783&1035 \\ \hline
\end{tabular}
}
\caption{Chinese usages: the second row represents the entries of the Chinese usages against letters of roman alphabet in
the serial order}
\end{table}
\begin{table}
\begin{center}
\resizebox{14cm}{3cm}{
\begin{tabular}{|l|l|l|l|l|l|l|}\hline
k & lnk & lnk/$lnk_{lim}$ & f & lnf & lnf/$lnf_{max}$ & lnf/$lnf_{next-next-max}$\\\hline
1 & 0 & 0        & 4058 & 8.31 & 1  & Blank\\\hline
2 & 0.69 & 0.239 & 2893 & 7.97 & 0.959 & Blank\\\hline
3 & 1.10 & 0.381 & 1981 & 7.59 & 0.913 & 1\\\hline
4 & 1.39 & 0.481 & 1775 & 7.48 & 0.900 & 0.986\\\hline
5 & 1.61 & 0.557 & 1692 & 7.43 & 0.894 &0.979\\\hline
6 & 1.79 & 0.619 & 1691 & 7.43 & 0.894 &0.979\\\hline
7 & 1.95 & 0.675 & 1035 & 6.94 & 0.835 &0.914\\\hline
8 & 2.08 & 0.720 & 982 & 6.89 & 0.829 &0.908\\\hline
9 & 2.20 & 0.761 & 783 & 6.66 & 0.801 &0.877\\\hline
10 & 2.30 & 0.796 & 751 & 6.62 & 0.797&0.872\\\hline
11 & 2.40 & 0.830 & 701 & 6.55 & 0.788&0.863\\\hline
12 & 2.48 & 0.858 & 439 & 6.08 & 0.732&0.801\\\hline
13 & 2.56 & 0.886 & 314 & 5.75 & 0.692&0.758\\\hline
14 & 2.64 & 0.913 & 306 & 5.72 & 0.688&0.754\\\hline
15 & 2.71 & 0.938 & 155 & 5.04 & 0.606&0.664\\\hline
16 & 2.77 & 0.958 & 57 & 4.04 & 0.486&0.532\\\hline
17 & 2.83 & 0.979 & 42 & 3.74 & 0.450&0.493\\\hline
18 & 2.89 & 1 & 1 & 0 & 0&0 \\\hline
\end{tabular}
}
\end{center}
\caption{Chinese usages: ranking, natural logarithm, normalisations}
\end{table}

\end{section}

\clearpage
\begin{center}
\large\bf{Lakher(Mara) language and the Graphical law}
\end{center}
\begin{abstract}
\noindent
We study Lakher to English dictionary. We draw in the log scale, 
number of words, nouns, verbs, adverbs and adjectives starting with an alphabet vs 
rank of the alphabet, both normalised.
We find that the graphs are closer to the curves of reduced 
magnetisation vs reduced temperature for various approximations of Ising model. 
\end{abstract}

\begin{section}
\noindent
In this module, we continue our study into the Lakher language. Lakher is a dialect 
of Lai that belongs to the central sub-group of Lakher languages,\cite{Lakher}. 
They are a Hill tribe of Malayan stock. They have migrated from Lakher Hills. 
Lakher is the Lushai name for the Mara 
tribe. Mara is the correct name for the people in their own language. 
According to the census of 1931, they were 6186 in number. Around 1949, 
their population was around 20,000. They were living in the 
South Lushai Hills, now in Mizoram, in the Lakher Hills of Burma, now 
Myanmar, and in the North Arakan Yoma Mountains. 
Around 1907, \cite{Lakher}, they were a 
head-hunting ferocious tribe. The author of the book,\cite{Lakher}, which we 
are using, landed among them at that time, painstakingly reduced 
the Lakher language to writing, rigorously composed Grammar and 
exhuastively developed Dictionary before he departed for heaven in the year, 
1944. A recent account on Lakher language is reference\cite{thesis}.  
The Lakher alphabet is composed of 25 entries.  This includes ten vowels. 
Of these ao, yu are diphthongs; o,\^{o} are pure sounds, not letters. The eight 
parts of speech are noun, pronoun, verb, adverb, adjective, preposition, conjunction 
and interjection. The number of entries, counted by us, under words and 
different parts of speech ala the work of 
Rev. Lorrain, \cite{Lakher} are as follows:
\begin{table}
\begin{center}
\resizebox{14cm}{4cm}{
\begin{tabular}{|l|l|l|l|l|l|l|l|l|l|l|l|l|l|l|l|l|l|l|l|l|l|l|l|l|l|l|}\hline
 letter&A&AW&Y&B&CH&D&E&H&I&K&L&M&N&Ng&O&\^{o}&P&R&S&T&U&V&Z \\ \hline 
 words&526&55&32&297&712&180&24&550&59&621&434&404&252&154&39&14&1064&361&588&853&25&243&198 \\ \hline
 nouns&306&53&15&220&397&100&5&295&46&381&299&287&127&104&28&11&522&233&470&596&12&169&118\\\hline
 verbs&118&2&13&72&274&52&4&162&9&134&98&100&78&40&6&2&530&114&99&230&8&55&53\\\hline
 adverbs&49&1&2&8&74&14&14&57&2&83&16&14&17&11&5&0&64&15&14&53&2&16&22\\\hline
 adjectives&80&0&7&31&88&51&2&116&6&71&56&84&49&29&1&2&216&58&112&122&6&14&37\\\hline
 pronouns&18&0&0&0&3&0&5&7&4&23&0&2&8&0&0&0&2&1&0&0&0&0&0\\\hline
 preposition&0&0&0&0&7&1&0&6&0&5&8&2&1&0&2&0&3&1&0&4&0&3&1\\\hline
 conjunction&1&0&0&0&10&1&0&13&0&9&1&0&2&0&0&0&8&0&0&4&0&2&0\\\hline
 interjection&1&1&0&0&0&0&0&1&0&0&0&0&0&0&1&0&0&0&0&6&0&0&0\\\hline
 \end{tabular}
}
\end{center}
\caption{Lakher(Mara): the second row represents the words and the later rows parts of 
speech of the Lakher(Mara) language against letters of roman alphabet in
the serial order}
\end{table}
\noindent
We describe how a graphical law is hidden within in 
the Lakher language, in this module.
We organise the module as follows. 
We explain our method of study in the section XXXII.
In the ensuing section, section XXXIII, we narrate our graphical results. 
Then we conclude about the existence of the graphical law 
in the section XXXIV. We acknowledge again in the section XXXV. 
In an adjoining appendix, section XXXVI, we give the datas for the Lakher 
language and datas for plotting comparators. We end up this module and 
the paper through a section, section XXXVII, on an elementary possibilty 
behind the law and a reference section.
\end{section}

\begin{section}{Method of study}
We take the Lakher dictionary,\cite{Lakher}.
Then we count the components,  
one by one from the beginning to the end, starting with different alphabets. 
We assort the alphabets according to their rankings. 
We take natural logarithm of both number of occurances of a component, denoted by $f$ 
and the respective rank, denoted by $k$. $k$ is a positive 
integer starting from one. 
Since each component has an 
alphabet, number of occurances initiating with it being very close to one or, 
one, we attach 
a limiting rank, $k_{lim}$, and a limiting number of occurances to each component. 
The limiting rank is just maximum rank 
(maximum rank plus one) if it is one (close to one) and 
the limiting number of occurance is one.
As a result both $\frac{lnf}{lnf_{max}}$ and $\frac{lnk}{lnk_{lim}}$ varies from
zero to one. Then we plot $\frac{lnf}{lnf_{max}}$ against $\frac{lnk}{lnk_{lim}}$.
\end{section}

\begin{section}{Results}
\noindent
We plot $\frac{lnf}{lnf_{max}}$ vs $\frac{lnk}{lnk_{lim}}$ 
for the components of the Lakher language(\cite{Lakher}). 
On the plots of words, nouns and adjectives we superimpose a 
Bragg-Williams curve in presence of little magnetic field, $c=0.01$,
as a comparator. For verbs and adverbs 
the Bragg-Williams line is used as visual best fit
\begin{figure}
\centering
\includegraphics[width=13cm,height=7cm]{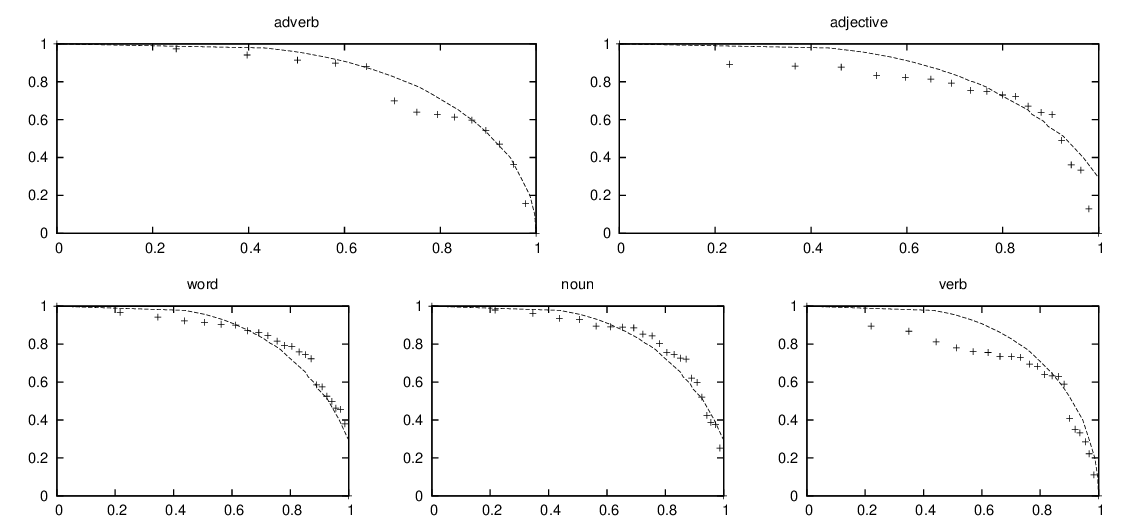}
\caption{Vertical axis is $\frac{lnf}{lnf_{max}}$ and horizontal 
axis is $\frac{lnk}{lnk_{lim}}$. The $+$ points represent the Lakher language 
components as represented by the titles. For words, nouns and adjectives fit curve 
is Bragg-Williams in presence of little magnetic field. For verbs and adverbs 
the Bragg-Williams line is used as comparator.}
\label{FigureL1}
\end{figure}
\begin{figure}
\centering
\includegraphics[width=13cm,height=7cm]{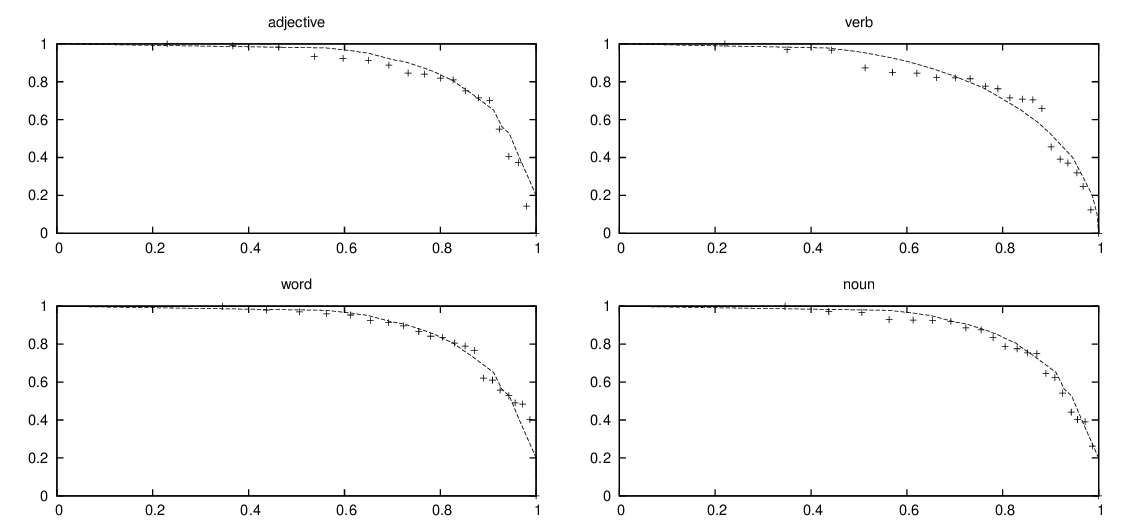}
\caption{The $+$ points represent the components as represented 
by the titles of the Lakher language.
Vertical axis is $\frac{lnf}{lnf_{nextnextmax}}$ 
and horizontal axis is $\frac{lnk}{lnk_{lim}}$. 
Fit curve is Bethe line for four neighbours for words, nouns and adjectives. 
For verbs it is Bragg-Williams line used as comparator 
and vertical axis is $\frac{lnf}{lnf_{nextmax}}$.}
\label{FigureL2}
\end{figure}
\noindent
We observe that the points belonging to words, nouns and adjectives are  
not matched with Bragg-Williams line, in presence of magnetic field,  
fully in fig.\ref{FigureL1}. We then ignore the letters with 
the highest and next highest number of occurances and redo the plot, 
normalising the $lnf$s with 
next-to-next-to-maximum $lnf_{nextnextmax}$, and starting from $k=3$. 
We see, to our surprise, that 
the resulting points almost fall 
on the Bethe line for $\gamma=4$ in fig.\ref{FigureL2}. Moreover, verbs are 
also not matched with Bragg-Williams line fully in fig.\ref{FigureL1}.
We then ignore the letters with 
the highest of occurances and redo the plot, 
normalising the $lnf$s with 
next-to-maximum $lnf_{nextmax}$, and starting from $k=2$. We notice that 
the resulting points almost fall on the Bragg-Williams line in fig.\ref{FigureL2}.
\noindent
Hence, we can put the five components of the Lakher languages 
in the following classification:
\begin{table}
\begin{center}
\resizebox{10cm}{.5cm}{
\begin{tabular}{|l|l|l|l|l|}\hline
word & noun & verb & adverb & adjective \\\hline
Bethe(4) &Bethe(4)& Bragg-Williams & Bragg-Williams  & Bethe(4)\\\hline      
\end{tabular}
}
\end{center}
\caption{classification of words and parts of speech of Lakher(Mara) language according to the underlying magnetisation curves}
\end{table}
\end{section}
\begin{section}{Conclusion}
From the figures (fig.\ref{FigureL1}-fig.\ref{FigureL2}), 
we observe that there are curves of magnetisation, behind words and four parts 
of speech of the Lakher language.
In other words, we conclude that the graphical law exists beneath the
Lakher language.

\noindent
Moreover, the associated correspondances are,\\
for adverbs, 
\begin{equation}
 \frac{lnf}{lnf_{maximum}} \longleftrightarrow \frac{M}{M_{max}},\nonumber\\
\end{equation}
for verbs,
\begin{equation}
 \frac{lnf}{lnf_{next-to-maximum}} \longleftrightarrow \frac{M}{M_{max}},\nonumber\\
\end{equation}
for words, nouns and adjectives 
\begin{equation}
\frac{lnf}{lnf_{next-to-next-to-maximum}}\longleftrightarrow \frac{M}{M_{max}},\nonumber\\
\end{equation}
and
\begin{equation}
lnk\longleftrightarrow T.\nonumber
\end{equation}

\end{section}

\begin{section}{Acknowledgement}
We would like to thank NEHU library
for allowing us to use (French, German and Abor-Miri) to English 
dictionaries; Chinese to English 
dictionary for contemporary usage,\cite{Chin} and Lakher to English 
dictionary,\cite{Lakher}.
We have used gnuplot for drawing the figures.
\end{section}

\newpage

\begin{section}{appendix}

\begin{table}
\begin{center}
\begin{tabular}{|l|l|l|l|l|l|l|}\hline
k & lnk & lnk/$lnk_{lim}$ & f & lnf & lnf/$lnf_{max}$ & lnf/$lnf_{next-next-max}$\\\hline
1 & 0 & 0        & 1064 & 6.97 & 1  & Blank\\\hline
2 & 0.69 & 0.217 & 853 & 6.75 & 0.968 & Blank\\\hline
3 & 1.10 & 0.346 & 712 & 6.57 & 0.943 & 1\\\hline
4 & 1.39 & 0.437 & 621 & 6.43 & 0.923 & 0.979\\\hline
5 & 1.61 & 0.506 & 588 & 6.38 & 0.915 &0.971\\\hline
6 & 1.79 & 0.563 & 550 & 6.31 & 0.905 &0.960\\\hline
7 & 1.95 & 0.613 & 526 & 6.27 & 0.900 &0.954\\\hline
8 & 2.08 & 0.654 & 434 & 6.07 & 0.871 &0.924\\\hline
9 & 2.20 & 0.692 & 404 & 6.00 & 0.861 &0.913\\\hline
10 & 2.30 & 0.723 & 361 & 5.89 & 0.845&0.896\\\hline
11 & 2.40 & 0.755 & 297 & 5.69 & 0.816&0.866\\\hline
12 & 2.48 & 0.780 & 252 & 5.53 & 0.793&0.842\\\hline
13 & 2.56 & 0.805 & 243 & 5.49 & 0.788&0.836\\\hline
14 & 2.64 & 0.830 & 198 & 5.29 & 0.759&0.805\\\hline
15 & 2.71 & 0.852 & 180 & 5.19 & 0.745&0.790\\\hline
16 & 2.77 & 0.871 & 154 & 5.04 & 0.723&0.767\\\hline
17 & 2.83 & 0.890 & 59 & 4.08 & 0.585&0.621\\\hline
18 & 2.89 & 0.909 & 55 & 4.01 & 0.575&0.610 \\\hline
19& 2.94&  0.925 &39 &3.66 & 0.525& 0.557\\\hline
20& 3.00& 0.943  &32 &3.47 & 0.498&0.528\\\hline
21& 3.04& 0.956  &25 &3.22 & 0.462&0.490\\\hline
22& 3.09& 0.972  &24 &3.18 & 0.456&0.484\\\hline
23& 3.14& 0.987  &14 &2.64 & 0.379&0.402\\\hline
24& 3.18&  1     &1  &0    & 0  & 0\\\hline
\end{tabular}
\end{center}
\caption{Lakher(Mara) words: ranking, natural logarithm, normalisations}
\end{table}
\clearpage
\begin{table}
\begin{center}
\begin{tabular}{|l|l|l|l|l|l|l|}\hline
k & lnk & lnk/$lnk_{lim}$ & f & lnf & lnf/$lnf_{max}$ & lnf/$lnf_{next-next-max}$\\\hline
1 & 0 & 0        & 596 & 6.39 & 1  & Blank\\\hline
2 & 0.69 & 0.217 & 522 & 6.26 & 0.980 & Blank\\\hline
3 & 1.10 & 0.346 & 470 & 6.15 & 0.962 & 1\\\hline
4 & 1.39 & 0.437 & 397 & 5.98 & 0.936 & 0.972\\\hline
5 & 1.61 & 0.506 & 381 & 5.94 & 0.930 &0.966\\\hline
6 & 1.79 & 0.563 & 306 & 5.72 & 0.895 &0.930\\\hline
7 & 1.95 & 0.613 & 299 & 5.70 & 0.892 &0.927\\\hline
8 & 2.08 & 0.654 & 295 & 5.69 & 0.890 &0.925\\\hline
9 & 2.20 & 0.692 & 287 & 5.66 & 0.886 &0.920\\\hline
10 & 2.30 & 0.723 & 233 & 5.45 & 0.853&0.886\\\hline
11 & 2.40 & 0.755 & 220 & 5.39 & 0.844&0.876\\\hline
12 & 2.48 & 0.780 & 169 & 5.13 & 0.803&0.834\\\hline
13 & 2.56 & 0.805 & 127 & 4.84 & 0.757&0.787\\\hline
14 & 2.64 & 0.830 & 118 & 4.77 & 0.746&0.776\\\hline
15 & 2.71 & 0.852 & 104 & 4.64 & 0.726&0.754\\\hline
16 & 2.77 & 0.871 & 100 & 4.61 & 0.721&0.750\\\hline
17 & 2.83 & 0.890 & 53 & 3.97 & 0.621&0.646\\\hline
18 & 2.89 & 0.909 & 46 & 3.83 & 0.599&0.623 \\\hline
19& 2.94&  0.925 &28 &3.33 & 0.521& 0.541\\\hline
20& 3.00& 0.943  &15 &2.71 & 0.424&0.441\\\hline
21& 3.04& 0.956  &12 &2.48 & 0.388&0.403\\\hline
22& 3.09& 0.972  &11 &2.40 & 0.376&0.390\\\hline
23& 3.14& 0.987  &5 &1.61 & 0.252&0.262\\\hline
24& 3.18&  1     &1  &0    & 0  & 0\\\hline
\end{tabular}
\end{center}
\caption{Lakher(Mara) nouns: ranking, natural logarithm, normalisations}
\end{table}
\clearpage
\begin{table}
\begin{center}
\begin{tabular}{|l|l|l|l|l|l|l|}\hline
k & lnk & lnk/$lnk_{lim}$ & f & lnf & lnf/$lnf_{max}$ & lnf/$lnf_{next-next-max}$\\\hline
1 & 0 & 0        & 530 & 6.27 & 1  & Blank\\\hline
2 & 0.69 & 0.220 & 274 & 5.61 & 0.895 & 1\\\hline
3 & 1.10 & 0.350 & 230 & 5.44 & 0.868 & 0.970\\\hline
4 & 1.39 & 0.443 & 162 & 5.09 & 0.812 & 0.967\\\hline
5 & 1.61 & 0.513 & 134 & 4.90 & 0.781 &0.873\\\hline
6 & 1.79 & 0.570 & 118 & 4.77 & 0.761 &0.850\\\hline
7 & 1.95 & 0.621 & 114 & 4.74 & 0.756 &0.845\\\hline
8 & 2.08 & 0.662 & 100 & 4.61 & 0.735 &0.822\\\hline
9 & 2.20 & 0.701 & 99 & 4.60 & 0.734 &0.820\\\hline
10 & 2.30 & 0.732 & 98 & 4.58 & 0.730&0.816\\\hline
11 & 2.40 & 0.764 & 78 & 4.36 & 0.695&0.777\\\hline
12 & 2.48 & 0.790 & 72 & 4.28 & 0.683&0.763\\\hline
13 & 2.56 & 0.815 & 55 & 4.01 & 0.640&0.715\\\hline
14 & 2.64 & 0.841 & 53 & 3.97 & 0.633&0.708\\\hline
15 & 2.71 & 0.863 & 52 & 3.95 & 0.630&0.704\\\hline
16 & 2.77 & 0.882 & 40 & 3.69 & 0.589&0.658\\\hline
17 & 2.83 & 0.901 & 13 & 2.56 & 0.408&0.456\\\hline
18 & 2.89 & 0.920 & 9 & 2.20 & 0.351&0.392 \\\hline
19& 2.94&  0.936 &8 &2.08 & 0.332& 0.371\\\hline
20& 3.00& 0.955  &6 &1.79 & 0.285&0.319\\\hline
21& 3.04& 0.968  &4 &1.39 & 0.222&0.248\\\hline
22& 3.09& 0.984  &2 &0.693 & 0.111&0.124\\\hline
23& 3.14&  1     &1 &0     & 0&0\\\hline
\end{tabular}
\end{center}
\caption{Lakher(Mara) verbs: ranking, natural logarithm, normalisations}
\end{table}
\clearpage
\begin{table}
\begin{center}
\begin{tabular}{|l|l|l|l|l|l|l|}\hline
k & lnk & lnk/$lnk_{lim}$ & f & lnf & lnf/$lnf_{max}$ & lnf/$lnf_{next-max}$\\\hline
1 & 0 & 0        & 216 & 5.38 & 1  & Blank\\\hline
2 & 0.69 & 0.230 & 122 & 4.80 & 0.892 & 1\\\hline
3 & 1.10 & 0.367 & 116 & 4.75 & 0.883 & 0.990\\\hline
4 & 1.39 & 0.463 & 112 & 4.72 & 0.877 & 0.983\\\hline
5 & 1.61 & 0.537 & 88 & 4.48 & 0.833 &0.933\\\hline
6 & 1.79 & 0.597 & 84 & 4.43 & 0.823 &0.923\\\hline
7 & 1.95 & 0.650 & 80 & 4.38 & 0.814 &0.913\\\hline
8 & 2.08 & 0.693 & 71 & 4.26 & 0.792 &0.888\\\hline
9 & 2.20 & 0.733 & 58 & 4.06 & 0.755 &0.846\\\hline
10 & 2.30 & 0.767 & 56 & 4.03 & 0.749&0.840\\\hline
11 & 2.40 & 0.800 & 51 & 3.93 & 0.730&0.819\\\hline
12 & 2.48 & 0.827 & 49 & 3.89 & 0.723&0.810\\\hline
13 & 2.56 & 0.853 & 37 & 3.61 & 0.671&0.752\\\hline
14 & 2.64 & 0.880 & 31 & 3.43 & 0.638&0.715\\\hline
15 & 2.71 & 0.903 & 29 & 3.37 & 0.626&0.702\\\hline
16 & 2.77 & 0.923 & 14 & 2.64 & 0.491&0.550\\\hline
17 & 2.83 & 0.943 & 7 & 1.95 & 0.362&0.406\\\hline
18 & 2.89 & 0.963 & 6 & 1.79 & 0.333&0.373 \\\hline
19& 2.94&  0.980 &2 &0.693 & 0.129& 0.144\\\hline
20& 3.00& 1  &1 &0 & 0&0\\\hline
\end{tabular}
\end{center}
\caption{Lakher(Mara) adjectives: ranking, natural logarithm, normalisations}
\end{table}
\clearpage
\begin{table}
\begin{center}
\begin{tabular}{|l|l|l|l|l|l|l|}\hline
k & lnk & lnk/$lnk_{lim}$ & f & lnf & lnf/$lnf_{max}$ \\\hline
1 & 0 & 0        & 83 & 4.42 & 1  \\\hline
2 & 0.69 & 0.249 & 74 & 4.30 & 0.973 \\\hline
3 & 1.10 & 0.397 & 64 & 4.16 & 0.941 \\\hline
4 & 1.39 & 0.502 & 57 & 4.04 & 0.914 \\\hline
5 & 1.61 & 0.581 & 53 & 3.97 & 0.898 \\\hline
6 & 1.79 & 0.646 & 49 & 3.89 & 0.880 \\\hline
7 & 1.95 & 0.704 & 22 & 3.09 & 0.699 \\\hline
8 & 2.08 & 0.751 & 17 & 2.83 & 0.640 \\\hline
9 & 2.20 & 0.794 & 16 & 2.77 & 0.627 \\\hline
10 & 2.30 & 0.830 & 15 & 2.71 & 0.613\\\hline
11 & 2.40 & 0.866 & 14 & 2.64 & 0.597\\\hline
12 & 2.48 & 0.895 & 11 & 2.40 & 0.543\\\hline
13 & 2.56 & 0.924 & 8 & 2.08 & 0.471\\\hline
14 & 2.64 & 0.953 & 5 & 1.61 & 0.364\\\hline
15 & 2.71 & 0.978 & 2 & 0.693 & 0.157\\\hline
16 & 2.77 & 1 & 1 & 0 & 0\\\hline
\end{tabular}
\end{center}
\caption{Lakher(Mara) adverbs: ranking, natural logarithm, normalisations}
\end{table}
\end{section}
\begin{section}{A plausible theoretical scenario} 
\noindent
To motivate the scenario, let us consider a collection of a 
set of people with similar interests. 
It may be a set of faculties or, a set of industrialists or, a set 
of students or, any other group.
At zero temperature there is no random kinetic energy between 
two parts. These two parts may be a topshot and the rest in case 
of faculties; a successful innovative individual and the rest in 
case of industrialists. Then all the rest club together against(very 
rarely for) the singularity. Their jealousies align perfectly along 
the same direction. In the opposite situation, when everyone is performing 
at more or, less equal level, or, "no super and sub" then all jealousies 
get cancelled. A highly congenial climate prevails. Interaction is high. 
Exchange is high. "Temperature" is high. Hence, we can reasonably consider 
a lattice of faculties or, industrialists, or, families etc with an Ising 
model built on that. The correspondance is as follows:
\begin{eqnarray}
jealousy\leftrightarrow spin, \nonumber \\
coupling\quad between \quad different\quad units \leftrightarrow J_{ij}, \nonumber \\
level\quad of\quad collective\quad activity \leftrightarrow temperature, \nonumber 
\end{eqnarray}
\noindent
Consequently, we get 
\begin{equation}
\frac{collective\quad jealousy}{number \quad of\quad units} \leftrightarrow magnetisation, \nonumber\\
\end{equation}
and a curve of magnetisation between 
$\frac{collective\quad jealousy}{number\quad of\quad units}$ \\
and  $level\quad of\quad collective\quad activity$. 

\noindent
Noun, verb, adverb, adjective and any word of a language is one or, other 
expression of jealousy. That's why we see underlying a language curves of 
magnetisation. People have time to be creative when activity is low. More 
number of words etc are generated then. If one set of people chooses one letter, 
another set of people chooses another letter, due to collective jealousy vs. 
collective jealousy at the zero temperature. 
That's why see ranking of letters in logarithmic scale is anlogous to temperature 
and different arrangement of letters along the ranking for different componenets 
or, languages.

\end{section}

\end{document}